\documentclass[useAMS,usenatbib]{mnras}

\usepackage{graphicx} 
\usepackage{times}
\usepackage{epsfig}
\usepackage{epstopdf}
\usepackage{amsfonts}
\usepackage{amsmath}
\usepackage{amsbsy}
\usepackage{bm}
\usepackage{url}
\usepackage{microtype}
\usepackage{rotating}
\usepackage{sidecap}
\usepackage{longtable}
\usepackage{float}
\usepackage{lscape}
\usepackage{rotating}
\usepackage{amssymb,xcolor}
\def\apj{ApJ}
\def\apjs{ApJ}
\def\mnras{MNRAS}
\def\aap{A\&A}
\def\apjl{ApJ}

\def\nat{Nature}

\def\apss{Ap\&SS}

\def\ssr{Space Science Reviews}

\usepackage{booktabs,caption}
\usepackage[flushleft]{threeparttable}
\usepackage{colortbl}
\newcommand{\SEM}[1]{{\color{cyan} #1}}

\definecolor{amethyst}{rgb}{0.6, 0.4, 0.8}
\definecolor{blue-violet}{rgb}{0.54, 0.17, 0.89}

\usepackage[T1]{fontenc}
\usepackage{ae,aecompl}

\usepackage{newtxtext,newtxmath}

\title[Radio-loudness in BHTs: evidence for an inclination effect]{Radio-loudness in black hole transients: evidence for an inclination effect}

\author[S.E.Motta]{Motta, S.E.$^{1}$, Casella, P.$^{2}$, Fender, R.$^{1}$\\
$^{1}$University of Oxford, Department of Physics, Astrophysics, Denys Wilkinson Building, Keble Road, OX1 3RH, Oxford, United Kingdom\\
$^{2}$ INAF, Osservatorio Astronomico di Roma, Via Frascati 33, I-00040, Monteporzio Catone, Italy\\
}
 
\date{Last updated 2018 xxxx xxxx;}

\pubyear{2018}

\begin{document}
\label{firstpage}
\pagerange{\pageref{firstpage}--\pageref{lastpage}}
\maketitle

\begin{abstract}
\noindent 
 
\noindent Accreting stellar-mass black holes appear to populate two branches in a radio:X-ray luminosity plane. We have investigated the X-ray variability properties of a large number of black hole low-mass X-ray binaries, with the aim of unveiling the physical reasons underlying the radio-loud/radio-quiet nature of these sources, in the context of the known accretion-ejection connection.
A reconsideration of the available radio and X-ray data from a sample of black hole X-ray binaries confirms that being radio-quiet is the more normal mode of behaviour for black hole binaries.
In the light of this we chose to test, once more, the hypothesis that radio loudness could be a consequence of the inclination of the X-ray binary. We compared the slope of the `hard-line' (an approximately linear correlation between X-ray count rate and rms variability, visible in the hard states of active black holes), the orbital inclination, and the radio-nature of the sources of our sample. We found that high-inclination objects show steeper hard-lines than low-inclination objects, and tend to display a radio-quiet nature (with the only exception of V404 Cyg), as opposed to low-inclination objects, which appear to be radio-loud(er). 
While in need of further confirmation, our results suggest that - contrary to what has been believed for years - the radio-loud/quiet nature of black-hole low mass X-ray binaries might be an inclination effect, rather than an intrinsic source property. This would solve an important issue in the context of the inflow-outflow connection, thus providing significant constraints to the models for the launch of hard-state compact jets.

\end{abstract}




\section{Introduction}\label{sec:intro}

Black-hole X-ray binaries (BHXBs) typically show a number of accretion states, defined based on their X-ray spectral and fast time-variability properties (see, e.g., \citealt{Belloni2016}). In the hard state (or low-hard state, LHS) the source X-ray spectrum is dominated by the emission from Compton up-scattering of soft seed photons, either produced in a cool geometrically thin accretion disc truncated at large radii, by synchrotron emission from hot electrons located close to the central black hole (e.g., \citealt{Poutanen2014}), or by synchrotron emission from a compact jet (e.g. \citealt{Markoff2010}). In this state, the power density spectra (PDS) of the source emission show high variability (rms $\gtrsim 20\%$) and - at relatively high fluxes - the presence of quasi-periodic oscillations (QPOs). In the hard state the X-ray count rate (which is a good proxy of the luminosity in this state) forms a tight correlation with the X-ray rms, which is referred to as \textit{hard-line} \citep{Munoz-Darias2011}.
In the soft state (or high-soft state, HSS) the emission is typically dominated by thermal emission from a geometrically thin, optically thick accretion disc extending down to the innermost stable circular orbit around the BH (\citealt{Shakura1973}). The fast time variability in the X-rays is low in this state (rms $\lesssim 5\%$), and the PDS typically show only a weak power law noise component and (rarely, again at relatively high fluxes) weak QPOs.
In between these two states lie the so-called \textit{intermediate} states (hard and soft intermediate states, HIMS and SIMS, respectively), where the energy spectra show the properties of both the LHS and the HSS evolving smoothly during the transitions between the LHS and HSS, but where the most striking changes in the time domain are observed (see, e.g., \citealt{Nespoli2003}).

\smallskip

\cite{Fender2001} established that all BHXBs in their hard states show flat radio spectrum emission ($\alpha \sim 0$, where the radio spectrum is described by $F \propto \nu^{\alpha}$), which is regarded as the signature of jets, and that is interpreted as synchrotron radiation from accelerated electrons in a magnetic field. In contrast, very little to no core radio emission is typically observed in the soft state (\citealt{Fender1999}, \citealt{Russell2011}), which led to the conclusion that a radio jet is absent in such states. However, based on the internal shock jet model (\citealt{Malzac2013}), \cite{Drappeau2017} has recently shown that the predicted radio emission of a jet in the soft state is consistent with the current soft-state upper limits. This implies that a compact ``dark'' jet could persist in the soft state with no significant differences in the kinetic power with respect to the hard-state jets.

Fundamental to the study of the connection between the accretion and the jet production mechanism have been the (quasi) simultaneous X-ray and radio observations of BHXBs, particularly in the hard state, during which sources typically evolve rather slowly compared to other states, and follow clear correlations.
Particularly well-known is the non-linear correlation between the X-ray and the radio luminosity, originally observed in GX 339-4 (\citealt{Hannikainen1998}) and then found in many other BHXBs by several authors in the last two decades. While this correlation was initially considered universal (\citealt{Gallo2003}) - which led to a number of works on the fundamental plane of black hole activity (see, e.g., \citealt{Merloni2003}, \citealt{Falcke2004}) - it more recently became clear that some BHXBs were considerably less luminous in the radio band than the canonical sources such as GX 339-4 (e.g., \citealt{Coriat2011}), thus populating a second track in the radio:X-ray plane.

The origin of the two tracks have so far been explained in terms of differences in the jet magnetic field (e.g., \citealt{Casella2009}), differences in the accretion flow radiative efficiency (\citealt{Coriat2011}), or differences in the contribution from an additional inner accretion disc (\citealt{Meyer-Hofmeister2014}). 
Attempts to identify the underlying cause of the \textit{radio-loud} and \textit{radio-quiet} nature of BHXBs have been, however, unsuccessful for years. \cite{Dincer2014} reported the first difference between the two populations, in addition to their different position in the radio:X-ray plane, based on a small sample of BHXBs (7, between radio-quiet and radio-loud). These authors found that radio-loud sources appear to have higher average X-ray rms variability then radio-quiet sources. Recently, \cite{Espinasse2018}, based on a larger sample of sources (17 in total) found that radio-loud BHXBs have significantly different radio spectra than the radio-quiet BHXBs, with the radio-loud sources characterized by a spectral slope $\alpha$ normally distributed around $\alpha$ = +0.2, and the radio-quiet sources characterized by a distribution with mean $\alpha$ = - 0.2. 
\cite{Espinasse2018} explored different possible explanations for such a difference and concluded that, once a dependence on the inclination angle $i$ of the jet to the line of sight 
is ruled out, a difference in the physics of the core jets would be the only remaining plausible responsible for the radio-loudness or quietness of a given source.

In this paper we investigate the X-ray variability and luminosity properties of radio-loud and radio-quiet sources in a large number of BHXBs observed by the Rossi X-ray Timing Explorer (RXTE) satellite. We find indications that - contrary to what has been concluded in the past - the radio-loud/quiet nature of BHXBs is a function of the orbital inclination. Similarly, we find that the slope of the \textit{hard-line} is to date the most reliable X-ray indication of the radio-loudness of a given source. 

\begin{figure*}
\centering
\includegraphics[width=1.0\textwidth]{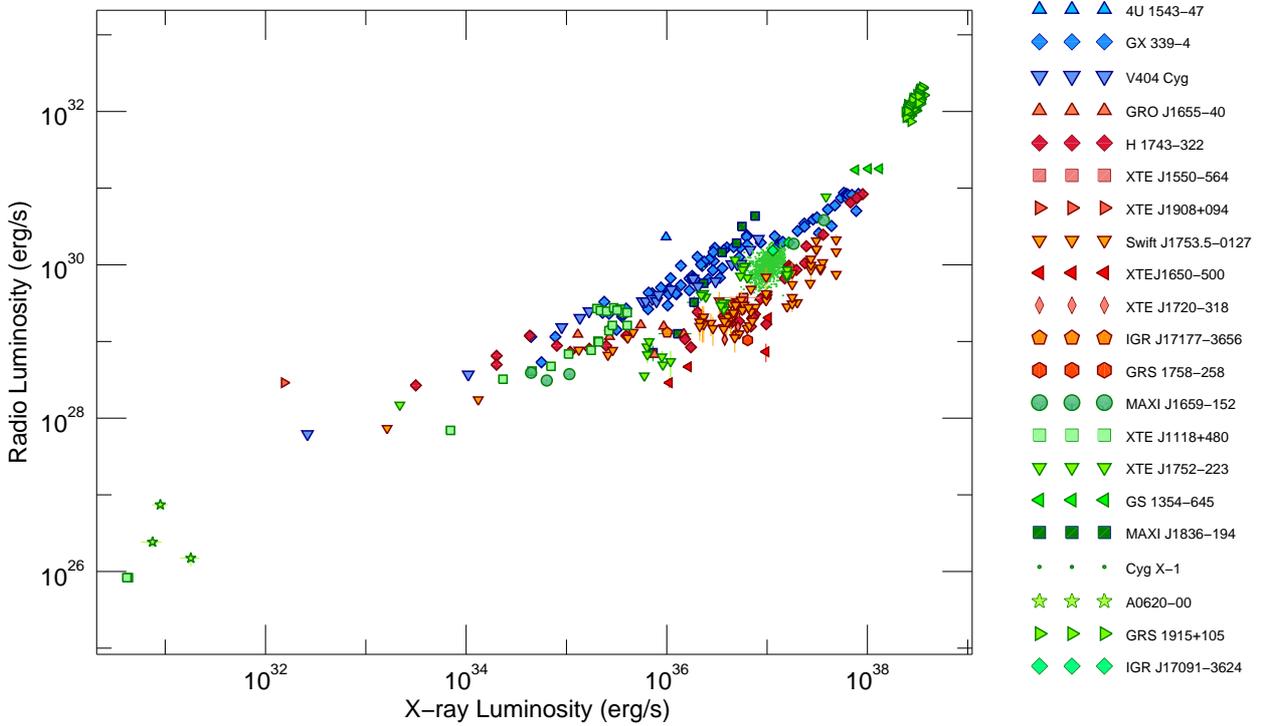}
\caption{Radio monocromatic luminosity versus X-ray luminosity correlation for all the sources of sample R. Points in different shades of blue and reds mark the sources generally classified as radio-loud and radio-quiet, respectively (see legend for details), while greens are used for radio-unclassified/undefined sources. Please note that in Fig. 1 of Espinasse \& Fender 2018 the radio luminosities are erroneously shifted by a factor $\approx 10^{10}$ that does not affect the results of that paper. }
\label{fig:LRLx}
\end{figure*}

\section{Sample selection and data Analysis}

We first collected information on the radio-loudness, X-ray variability and orbital inclination of as many BHXBs as possible. The sample selection procedure is described below, and Table \ref{tab:log} summarises the properties of all the sources we considered in our analysis. For the sake of completeness, we report in this table all the sources for which we could determine {\it at least one} of the aforementioned three properties.

\subsection{Radio-loudness}
We searched the literature for information on the radio-loudness of known BHXBs. We collected radio and X-ray simultaneous measurements from 21 sources in total, which constitute our \textit{\textbf{sample R}} (radio-selected, see Tab. \ref{tab:log}, column \emph{sample}). The radio luminosities L$_{\rm R}$ used in the following are obtained by converting the flux densities in luminosity using the distances reported in Tab. \ref{tab:log}, and integrating them up to 5GHz assuming a flat radio spectrum. Following \cite{Espinasse2018}, the X-ray luminosities L$_{\rm X}$ are obtained from X-ray fluxes in the (1-10) keV energy band, either directly measured or inferred based on the fluxes from different bands taking a photon index of $\approx$1.6 (taken from the literature, or assumed when not available).

Sample R includes: 4U 1543--47, GX 339--4, MAXI J1836--194, GRO J1655--40, H1743--322, XTE J1550--564, Swift J1753.5--0127, XTE J1650--500, XTE J1908+094, XTE J1720--318, MAXI J1659-152, XTE J1118+480, XTE J1752--322, GS1354-645, V404 Cyg,  IGR J17177--3656, Cyg X--1, GRS1758--258, A0620-00, GRS 1915+105, IGR J17091-3624 (see Tab. \ref{tab:log}).
In Fig. \ref{fig:LRLx} we show the radio:X-ray correlation for all the sources of sample R, where we plot the radio luminosity versus the X-ray luminosity. Different symbols/colours mark different sources (indicated in the plot legend). 

In order to define the radio-loudness of all sources in sample R, we first note that the well-known dual behaviour in Fig. \ref{fig:LRLx} is limited to the luminosity range (roughly $10^{35.5} < L_{X} < 10^{37.5} erg/s$) in which the sources populate two roughly distinct branches \citep[see][for a statistical analysis of this apparent dualism]{Gallo2014}. Thus, the radio-loudness of sources populating the diagram \textit{outside} this luminosity range cannot be defined. This is the case, for example, of XTE J1118+480, which has been historically classified as radio-loud, even though it never reached the radio and X-ray luminosities where radio-quiet sources appear to deviate from radio-loud ones in the radio:X-ray plane. The same is true for A0620-00, which was only observed in the hard state at very low luminosities. GS 1354--645, GRS 1915+105,  and IGR J17091--3624, instead, all sit at high radio and X-ray luminosities, where the radio-quiet and radio-loud branches meet. We note that Cyg X--1 was classified as radio-loud by \cite{Espinasse2018} because its spectral index was positive for the three observations that these authors considered, but its classification is uncertain.
XTE J1752-322, suffers from a large distance uncertainty, which prevents a solid radio classification. In addition, this source's radio:X-ray track crosses both the radio-loud and radio-quiet branches, and is not obviously consistent with either. The same is true for MAXI J1836-294, which was found to mostly lie on the radio-loud branch, but not following the trend of either radio-loud or radio-quiet sources \citep{Russell2015}, even considering the large uncertainty on its distance. 

Based on the above, the radio classification of XTE J1118+480, GS 1354--645, GRS 1915+105, XTE J1752--322, MAXI J1836-294, A0620-00 and Cyg X--1  remains dubious. Among the remaining sources in sample R, the radio-loud sources are only 4U 1543--47, GX 339--4 and V404 Cyg, while the radio-quiet sources are GRO J1655--40, H1743--322, XTE J1550--564, Swift J1753.5--0127, XTE J1650--500, XTE J1908+094, XTE J1720--318, IGR J17177--3656 and GRS1758--258.

\subsection{X-ray variability}
Next, we searched the RXTE archive and we selected all the BHXBs that showed a clear LHS (see, e.g., \citealt{Belloni2016} for details on the state classification). 
We selected only observations with average count rate per PCU higher than 5 counts/s and exposure larger than $\sim 700 s$ (i.e., five 128s-long time intervals, in order to combine at least 5 PDS to produce a good S/N average PDS, see below) which allowed us to obtain a solid measurement of the rms. 
For each RXTE observation we produced PDS every 128s in the $\approx\,2-100$ keV energy band (0-249 Absolute channel range) with a Nyquist frequency of 1024Hz, which we averaged to obtain one PDS per observation. From each average PDS we then measured the fractional rms in the 0.01-64 Hz frequency band.

We excluded from the subsequent analysis a few sources, for which we did not find observations meeting our selection criteria. In the case of GRS 1758--258 and V404Cyg, the RXTE data showed average count rates lower than our threshold of 5 counts/s. In the case of IGRJ17091--3624, Cyg X--1, GRS 1915+105 and 4U1630--47, we could not unambiguously identify a hard state consistently with the other sources. In the case of 4U1630--47 the high absorption in the direction of the source prevents selecting unambiguously hard state observations. 
For GRS 1915+105 things are further complicated by the fact that the source is always very variable in both the X-ray and the radio band. Hence, it is difficult to measure the rms in a way that would make it comparable with that from other systems. 
We also note that RXTE did not observe XTEJ1859+226 and MAXIJ1543--564 during their hard states.

All the remaining sources constitute our  \textit{\textbf{sample X}} (X-ray selected, see Tab. \ref{tab:log}, column \emph{sample}), in which the radio-loud sources are: 4U1543--47 and GX339--4; the radio-quiet sources are: GROJ1655--40, H1743--322, XTEJ1550--564, Swift J1753.5--0127, XTEJ1650--500, XTEJ1908+094, and XTEJ1720--318; the unclassified sources are: MAXIJ1659--152, XTEJ1748--288, XTEJ1817--330, XTEJ1118+480, XTEJ1752--223, MAXIJ1836--194, GS1354--645 and Swift J1842.5-1124.

For all the sources of sample X, we also measured the count rate in the 2-15 keV energy band, which we then rescaled by the sources distance using the values reported in the literature (see Tab. \ref{tab:log}). 
Finally, we used the fractional rms values and the rescaled count rates to build the rms-intensity diagrams (\citealt{Munoz-Darias2011}) for each source of sample X, which - given our selection criteria - only contain the hard-line(s). 

\subsection{Inclination }\label{Sec:inclination}

Finally, we searched the literature for information on the orbital inclination of known BHXBs, including those in our samples R and X. In all the cases where the inclination of a given source has not been directly measured through optical photometry, we inferred an approximate inclination based on other source properties known to depend on the inclination. These are: (i) the presence/absence of absorption dips or eclipses in the sources light curves (see, e.g., \citealt{Motta2015}, appendix A);  (ii) the presence/absence of absorption features ascribed to the launch of equatorial winds (\citealt{Ponti2012}); (iii) the shape of the HID (\citealt{Munoz-Darias2013}); (iv) the amplitude of low frequency QPOs (\citealt{Motta2015}). 
We classify as "high-inclination" sources those whose orbital axis is close to the perpendicular to the line of sight, with an inclination angle $i$ above  $\approx$65$^{\circ}$. 

A word must be spent on a few sources of the sample. \cite{Neustroev2014} showed that the inclination of Swift J1753.5--0127 is strictly larger than 40$^{\circ}$ based on the strong photometric and spectroscopic orbital variability seen in Optical and UV. The absence of eclipses and dips, however, imposes also an upper limit to the inclination of the system. In general, such upper limit would be $\approx$70$^{\circ}$, considering the size of the accretion disc and of the disc bulge (see, e.g., \citealt{Motta2015}, Appendix A). However, in case of very small accretion discs (expected for short orbital periods sources), the size of the bulge is what determines the minimum inclination angle that would provoke absorption dips. The bulge can have an angular size as small as $\approx$10$^{\circ}$ (\citealt{White1982a}). Given the very short orbital period of Swift J1753.5--0127, the above implies that the upper limit to the inclination is $\approx$80$^{\circ}$.

The orbital inclination of MAXI J1659-152 has not yet been measured in quiescence. While the HID shows the typical spikes that are suggestive of an high inclination source (see, e.g., \citealt{Munoz-Darias2013}), the overall  amplitude of the type-C QPOs observed in this source, clearly points to a more intermediate inclination  (see \citealt{Motta2015}, Fig. 6). We therefore conservatively assumed an intermediate inclination for this source, ranging from 30$^{\circ}$ to 70$^{\circ}$.

We note that in \cite{Motta2015} we assumed an inclination of $<49^{\circ}$ for XTE J1752-322, based on the inclination of the radio jet (see \citealt{Miller-Jones2011}), assuming that the radio jet is perpendicular to the accretion disc. Such an inclination is, however, relative to the \emph{inner} disk, and might be different from the orbital inclination. Since the orbital inclination can be measured directly in a number of sources, while the inner disc inclination can only be measured in a few systems and not directly, for the sake of consistency here we consider only orbital inclinations. Therefore, we assume for XTE J1752-322 a conservative orbital inclination of $5-60^{\circ}$ based on the same criteria we applied to the other sources we considered. The same reasoning applies to XTE J1908+094, for which the jet-to the line of sight angle has been estimated to be $>79^{\circ}$ (\citealt{Rushton2017}). For this source we therefore assumed an inclination between 65$^{\circ}$ and 80$^{\circ}$ (with the upper limit set based on the lack of X-ray eclipses).

\begin{table*}
\caption{The sources considered in this work. We coloured in blue table cells containing a \textit{radio-loud} statement, low inclination values/statements, and a steep hard-line slope value. Similarly, we coloured in red cells containing a \textit{radio-quiet} statement, high inclination values/statement (where high is $> $65-70$^{\circ}$ and low is $< $60$^{\circ}$), or a shallow hard-line slope value (where steep corresponds to $>$ 1.07). 
We marked in purple cells containing inclination values that qualify as intermediate or that are affected by uncertainties large enough to prevent a definitive classification. For each source we give different references: the reference in italic refers to the radio information, the second (and third and fourth, when present) references contain information on the distance and inclination of a given system, respectively, unless either the inclination or distance is unknown, or unless the same reference provides both distance and inclination. }\label{tab:log}
  \begin{threeparttable}

 \centering
\begin{tabular}{p{0.1cm} p{0.5cm} p{2.2cm} | p{1.3cm} | p{1.2cm} | p{1.2cm} | p{1.7cm} | p{4.5cm} p{0.3cm} p{0.5cm} }

\arrayrulecolor{black} \hline																						
\arrayrulecolor{black} \hline																						
  & Sample		&	Source			&	Distance	&	Radio $^{\dagger}$				&		Inclination						&		Hard-line slope						&	Comments							&	Ref.		&	RXTE	\\
  &     		&					&	[kpc]		&	Loud/quiet						&		[Degrees]						&		(log-log fit)						&										&				&	Obs.*	\\
\arrayrulecolor{black} \hline												
\arrayrulecolor{black} \hline												
1  & R, X		&	4U1543--47		&	7	-	8	& \cellcolor{blue!15}Loud			&	\cellcolor{blue!15}	19	 -	22		&	\cellcolor{blue!15}	1.308$\pm$0.003 	&											& (a)			&	22	\\
\arrayrulecolor{gray} \hline																					
2  & R, X		&	GX339--4		&	7.5 - 9.3	& \cellcolor{blue!15}Loud			& \cellcolor{blue!15}	40	-	60		& \cellcolor{blue!15}	1.360$\pm$0.003		&	Different hard-line slope on out-		&	(b)			&	532	\\
   & 			&					&				& \cellcolor{blue!15}				& \cellcolor{blue!15}					& \cellcolor{blue!15}	1.272$\pm$0.003		&		burst rise and decays (different 	&				&		\\
   &			&					&				& \cellcolor{blue!15}				& \cellcolor{blue!15}					& \cellcolor{blue!15}	1.155$\pm$0.001		&		hard-lines for different decays).	&				&		\\
\hline																										
3  & R, X		&	GROJ1655--40	&	3	-	3.4	& \cellcolor{red!15}Quiet			& \cellcolor{red!15}	69	-	71		& \cellcolor{red!15}	0.971$\pm$0.001		&											&	(c) 			&	74	\\
\hline																					
4  & R, X		&	H1743--322		&	7.7	-	9.3	& \cellcolor{red!15}Quiet			& \cellcolor{red!15}	72	-	78		& \cellcolor{red!15}	0.804$\pm$0.001		&											&	(d) 			&	225	\\
\hline																					
5  & R, X		&	XTEJ1550--564	&	4	-	5	& \cellcolor{red!15}Quiet			& \cellcolor{red!15}	71	-	78.5	& \cellcolor{red!15}	0.995$\pm$0.001		&Different hard-line slope on out-			&	(e) 			&	111	\\
	&   		&					&				&	\cellcolor{red!15}				&	\cellcolor{red!15}					& \cellcolor{red!15}	0.957$\pm$0.001		& burst rise and decay. 					&				&		\\
\hline

6  & R, X		&	XTEJ1908+094	&	7.8 - 13.6		& \cellcolor{red!15}Quiet		&	\cellcolor{red!15} 65-80			& \cellcolor{red!15}	1.063$\pm$0.01 		&	$i$ inferred from jet-angle from line of sight. 		&	(f) 			&	202	\\
\hline																					
																																												
7  & R, X		&SwiftJ1753.5--0127	&	4 - 8	    & \cellcolor{red!15}Quiet			& \cellcolor{blue-violet!45}	40	-	80	& \cellcolor{red!15}	0.976$\pm$0.001	&	Short orbital period implies i $\lesssim$ 80 &	(g) 		&	281	\\
\hline																					
8  & R, X		&	XTEJ1650--500	&	2	-	3.3	& \cellcolor{red!15}Quiet			& \cellcolor{blue-violet!45}	47	-	70	& \cellcolor{red!15}	1.044$\pm$0.003	&											&	(h) 		&	19	\\																						
\hline																				
																																					
\hline																													
9  & R, X		&	XTEJ1720--318	&	3 - 10			& \cellcolor{red!15}Quiet		&		 -								& \cellcolor{red!15}	0.990$\pm$0.004 	&											&	(i) 			&	34	\\

\hline																					

\hline																																												
10  & R, X		&	MAXIJ1659--152	&	?				&	{\it undef.}				& \cellcolor{blue-violet!45} 30 - 70	& \cellcolor{blue!15}	1.094$\pm$0.003		&	$i$ inferred from type-C QPO amplitude	&	(j) 			&	8	\\
\hline																					
11  & X			&	XTEJ1748--288	&	?				&	?							& \cellcolor{red!15}	High			& \cellcolor{red!15}	$0.607\pm0.003$		&	$i$ inferred from presence of dips		&	(k) 			&	13	\\
\hline																					
12  & X			&	XTEJ1817--330	&	1	-	5		&	?							& \cellcolor{blue!15}	5-60				& \cellcolor{blue!15}	1.24$\pm$0.01	 	&	$i$ inferred from HID shape, no dips, no wind-related absorption &	(l) (m)	&	22	\\
\hline																					
13  & R, X		&	XTEJ1118+480	&	1.62 - 1.82		&	{\it undef.}				& \cellcolor{red!15}	68	-	79		& \cellcolor{red!15}	$1.064\pm0.006$	 	& Different hard-line slopes above			&	(m) 		&	79	\\
	&    		&	             	&	            	&	 					   		& \cellcolor{red!15}	           		& \cellcolor{red!15}	$0.591\pm0.004$	 	&	and below 60 counts/s.					&				&		\\

\hline																					
14  & R, X		&	XTEJ1752--223	&	3.5	- 8			&	{\it undef.}				& \cellcolor{blue!15} 5-60				& \cellcolor{blue!15}	1.077$\pm$0.003	 	&		$i$ inferred from HID shape, no dips, no wind-related absorption &	(n) (o)	&	61	\\
\hline																					
15  & R, X		&	GS1354--645$^{\ddagger}$		&	25	- ?			&	{\it undef.}				& \cellcolor{red!15}	73	-	77		& \cellcolor{red!15}	1.015$\pm$0.005  	&											&	(o)			&	7	\\

\hline																					

16  & X			&Swift J1842.4-1124 &	 ?				&	 ?							& \cellcolor{red!15}High 				&  \cellcolor{red!15}	0.996$\pm$0.004     & $i$ inferred from HID shape. Possible dips.&	- 			&	14	\\

\hline																		
17  & R, X		&	MAXIJ1836--194	&	4	-	10		& \SEM{\emph{undef.}}			& \cellcolor{blue!15}	4	-	15		& \cellcolor{blue!15}	1.173$\pm$0.003		&											&	(p)			&	62	\\

\hline																				
																							
\hline																

18  & R			&	V404 Cygni		&	2.25 - 2.53		& \cellcolor{blue!15}Loud	    & \cellcolor{red!15}	66	-	70		&		 -									&	Extremely High local absorption and variable inclination jet	&	(q)	&	-	\\
\hline																					
19  & R			&	IGRJ17177--3656	&	8				& \cellcolor{red!15}Quiet	    &	\cellcolor{red!15}	High			&		 -									& $i$ inferred from very high local absorption.						&	(r)	&	 2	\\
\hline																					
																							
\hline
20  & R			&	CygX--1			&	1.75 - 1.98		& {\it undef.}	&		\cellcolor{blue!15} 26 - 28						&		 -									&																	&	(s)	&	 -	\\
\hline																					
21  & R			&	GRS1758--258	&	8.5				& \cellcolor{red!15}Quiet		&		-								&		 -									&																	&	(t)	&	 -	\\
\hline																					
22  & -			&	4U1630-47		&	?				&	?							& \cellcolor{red!15}	High			&		 -									& $i$ inferred from HID shape and presence of dips					&	(u)	&	 -	\\
\hline																					
23  & R			&	A0620--00		&	0.8 - 1.6		&	{\it undef.}				& \cellcolor{blue!15}	50	-	52		&		 -									&																	&	(v)	&	 -	\\
\hline						

24  & -			&	XTEJ1859+226	&	4.4	-	14		&	?							& \cellcolor{red!15}	60	-	90		&		 -									&																	&	(w)	&	 -	\\
\hline																					
25  & -			&	MAXIJ1543-564	&		?			&	?							& \cellcolor{red!15}	High			&		 -									&	$i$ inferred from HID shape (spikes) and based on type-C QPO amplitude. &	(x)	&	 -	\\
\hline																					
26  & -			&	4U1957+115		&		?			&	?							& \cellcolor{blue!15}	Low				&		 -									&	$i$ inferred from HID shape, no dips, no wind-related absorption features			&	(y)	&	 -	\\
\hline																					
27  & R			&	GRS 1915+105	&	7 - 10.6		&		{\it undef.}			&  	\cellcolor{blue!15} 55 - 65			&		 -									&	$i$ based inferred based on jet-line of sight angle	&	(z)	&	 -	\\
\hline																					
28  & R			&	IGR J17091-3624	&	11 - ?			&		{\it undef.}			&  	-									&		 -									&																		&	(aa)	&	 -	\\
\hline																					
\arrayrulecolor{black} \hline																				

\arrayrulecolor{black} \hline


\end{tabular}
\begin{tablenotes}
\footnotesize
\item (a) \emph{\cite{Gallo2003}}, \cite{Park2004}, \cite{Orosz2002};  
(b) \emph{\cite{Corbel2013}}, \cite{Parker2016}, \cite{Munoz-Darias2008};  

(c) \emph{\cite{Gallo2003}, \cite{Calvelo2010}, \cite{Coriat2011}}, \cite{Hjellming1995}, \cite{Kuulkers1998}, \cite{Greene2001};  
(d) \emph{\cite{Coriat2011}}, \cite{Steiner2012a}, \cite{Corbel2005}, \cite{Homan2005b};  
(e) \emph{\cite{Gallo2003}}, \cite{Orosz2011}, \cite{Homan2001};  
(f) \emph{\cite{Miller-Jones2011}}, \cite{Curran2015}, \cite{Rushton2017};  
(g) \emph{\cite{Soleri2010},\cite{Rushton2016}, \cite{Plotkin2017}}, \cite{CadolleBel2007}, \cite{Neustroev2014};  
(h) \emph{\cite{Corbel2004}}, \cite{Homan2006}, \cite{Orosz2004};  
(i) \emph{\cite{Brocksopp2005}}, \cite{Chaty2006};  
(j) \emph{\cite{Jonker2012}}, \cite{Kuulkers2013}, \cite{Motta2015};  
(k) \cite{Motta2015};  
(l) \cite{Sala2007}, \cite{Ponti2012};  
(m) \emph{\cite{Gallo2014}}, \cite{Gelino2006}, \cite{Khargharia2013};  
(n) \emph{\cite{Brocksopp2013}}, \cite{Ratti2012}, \cite{Miller-Jones2011};  
(o) \emph{\cite{Gallo2003}}, \cite{Casares2009}, \cite{El-Batal2016};  
(p) \emph{\cite{Russell2015}}, \cite{Russell2014}, \cite{Russell2014};  

(q) \emph{\cite{Corbel2008}}, \cite{Miller-Jones2009}, \citealt{Khargharia2010};  
(r) \emph{\cite{Paizis2011}};  
(s) \emph{\cite{Gallo2003}}, \cite{Reid2011}, \cite{Orosz2011a}, \cite{Brown2018};  
(t) \emph{\cite{Gallo2003}}, \cite{Keck2001}, \cite{Main1999};  
(u) \cite{Kuulkers1998}, \cite{Tomsick1998};  
(v) \emph{\cite{Gallo2006}, \cite{Dincer2018}}, \cite{Jonker2004}, \cite{Cantrell2010};  
(w) \cite{Hynes2002}, \cite{Corral-Santana2011};  
(x) \cite{Munoz-Darias2013}, \cite{Motta2015};  
(y) \cite{Ponti2012};  
(z) \emph{\cite{Rushton2010}}, \cite{Reid2014}; 
(aa) \emph{\cite{Rodriguez2011}}, \cite{Altamirano2011}.

\item ${\dagger}$ We mark as {\it undefined} the radio nature of BHXBs for which simultaneous radio and X-ray flux measurements are available, but where the radio loudness cannot be determined unambiguously (see Sect.~2.1). A question mark identifies BHXBs for which there are no (quasi-)simultaneous radio/X-ray measurements in the LHS.
\item * Number of RXTE Observations taken during the hard state considered in this work. 

\item ${\ddagger}$ We note that \cite{Gandhi2018} reports a distance of $\sim$kpc for GS1354--645, based on \emph{Gaia} parallax measurements. This would essentially place it at the very low-luminosity end of the radio:x-ray correlation, rather then at the high luminosity end of it, thus not affecting our results. 
\end{tablenotes}
\end{threeparttable}

\end{table*}

\section{Results}

The first result of our analysis of the radio:X-ray flux correlation of BHXBs, is a re-definition of the radio-loud/quiet nature of some known sources. While in the literature it is often found that all sources not deviating from the standard track are considered "radio-loud", we show here that for at least three sources this is inappropriate, as their radio and X-ray luminosities never reach the luminosity range where the correlation broadens substantially, forming the radio-loud and radio-quiet tracks. Thus, we classify the radio-loudness of these three sources as "undefined", which effectively equalizes them to those sources for which no (quasi-)simultaneous radio and X-ray measurements have been reported so far.

\subsection{X-ray variability properties}

Given the recent marginal evidence, reported by \cite{Dincer2014}, on the radio-loud sources being on average more variable in X-rays, we explored this correlation in sample X.
Once the radio nature of a source (loud or quiet) is known, the dependence of the properties of such a source on the radio nature can be investigated without the need for a simultaneous radio measurement, which allows us to use all the RXTE observations available for all sources in our sample X. 

We inspected the X-ray count rate and fractional rms bivariate distributions for the sources of sample X, dividing them in radio-loud, radio-quiet and (radio-)unclassified sources. Figure \ref{fig:rms_vs_rate_cont_map} shows the X-ray count rate and rms bivariate distributions for all the radio-loud sources (blue), all the radio-quiet sources (red) and all the radio unclassified sources (green). In the left panel of Fig. \ref{fig:rms_vs_rate_cont_map} the radio-loud sample includes GX 339-4, while in the right panel GX 339-4 is excluded, which reduces the radio loud sample to 4U1543-47 only. The histograms plotted along the axis of the main panels (colour-coded according to the above criteria) correspond to the marginal count rates (y-axis) and rms (x-axis) distributions. 
Similar plots for each individual source of sample X are shown in App. \ref{App:fig}, fig. \ref{fig:individual_maps} (where, for clarity, we show colour maps instead of contour plots). We remind the reader that the hard-state X-ray spectra of a given source typically show only normalization changes in the RXTE/PCA energy band (4-20 keV), with minor slope variations, thus the count rate is a good proxy of the X-ray flux.
 
From Figure \ref{fig:rms_vs_rate_cont_map} it is evident that, while overlapping significantly, the bivariate distributions for radio-loud and radio-quiet sources are significantly different. The radio-quiet sample is not dominated by any single source, even though most of the rms and count rate measurements come from four sources out of nine. The radio-loud sample, instead, is extremely small (two sources), and it is obviously dominated by GX 339-4 (532 observations out of 554). It follows that based on our data we cannot establish whether there is a significant difference between the bi-variate distributions of radio-loud and radio-quiet sources in general, or whether the difference apparent in Fig. \ref{fig:rms_vs_rate_cont_map} is rather to be ascribed to GX 339-4 being dissimilar to the sources of the radio-quiet sample. Interestingly, and somewhat unexpectedly, the bivariate distributions of each individual source is in most cases significantly different from the combined distribution of the remaining sources of the same radio-class. The same is true for the rms and rate distributions, and for each radio-class.

\subsection{Hard-line fits}

\begin{figure*}
\centering
\includegraphics[width=0.48\textwidth]{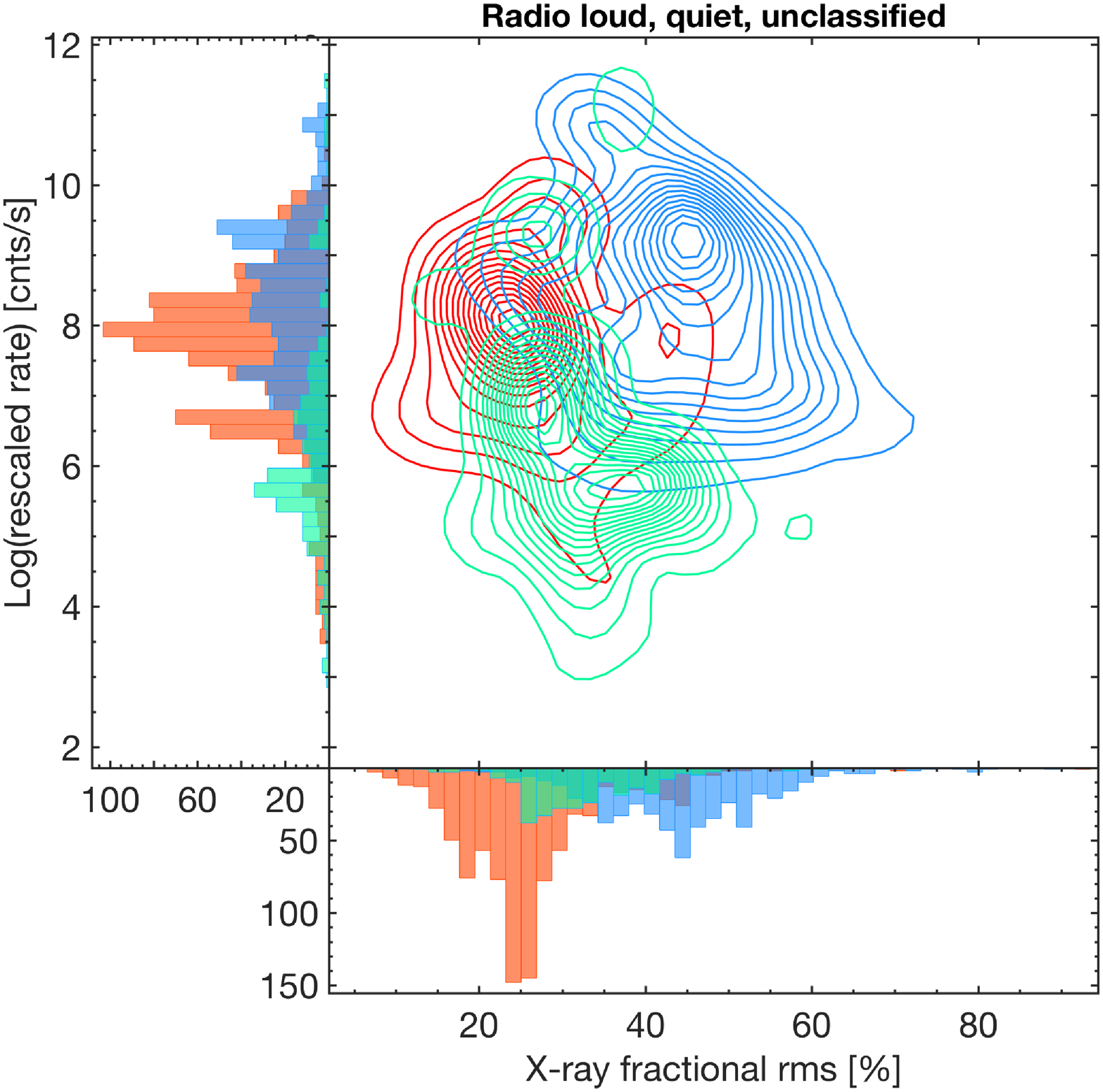}
\includegraphics[width=0.48\textwidth]{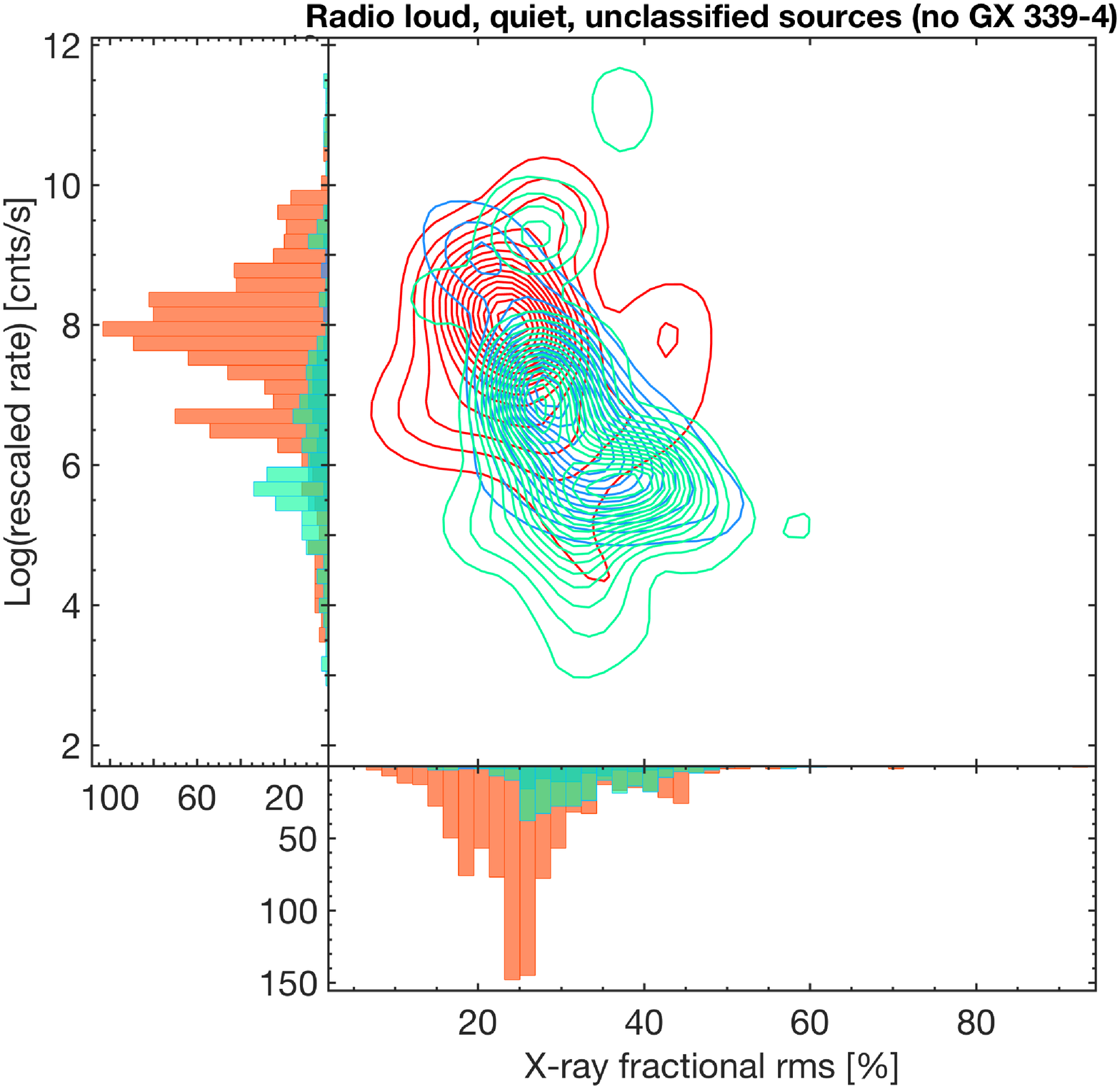}
\caption{Comparison between the bivariate distributions showing rate versus fractional rms for radio-loud sources (blue contours, corresponding to 2 sources and 554 measurements), radio-quiet sources (red contours, corresponding to 7 sources and 946 measurements) and radio-unclassified sources (green contours, corresponding to 8 sources and 266 measurements). The histograms on the x-axis and on the y-axis represent the fractional rms and count rate (rescaled by the source distance) marginal distributions, respectively. 
\textit{Left panel: } entire radio-loud sample compared with the entire radio-quiet and radio-unclassified samples. \textit{Right panel: } radio-loud sample excluding GX 339--4 compared with the entire radio-quiet and and radio-unclassified samples.}
\label{fig:rms_vs_rate_cont_map}
\end{figure*}

Both GX 339-4 and 4U1543--47 - our radio-loud sample - appear to be overall more X-ray variable than the radio-quiet sources, as reported by \cite{Dincer2014} based smaller sample of systems.
We note, however, that extending this property to radio-loud sources as a class requires great care, as our samples do suffer from severe small population issue.
Additionally, the details of the outburst evolution of each source, and the bias introduced by the uneven RXTE coverage of each outburst might be largely affecting the overall shape of the rms distributions shown in Fig. \ref{fig:rms_vs_rate_cont_map}. 
For example, some sources are known to spend more time than others in a particular state, a behaviour that is not necessarily linked to the radio-loudness nature of the sources themselves. Consider, for instance, the case of XTE J1752--223 that spent a long time in a bright LHS during its 2011 outburst \citep{Munoz-Darias2011}, as opposed to H1743--322, which always shows very fast LHSs (see, e.g., \citealt{Motta2010}). Similarly, for a variety of reasons (in this case definitely not linked to the radio-loudness nature of the source), RXTE covered particular outburst phases more densely than others. As an example, the (typically slow) decay phase of many sources has been monitored extensively, resulting in a dense coverage of the low-luminosity end of the LHS, while the rise phase was typically covered in a much sparser way, mostly due to fast outburst starts. Finally, the overall rescaled count rate\footnote{We remind the reader that we corrected the count rates by each source's distance. } distributions are  significantly affected by the large uncertainties on the distance of some source. 

All these effects might be at the origin of the apparent differences between the rate-rms distributions, both when comparing the three source classes (radio-loud/quiet/unclassified sources) against each other, or when comparing each individual source with the distribution formed by all the other sources of the same class (see Appendix \ref{App:fig}, Fig. \ref{fig:individual_maps}). These biasses cannot be easily taken into account and corrected for, which essentially makes it impossible to determine whether the apparent differences among the rsm, rate and bivariate distributions from the radio-loud, quiet and unclassified source are intrinsic to the sources themselves, or are due to observational biases.

In order to mitigate the effects of these biases, we looked for an X-ray variability related measurable that was the least affected by them. We found that the slope of the hard-line is a very suitable choice, as it is not strongly affected by sampling and by the detailed outburst evolution of a given source, and it is independent from the source distance.
Thus, 
we measured the slope of the hard-line that each source tracks in the rate-rms plane \citep{Munoz-Darias2011}. For each source, we fit such hard-line(s) in Log-Log space with a linear fit of the form $y = A + Bx$ (where $y$ is Log(count rate) and x is Log(rms)). Parameter $A$ is always consistent with 0, while parameter $B$ corresponds to the hard-line slope, which we report in Tab. \ref{tab:log}. The fits to the hard-lines for each source are shown in Appendix \ref{App:fig}, Fig. \ref{fig:individual_hardlines}.  
As already noted by \cite{Munoz-Darias2011}, some sources show more than one hard-line, typically (but not uniquely) corresponding to different stages of an outburst (i.e., outburst rise and decay), with slightly different slopes. This is the case for GX 339--4 and XTE J1550--564, for which we report the slopes of each hard-line individually. XTE J1118+480, unlike all the other sources, shows one single \emph{broken} hard-line. This results in two different slopes at high and low count rates (i.e. above and below $\approx$60 counts/s), which we report separately in Tab. \ref{tab:log}. We do not report any value for IGRJ17177--3656 since we could not constrain the slope of the hard-line due to very low number statistics. 

\begin{figure}
\centering
\includegraphics[width=0.50\textwidth]{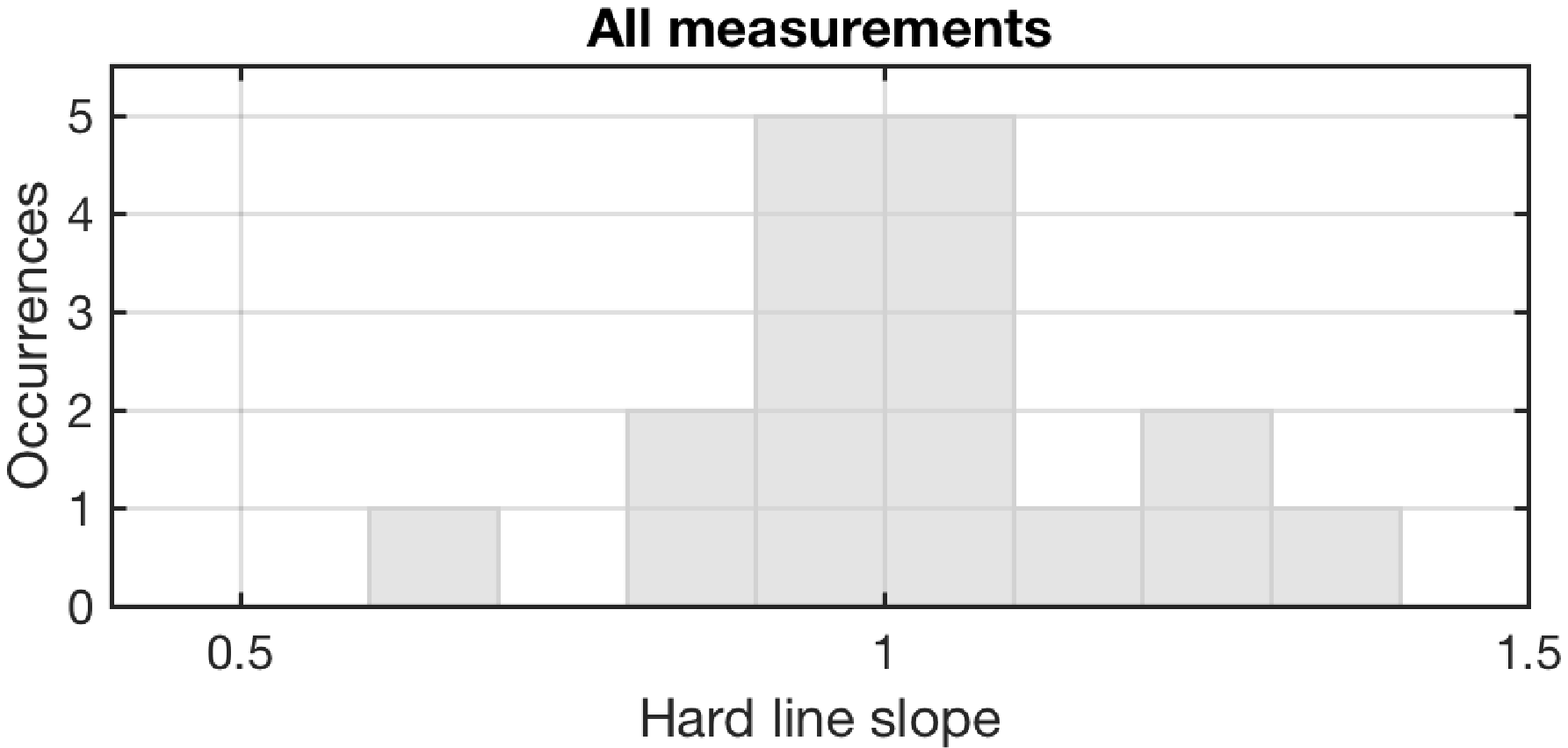}
\includegraphics[width=0.50\textwidth]{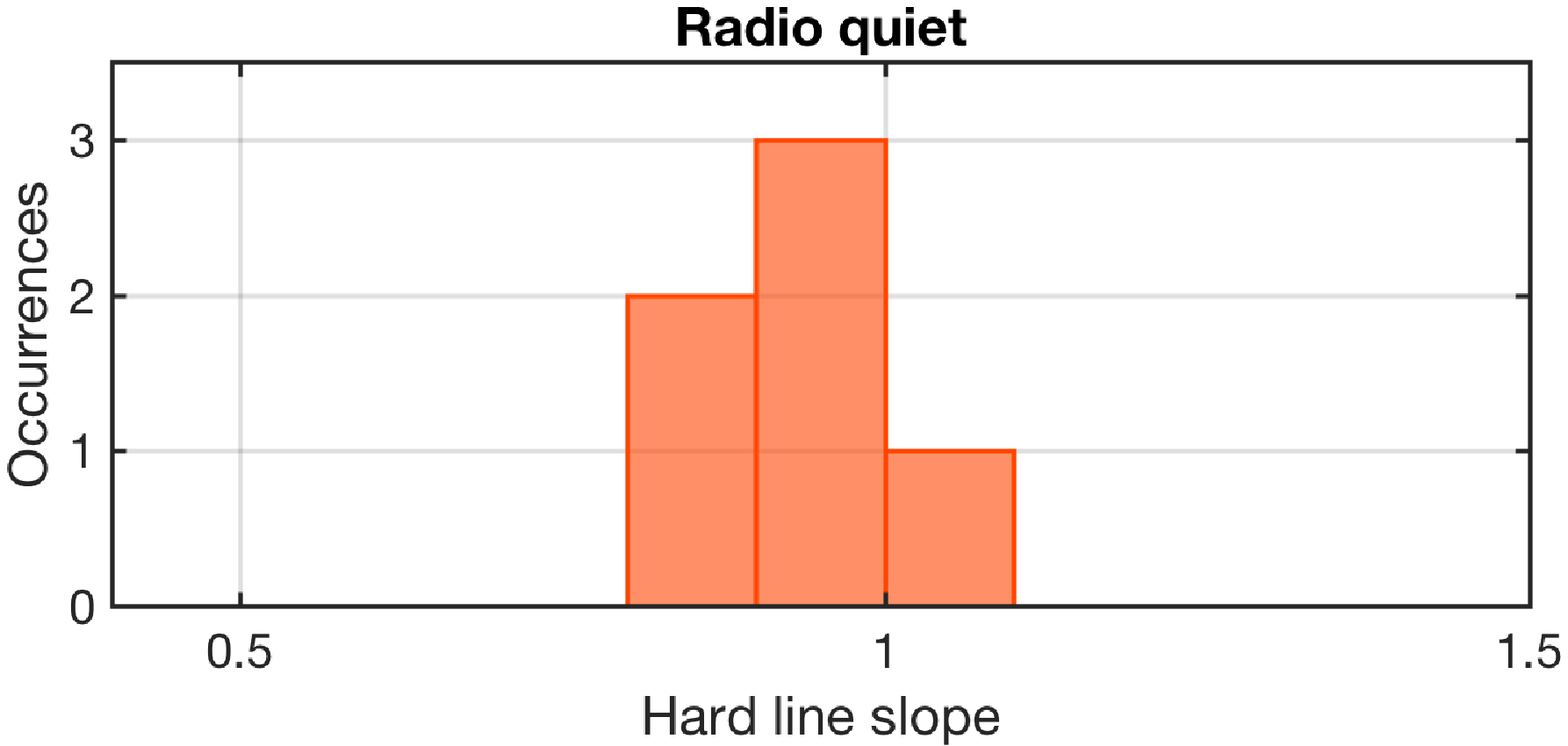}
\includegraphics[width=0.50\textwidth]{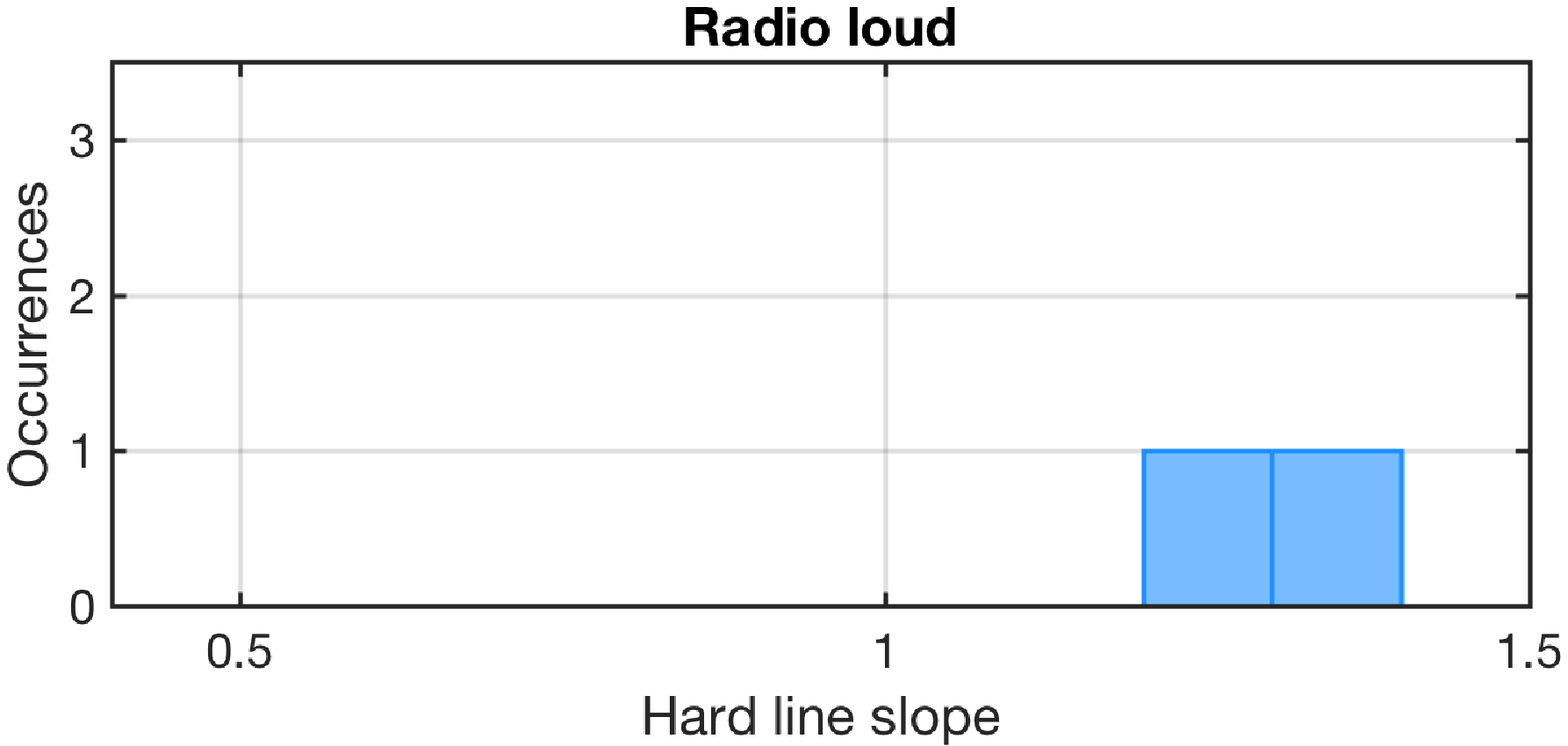}
\includegraphics[width=0.50\textwidth]{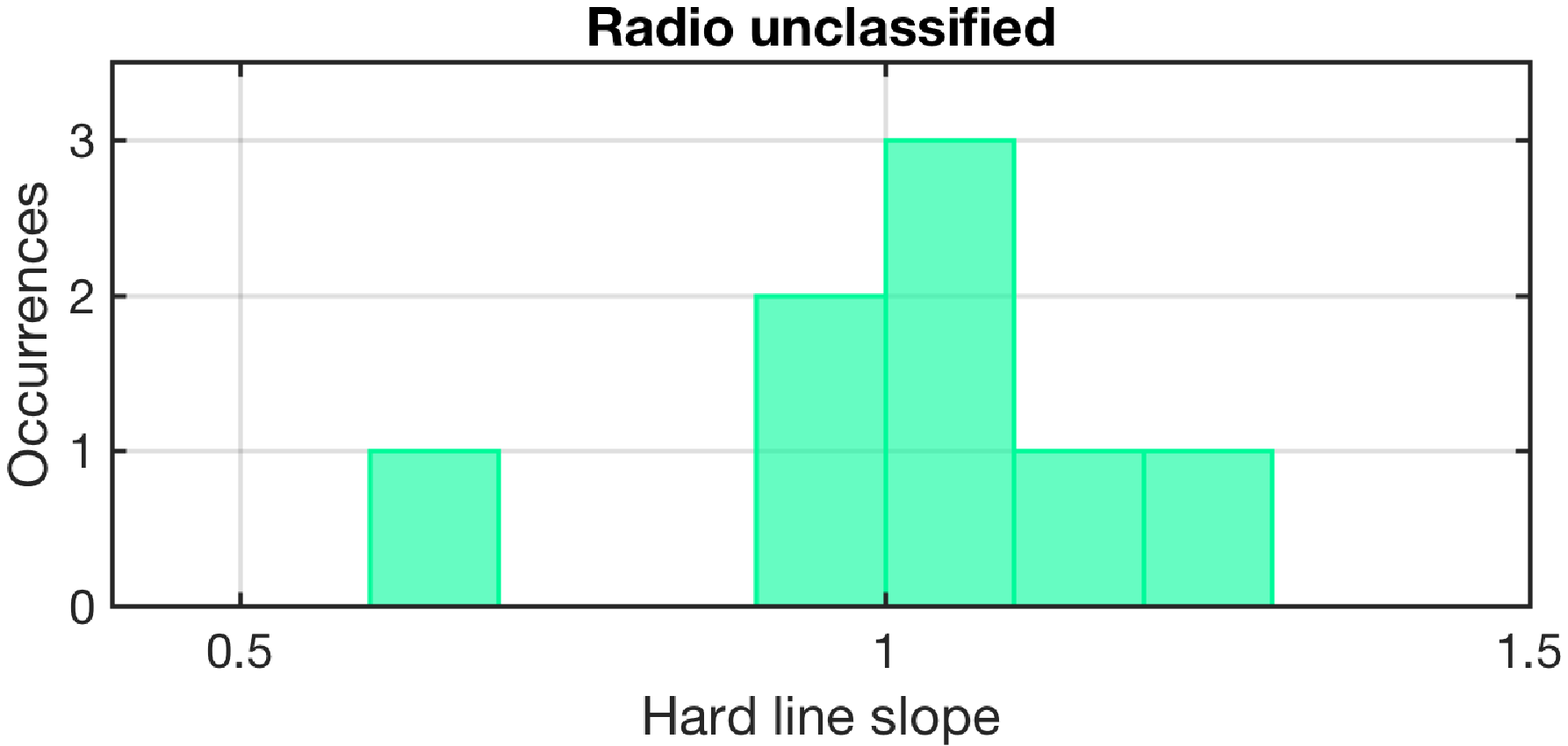}
\caption{Hard-line slope distributions for the sources of sample X. \textit{Top panel:} overall distributions of the hard-line slopes. \textit{Bottom panels:} sources divided (from top to bottom) in radio-quiet (red), radio-loud (blue) and radio-unclassified (green).  Note that not all the sources of sample X are included in the plots, since for a few sources we could not constrain the slope of the hard-line due to the low number statistics. For the sources where we fitted more than one hard-line, we took the median of the values.}
\label{fig:hard_line_slopes}
\end{figure}
\begin{figure}
\centering
\includegraphics[width=0.47\textwidth]{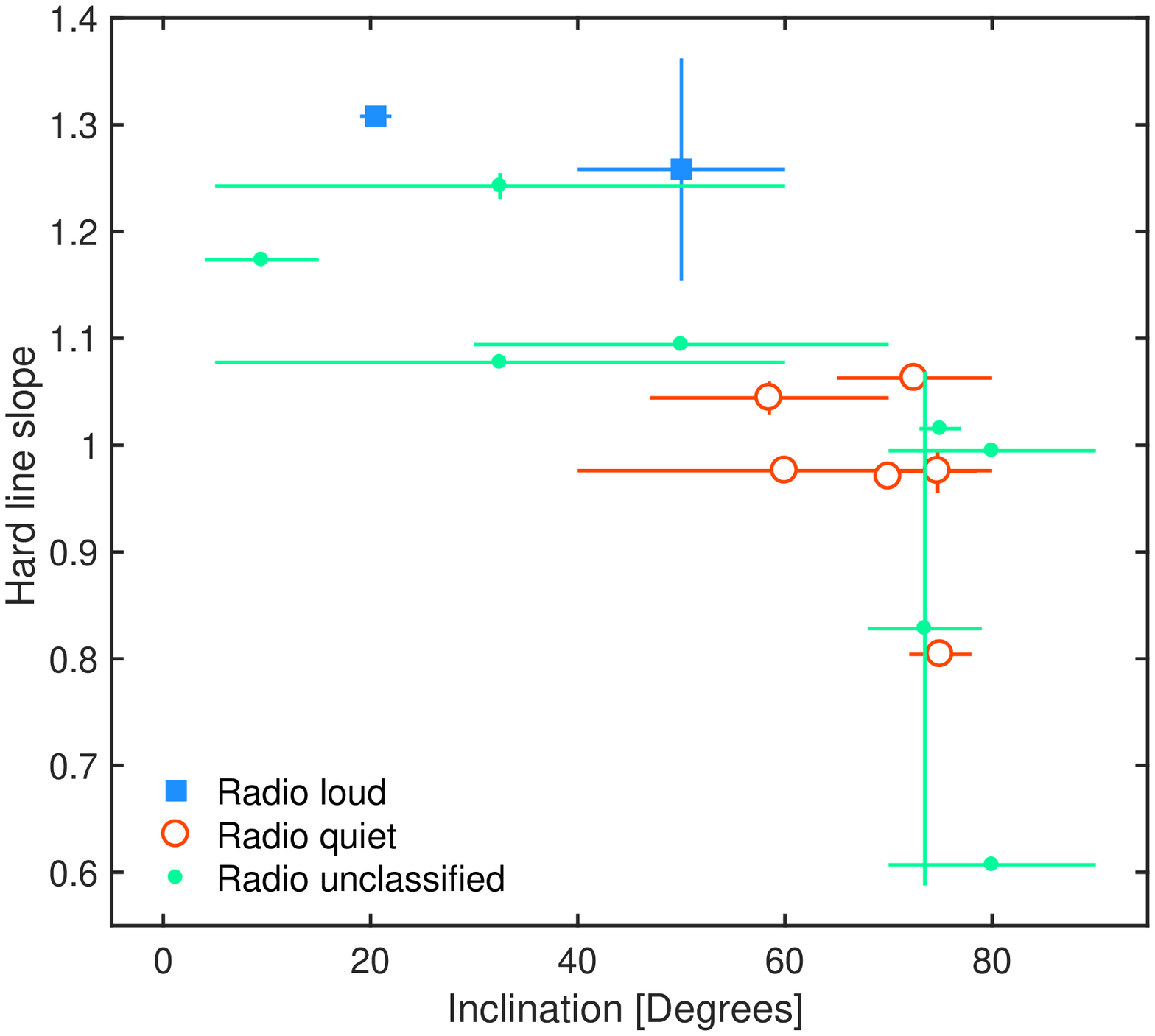}
\caption{Hard-line slope as a function of orbital inclination. For the sources where more than one hard-line is found, we combined the measurements in order to obtain one point per source. We assumed an inclination between 5 and 60 degrees, and 70 and 90, for the cases where we state low and high inclination, respectively, in Tab. \ref{tab:log}. The red open circles, blue squares and green dots mark radio-quiet, loud and unclassified source, respectively. Note that there is no inclination measurement of one of the radio-quiet sources (XTEJ1720-318), which is excluded from this figure, but included in Fig. \ref{fig:hard_line_slopes}.}
\label{fig:comparison} 
\end{figure}




In Figure \ref{fig:hard_line_slopes} we plot the distribution of the hard-line slopes for radio-loud sources (blue), radio-quiet sources (red), and radio-unclassified sources (green). For the sources where we detected more than one hard-line, we combined the hard-line slopes into one single value by taking the median. We thus obtained 17 hard-line slope measurements. Of these, 9 measurements come form sources with known radio nature (2 radio-loud and 7 radio-quiet), and 8 come from sources with undefined radio nature. Unfortunately, the small size of the samples we are considering does not allow us to directly test whether the hard-line distributions for different radio classes are consistent with being drawn from the same overall distribution.

In Figure \ref{fig:comparison} we show the hard-line slope versus the inclination for the 16 sources for which we have an estimate/measurement of both parameters. Points are colour-coded as in Fig. \ref{fig:hard_line_slopes}, i.e., radio-loud sources are plotted in blue, radio-quiet sources are plotted in red, and radio unclassified sources are plotted in green. Note that there is no inclination measurement of one of the radio-quiet sources (XTEJ1720-318), which is therefore absent in Fig. \ref{fig:comparison}.
In order to test the significance of the correlation visible in Fig. \ref{fig:comparison}, while taking into account the large uncertainties on the sources' inclination, we run a Monte-Carlo simulation to obtain 10$^{6}$ hard-line vs. inclination sets of values based on our dataset (see \citealt{Curran2015a}). We then estimated the average Spearman's rank correlation coefficient\footnote{The Spearman's rank correlation coefficient is a non-parametric - thus model independent - measure of rank correlation, which assesses how well the relationship between two variables can be described using a monotonic function.} and its uncertainty.
Under the null-hypothesis probability of the independence of the slope of the hard-line and the source inclination, we obtain a  Spearman's rank correlation coefficient $\rho = -0.80 \pm 0.03$ - that denotes significant (anti-)correlation - and an associated p-value of $p = 4.7\times 10^{-4}$, which allows us to reject the null-hypothesis probability.

\section{Discussion}

We have analysed the X-ray variability properties of a large sample of BH LMXBs, studying their rms and intensity (count rate rescaled by the source distance) distributions, bivariate rms and rate distributions, as well as the slope of the hard-line(s) in their X-ray rms-intensity plane.  
We now compare the three properties we have collected/measured for all sources in our sample X, namely: the radio-loudness, the slope of the hard-line and the orbital inclination. 

Under the assumption that the small number of radio-loud sources in our sample is representative of the behaviour of radio sources as a class, our results indicate the presence of a dependence of the hard-line slope and of the radio-nature of a given source on its orbital inclination. Such a dependence, however, must be considered carefully, as our source samples clearly suffer from small-population issues. Tests on different, independent samples are needed to confirm our findings. In particular, specifically in the case of radio loud sources, more data are required to exclude that the dependencies above are to be ascribed to characteristics intrinsic to the established radio-loud sources in our sample (GX 339-4, 4U1543-47, and possibly MAXI J1836-194, see below), rather then to radio-loud sources as a class. In the following we discuss in more detail our findings, which are summarized (and colour coded) in Tab. \ref{tab:log}.

From Table \ref{tab:log} it appears that radio-quiet sources tend to have shallower hard-lines than the two radio-loud sources for which we could measure the hard-line. Similarly, radio-quiet sources appear to be all consistent with being observed at high-inclination, as opposed to radio-loud sources, which are mostly low-inclination systems. 
The relatively large uncertainties on the orbital inclination of many sources, however, make it difficult to identify a clear correlation between inclination and radio-loudness. Therefore we analysed the relation between the orbital inclination and the hard-line slope, instead. Figure \ref{fig:comparison} confirms the tendencies identified above, showing a significant anti-correlation between orbital inclination and hard-line slope. By colour-coding the information on the radio-loudness in this correlation, we can see how radio-loud and radio-quiet sources tend to populate two separate regions of the parameter space. It also appears clear that the hard-line slopes of radio "unclassified" sources cover a rather large range, suggesting that their radio nature is probably diverse. 
Overall, these results highlight that both the hard-line slope and the radio-nature of a given system are function of its orbital inclination, with radio-loud sources having a lower inclination and steeper hard lines than radio-quiet sources. 

Of the two sources for which we could not measure the hard-line slope, but for which we have information on the inclination and on the radio-loudness, one - IGR 17177-3656 (line 19 of Tab. \ref{tab:log}) - is radio-quiet, and shows evidence for being at high inclination, thus supporting the correlation above.
The other is the only source that seems to violate the radio-loudness - inclination - hard-line slope correlation described above:  V404 Cyg (line 18 in Tab. \ref{tab:log}). Even though V404 Cyg was one of the first few sources for which the radio:X-ray correlation was found \citep{Gallo2003}, it has been recently shown to be affected by heavy and variable local absorption, which strongly influences the measured X-ray flux from the source \citep{Zycki1999, Sanchez-Fernandez2017, Motta2017a,Motta2017b}. \cite{Motta2017} noted that the large difference between observed and measured X-ray flux should be carefully taken into account, since while the X-ray emitting region could be at times almost completely obscured, the radio emitting region is most likely always visible, as it is probably located millions of gravitational radii away from the accretion disc mid-plane. This implies that V404 Cyg could be much more \textit{X-ray bright} than it appears in the radio:X-ray correlation (note that data from V404 Cyg come mostly from Ginga, and the data collected in 2015 during its last large outburst do not allow to consistently measure the X-ray rms). If this was really the case, it is still remarkable how the radio versus X-ray data for this source lines-up with the radio-loud radio:X-ray correlation, and in particular how similar it is to that of the best-studied source of sample R, GX 339-4.
Additionally, the inclination measurement of V404 Cyg comes from spectro-photometric observations performed in the optical band (\citealt{Casares1992}), and refers to the orbital inclination of the system. However, V404 Cyg has shown transient jets with highly varying angle w.r.t. the line of sight (with a maximum variation of $\sim$45$^{\circ}$, implying a disc axis-jet misalignment of more than $\sim$20$^{\circ}$), perhaps due to the jet being linked to the relativistic  precession of the inner portion (tens to hundreds of R$_{g}$) of the accretion flow (Miller-Jones et al. in prep.). This suggests that the inner disc and/or the jet in V404 Cyg could be strongly misaligned with respect to the orbital plane, which could result in a jet to the line of sight inclination much smaller than the $\approx 67^{\circ}$ inferred from the orbital inclination. This could, of course, occur for other sources too, but we note that many of the radio and X-ray properties of V404 Cyg, when studied in detail, are atypical.


In order to estimate the significance of a link among radio-loudness, inclination and hard-line slope, we performed a set of Fisher's exact tests. This kind of test is appropriate for small samples and it is commonly used in the analysis of contingency tables\footnote{A contingency table is a type of table in a matrix format that displays the (multivariate) frequency distribution of two or more variables.}. 
We tested the null hypothesis that the variables (or categorical features) radio-loudness, inclination, and hard-line slope, are unrelated and the distributions of cases we found is a result of random features combinations.
To better exploit the information from our (small) samples, we performed three separate tests, testing the three variables above in pairs by means of 2x2 contingency tables, instead of performing a 3-ways Fisher's tests, which would allow to test the three variables above simultaneously, but forcing us to consider only 6 sources out of 19 (i.e., those for which we have information on all the three categorical features). 
The Fisher's exact test is used in discrete statistics, thus in order to test the radio-loudness (treated as a bi-modal variable in this work) against hard-line slopes and source inclinations, we  divided our samples into discrete groups in terms of inclination (i.e. low vs. high inclination, with high inclination $i > 65^{\circ}$, see Sec. \ref{Sec:inclination}) and hard-line slopes (i.e. steep vs. shallow hard line, where steep means hard-line slope $> 1.07$). Note that especially the steep vs. low hard-line slope threshold is - to some extent - set arbitrarily, and was determined based on the based on the hard-line distribution and after comparison with the source inclination distribution.
Given its peculiarities, we initially excluded V404 Cyg from these tests. 
We obtained the following p-values for three different categorical features combinations\footnote{Note that we arbitrarily divided the hard lines in shallow and steep taking 1.07 as threshold slope.}:
(i) radio nature and inclination (considering 7 sources): $p = 0.048$;
(ii) radio nature and hard-line slope (considering 9 sources): $p = 0.028$;
(iii) hard-line slope and inclination: (considering 12 sources) $p = 0.001$.
Assuming a significance level of $\alpha = 0.05$, we can reject the null-hypothesis probability in all the above three cases. This allows us to asses that the apparent correlation radio-loudness - inclination - hard-line slope highlighted by our findings is likely not coincidental. 
The addition of V404 Cyg influences only the result of test (i) and - predictably - reduces the p-value to 0.1. 
We note, however, that the result of test (i) is likely mostly driven by the lack (already discussed above) of established radio-loud sources. For instance, we excluded MAXI J1836-194 from the radio-loud sources because, despite being mostly consistent with the radio-loud branch in the radio:X-ray plane, the slope of its evolution track is significantly steeper then that of the other radio-loud sources. The addition of MAXI J1836-194 (which remains the least ambiguous of the radio-undefined sources) to the radio loud sample, brings back the p-value from test (i) to $p = 0.048$ (which becomes $p = 0.018$ when V404 Cyg is excluded).

\smallskip

V404 Cyg aside - i.e., caveat very large disc-jet misalignments - our finding allows us to make a few considerations. 

\begin{itemize}

\item Since both the hard-line slope and the radio-loudness seem to correlate with the orbital inclination, it is reasonable to conclude that the differences observed (i.e. X-ray rms and radio luminosity increasing faster/slower with the X-ray rate) are not intrinsic to the sources, but are likely caused by the different line of sight of the observer. The orbital inclination is a continuous function that peaks at about 60$^{\circ}$ and Fig. \ref{fig:hard_line_slopes} (top panel) shows that the overall distribution of hard-line slopes is also continuous and single-peaked, with a maximum at $\approx 1$. 
This is not surprising and it is consistent with a geometric origin of the differences in hard-line slope. Similarly, it is possible that also the apparent dichotomy of radio-loud versus radio-quiet sources in Fig. \ref{fig:LRLx} and in Tab. \ref{tab:log} is in reality the result of a continuous distribution. This is supported by the analysis reported in \cite{Gallo2014} which shows, based on 
clustering and linear regression analysis, that radio-loud and quiet BHXBs are marginally consistent with being one single populations. We note however that - depending on the factors that cause the dependence on the inclination - the radio-loudness distribution could effectively appear as being discontinuous. 

\item The dependence of the hard-line slope with the inclination is not surprising, given the known dependence of the type-C QPO amplitude on the source inclination \citep{Motta2015}. The QPO amplitude in high-inclination sources -  when in the hard state - increases with frequency (thus with rate), while it remains roughly constant in low-inclination sources. The amplitude of the broad-band noise, instead, only shows a mild dependence on the source inclination (opposite with respect to that observed for type-C QPOs), and decreases monotonically with frequency (rate) roughly equally in all sources. Since the hard-line slope by construction strictly depends on the rms properties, we do expect the observed behaviour, i.e. the hard-state rms of high-inclination sources becoming more and more dominated by the QPO as the rate increases, causing a steeper hard-line. It still remains to be explained why the QPO amplitude \textit{intrinsically} increases in the hard state while the broad-band noise amplitude decreases. We can speculate that this is caused by the approaching (i.e. decreasing inner truncation radius) thin Shakura-Sunyaev disc, which obscures more and more the precessing inflow, because of its increasing relative thickness with respect to the inner flow, thus increasing the apparent amplitude of the oscillation. A detail modelling of the system, including relativistic effects such as light-bending, is needed to test this hypothesis, but is beyond the scope of this work.

\item The dependence of the radio-loudness on the inclination is also not surprising, as (i) the emission relativistic outflows will not be isotropic due to relativistic aberration, and (ii) different jet inclinations to the line of sight will result in different jet spectra (see, e.g, \citealt{Hjellming1988}).
In a population of randomly oriented BHBs, very small (fixed across the sources) Lorentz factors would produce limited scatter in the radio:X-ray correlation (i.e., the observed radio emission is only weakly boosted, see e.g., \citealt{Gallo2003}). However, as shown by \cite{Heinz2004} the same is true for high Lorentz factors, which would also induce low scatter since relativistic beaming could easily quench radio emission at high angles from the jet direction. The accurate measurement of {\em one} hard state jet speed would be enough to resolve this.
In any case, a Lorentz factor $\Gamma \approx$2 would be enough to explain the observed spread of approximately one order of magnitude in radio luminosity for jets with angles to the line of sight ranging between 30$^\circ$ and 70$^\circ$. As already noted by \cite{Heinz2004}, it must be stressed that it is rather unlikely that the bulk Lorentz factors of all the sources in the radio:X-ray plane are identical. Quantifying how relativistic aberration affects the scatter in the radio:X-ray correlation would require assumptions on the Lorentz factors distributions, which would understandably heavily affect the resulting predictions.

The above would explain why high inclination sources are more radio-quiet then low-inclination ones. However, it does not fully explain the morphology of the radio:X-ray correlation, and in particular the fact that the radio-quiet branch re-joins the radio-loud branch at the top right of the radio:X-ray plane. It might be possible to explain this behaviour in terms of jet acceleration (see, e.g., \citealt{Soleri2011}) - i.e., the jet should first accelerate with increasing accretion rate/X-ray luminosity (inducing the split of the correlation in radio-quiet and radio-loud branches), then decelerate at \textit{higher} rates/luminosities in order to allow the radio-quiet and loud branch to meet again at high X-ray and radio luminosities. The fine-tuning required to make this scenario viable is, however, not realistic. Additionally, this interpretation fails to explain why radio-quiet and radio-loud sources show different radio spectral indices, as reported by \cite{Espinasse2018}.

There is an alternative to the assumption that the jet spectrum appears \emph{intrinsically} different at different viewing angles. An ionised equatorial outflow with density decreasing with distance from the black hole and absorbing the jet emission, could explain the difference in spectral index in radio-loud (low-inclination) and radio quiet (high-inclination) systems reported by \cite{Espinasse2018}. In high-inclination sources the emission from regions closer to the black hole (emitting at higher frequencies) would be more absorbed than the emission from regions further away (peaking at lower frequencies), resulting in a negative, optically-thin like spectrum, as opposed to the $\approx$ flat radio spectrum seen in low-inclination sources. 
In this scenario, the properties of the absorbing medium should depend on luminosity/accretion rate, in order to explain why the radio-loud/quiet dichotomy is apparent only in a certain luminosity range. For instance, at low accretion rates the medium could be rarefied and optically thin, but its density could increase with accretion rate, causing partial absorption of the radio photons (which would explain the radio properties of the radio-quiet branch). At even higher accretion rates, the jet could become fast enough that all the emitting regions are shifted further away from the black hole, thus reducing the effects of the absorption, and consequently inducing the re-unification of the loud and quiet branches at the highest luminosities. Clearly, fine tuning on the properties of the absorbing medium (e.g., density, temperature profile) and their evolution is needed for this scenario to be viable.

There is also the possibility that radio-quiet sources are in reality \emph{X-ray bright}. It is already known that the shape of the X-ray HID does depend on the orbital inclination \citep{Munoz-Darias2013}, which suggests that the X-ray emission is not isotropic. In order to test this possibility, a comprehensive analysis of the spectral properties of a complete sample of sources is needed, which will be presented in a future work.  

Finally, it is possible (and indeed probable) that a mix of all the aforementioned factors (e.g., jet spectrum varying with viewing angle, different/varying jet speed, presence of a partial-covering absorbing medium, anisotropic X-ray emission) could play a role in producing the observed behaviour in the radio:X-ray plane. A detailed modelling of the system and of its geometry and physics is needed to test this hypothesis, and is again beyond the scope of this manuscript. 

\item Based on the discovered correlations, we can attempt a radio-classification of the unclassified sources based on the slope of the hard-line (and on the inclination, when known). Under the reasonable assumption that all sources follow the same inclination vs. hard-line slope correlation, we can predict that - if observed at the right luminosities - XTE J1748--288, XTE J1118+480, GS 1354--645 and Swift J1842.4-1124 will likely appear to be radio-quiet sources, while MAXI J1659--152 and XTE J1817+330 are likely to appear radio-loud(er). 
Following this reasoning, both MAXI J1836-194 and  XTE J1752--223 should be classified as radio loud sources (therefore they should be placed at the upper end of their estimated distance ranges). However, their peculiar evolution in the radio:X-ray plane (i.e. their steeper radio:X-ray correlation slope) remains unexplained.
\end{itemize}

\section{Conclusions}

We have analysed a large sample of black-hole low-mass X-ray binaries, comparing their radio and X-ray variability properties with their orbital inclination. 

We found that only two of the sources historically classified as radio-loud (GX 339-4 and 4U1543-47) can be confidently considered as such. Most black-hole X-ray binaries where the radio:X-ray correlation has been measured are to be classified as radio-quiet, when their radio nature can be established at all. It follows that the radio-loud behaviour 
is rarer then commonly believed.

A dependence of the radio loudness of black hole binaries on the orbital inclination cannot be currently excluded. In particular, the high-inclination sources we considered appear to be systematically radio-quiet, with the exception of V404 Cyg, which notably distinguished itself as atypical in many ways. Low-inclination sources, instead, appear to be radio-bright, even though the small size of the sample currently available limits our ability to establish a definitive dependence. The likely existence of a link between radio-loudness and orbital inclination therefore suggests that the radio-loudness of black hole X-ray binaries is a geometric effect due to the inclination of a source with respect to the observer's line of sight combined with Doppler boosting, and thus probably not linked to any difference intrinsic to the accretion flow.

This interpretation is supported by our discovery that in high-inclination sources the X-ray rms increases faster with the luminosity/accretion rate than in low-inclination sources, which implies that the slope of the hard-line (anti-)correlates with the orbital inclination. Hence, we argue that the slope of the hard-line could be adopted as an indicator of the orbital inclination of a source (similarly to the presence/absence of disc-winds or absorption dips, or the amplitude of QPOs). The hard-line slope is also found to correlate well with the radio-nature of the black-hole X-ray binaries we considered, which makes it the most reliable X-ray indication of the radio-loudness of a given source to date.  


\section*{Acknowledgements}

The authors would also like to thank Martin Henze and Luigi Stella for useful discussions, and the anonymous referee, which provided very useful comments that helped improving this paper. 
SEM also acknowledges Arash Bahramian and  Tony Rushton for useful discussion and for making their data pubblicly available (https://doi.org/10.5281/zenodo.1252036). 
SEM acknowledges the Science and Technology Facilities Council (STFC) for financial support.




\bibliographystyle{mnras.bst}
\bibliography{biblio} 
\newpage

\appendix
\onecolumn

\section{Additional figures}\label{App:fig}

\begin{figure}
\centering
\caption{Bivariate distribution showing rate versus fractional rms for the individual sources of sample X. Differently from Fig. \ref{fig:rms_vs_rate_cont_map}, for clarity's sake we use colour maps instead of contour plots. Brighter colours indicate higher bin values. Note that plots are scaled differently to facilitate inspection.}
\label{fig:individual_maps}
\begin{tabular}{c c c}
\includegraphics[width=0.3\textwidth]{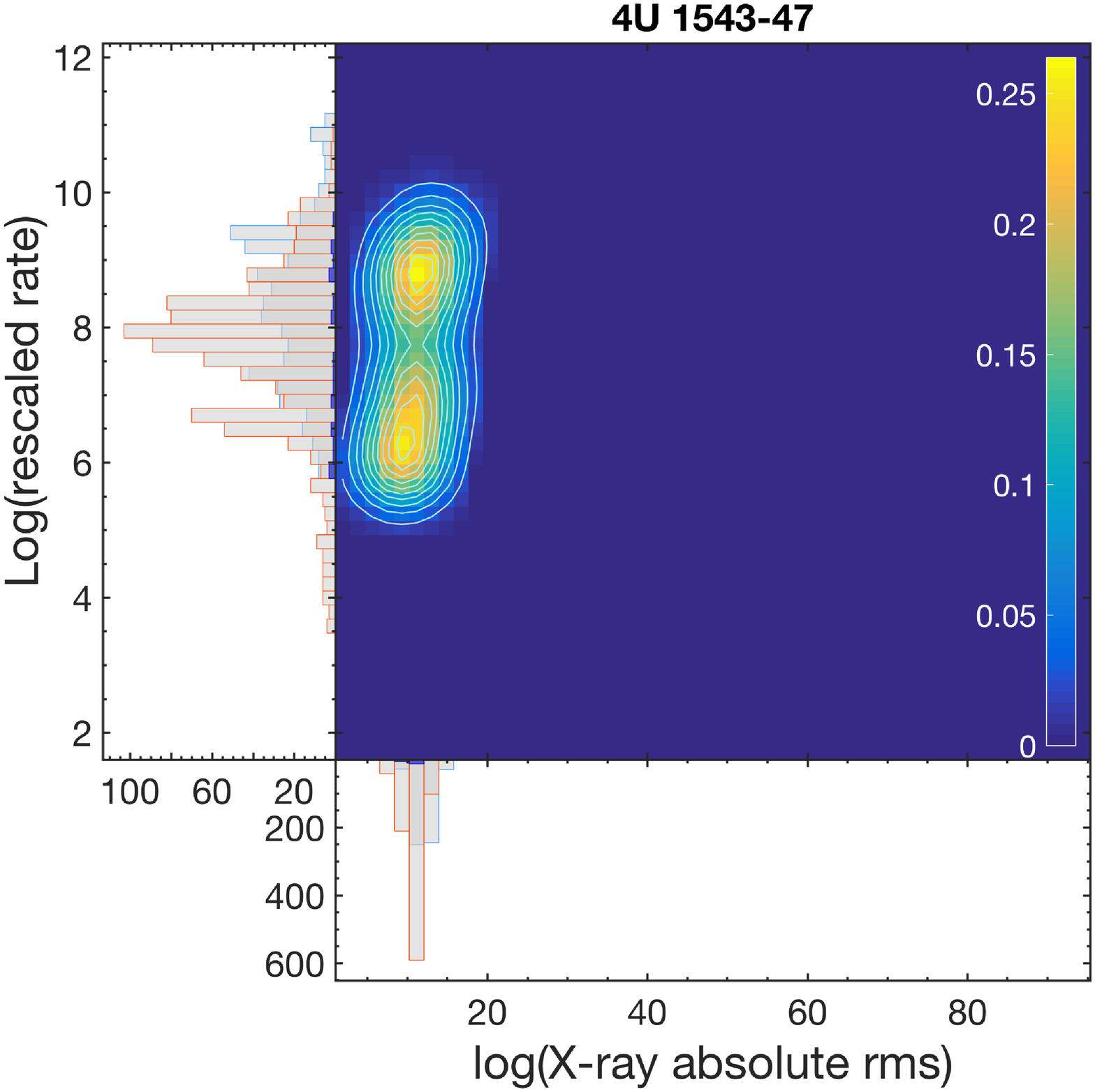} &
\includegraphics[width=0.3\textwidth]{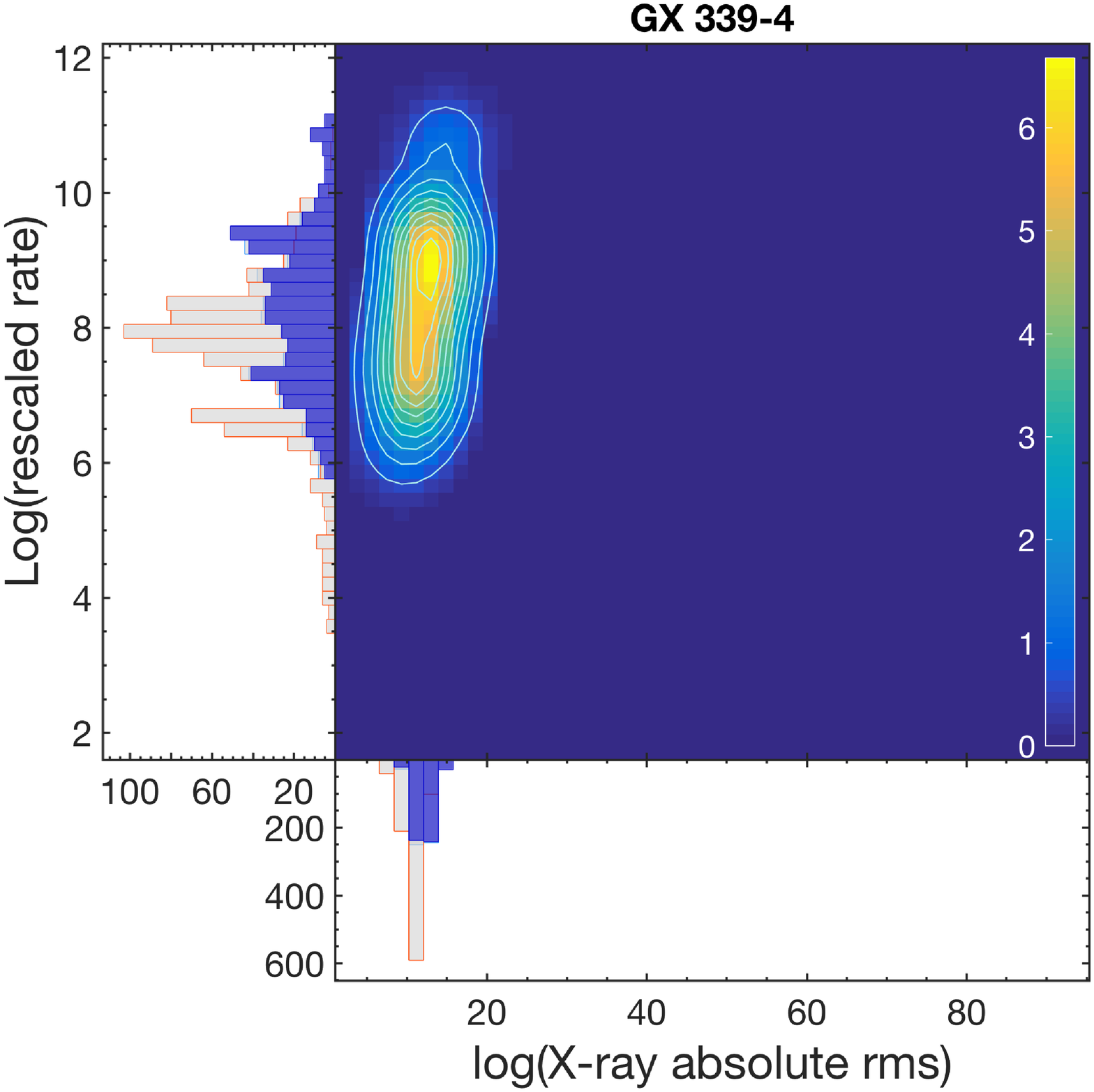}  &
\includegraphics[width=0.3\textwidth]{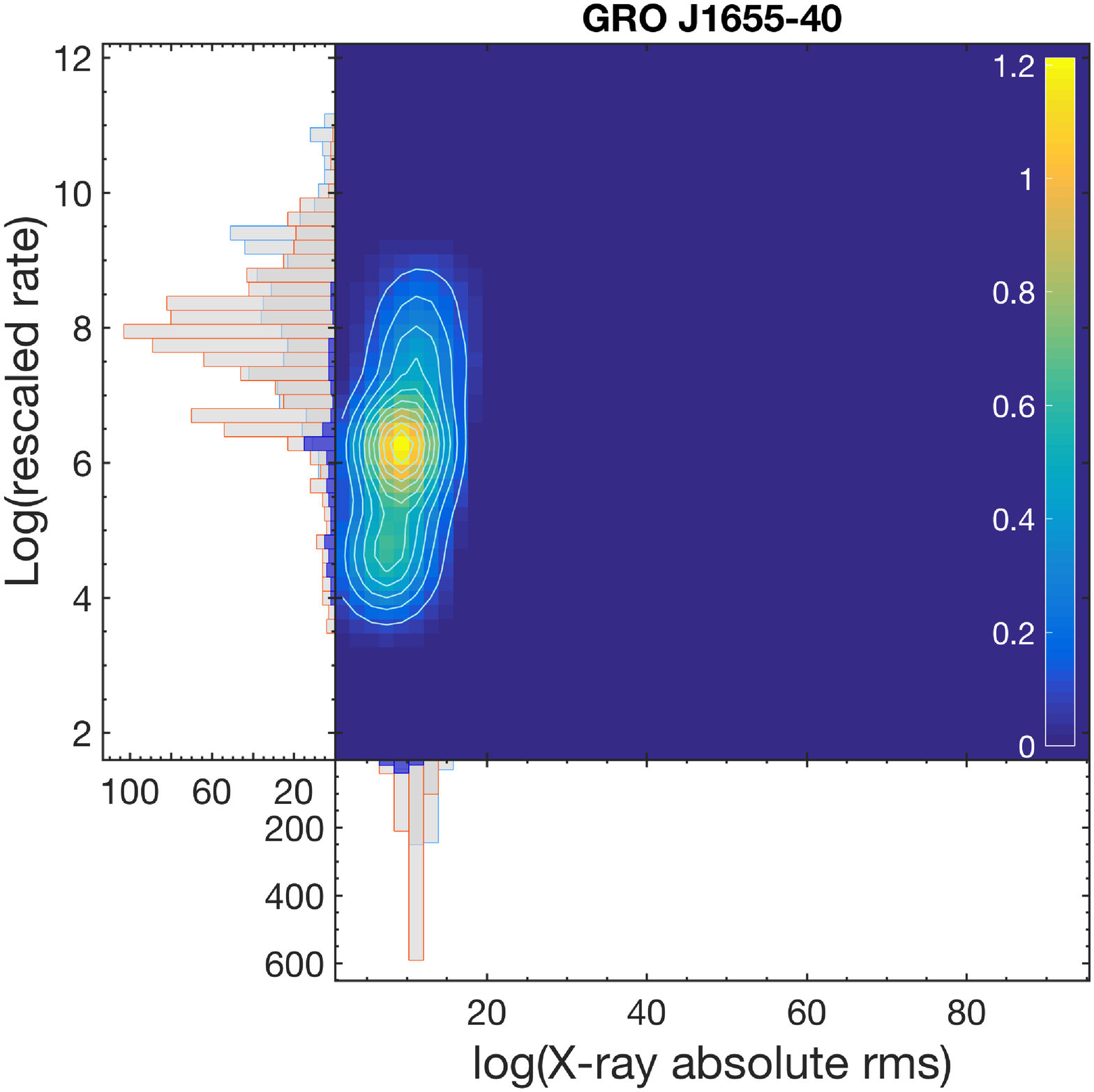} \\

\includegraphics[width=0.3\textwidth]{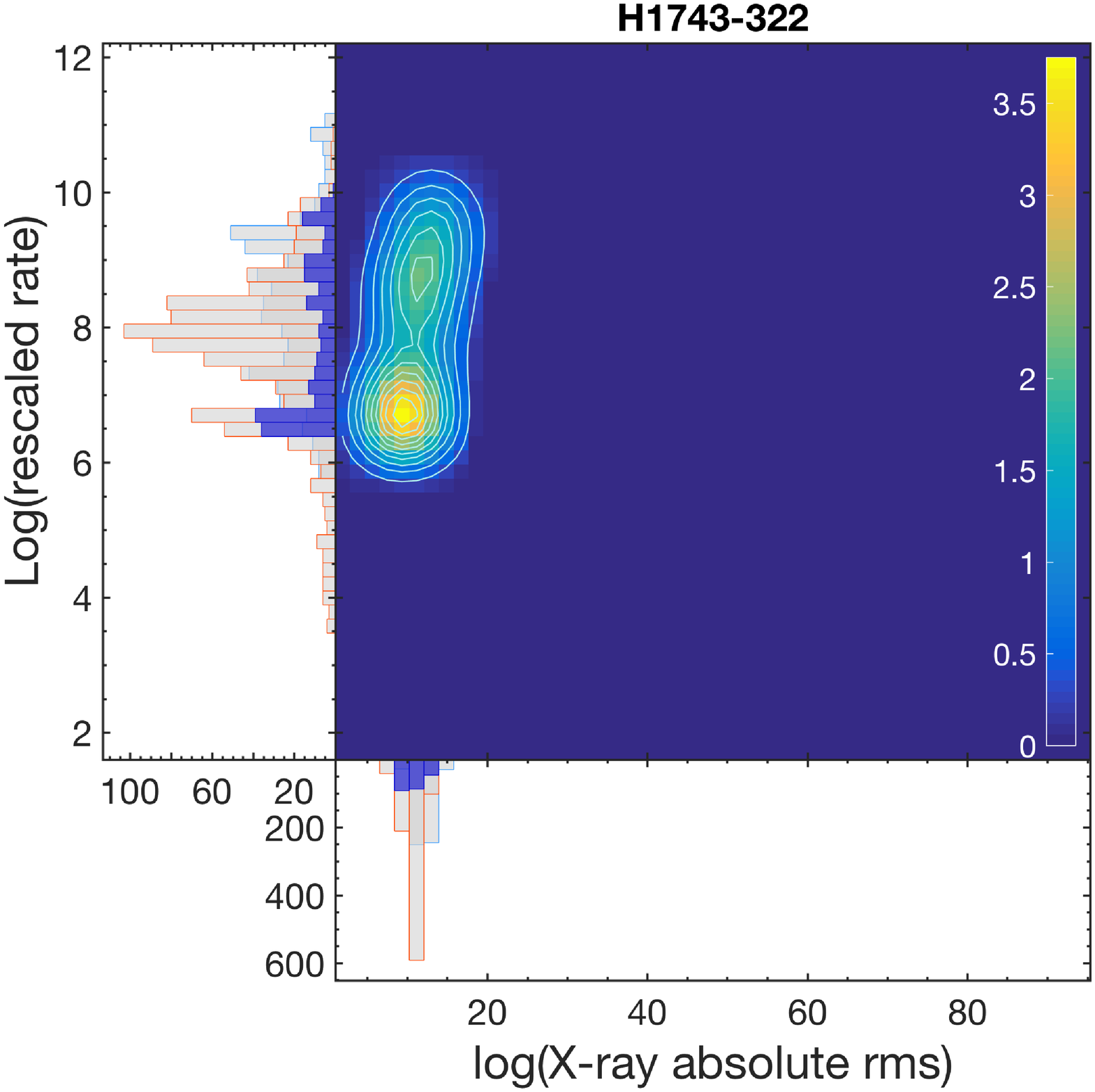} &
\includegraphics[width=0.3\textwidth]{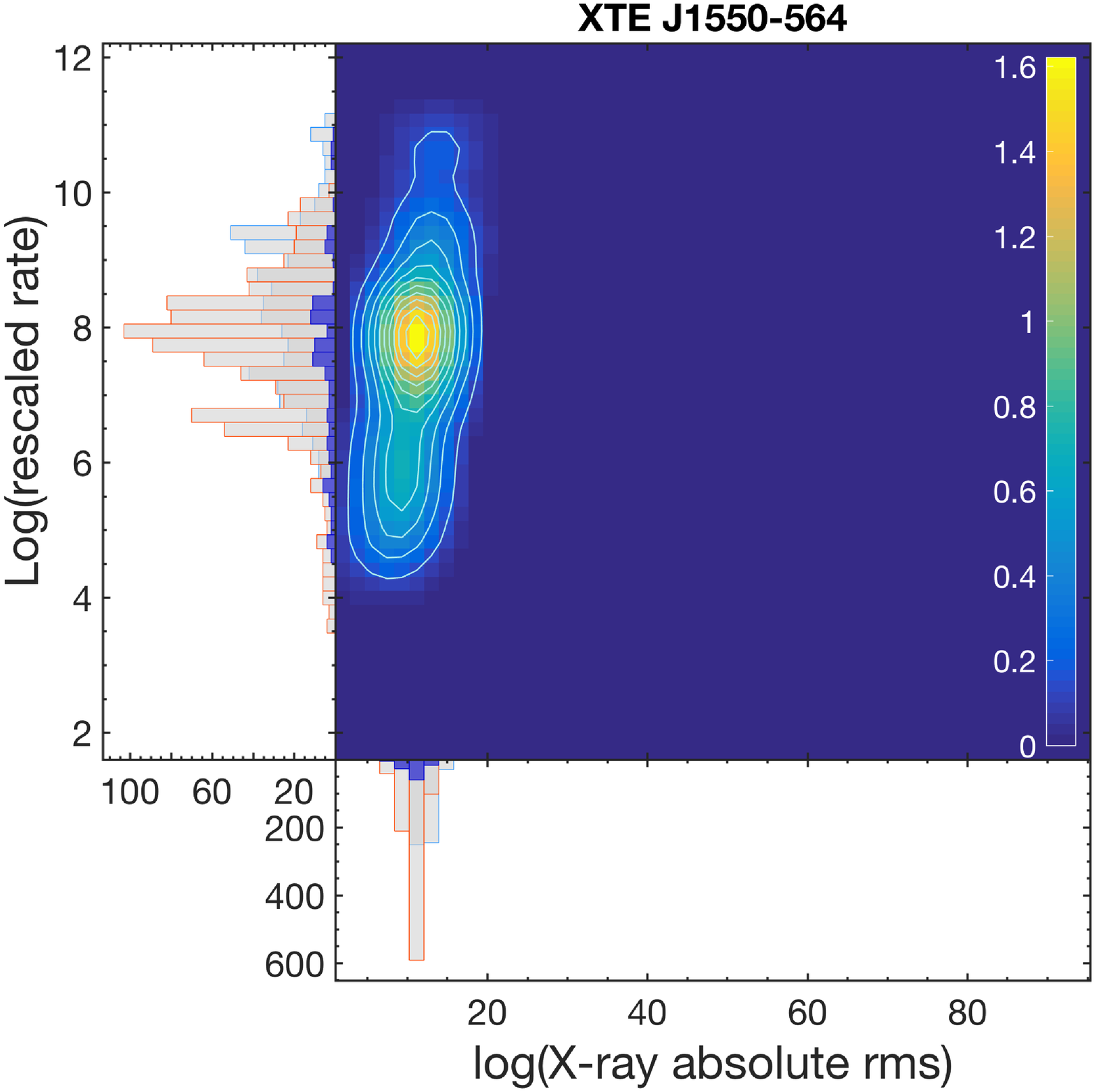} & 
\includegraphics[width=0.3\textwidth]{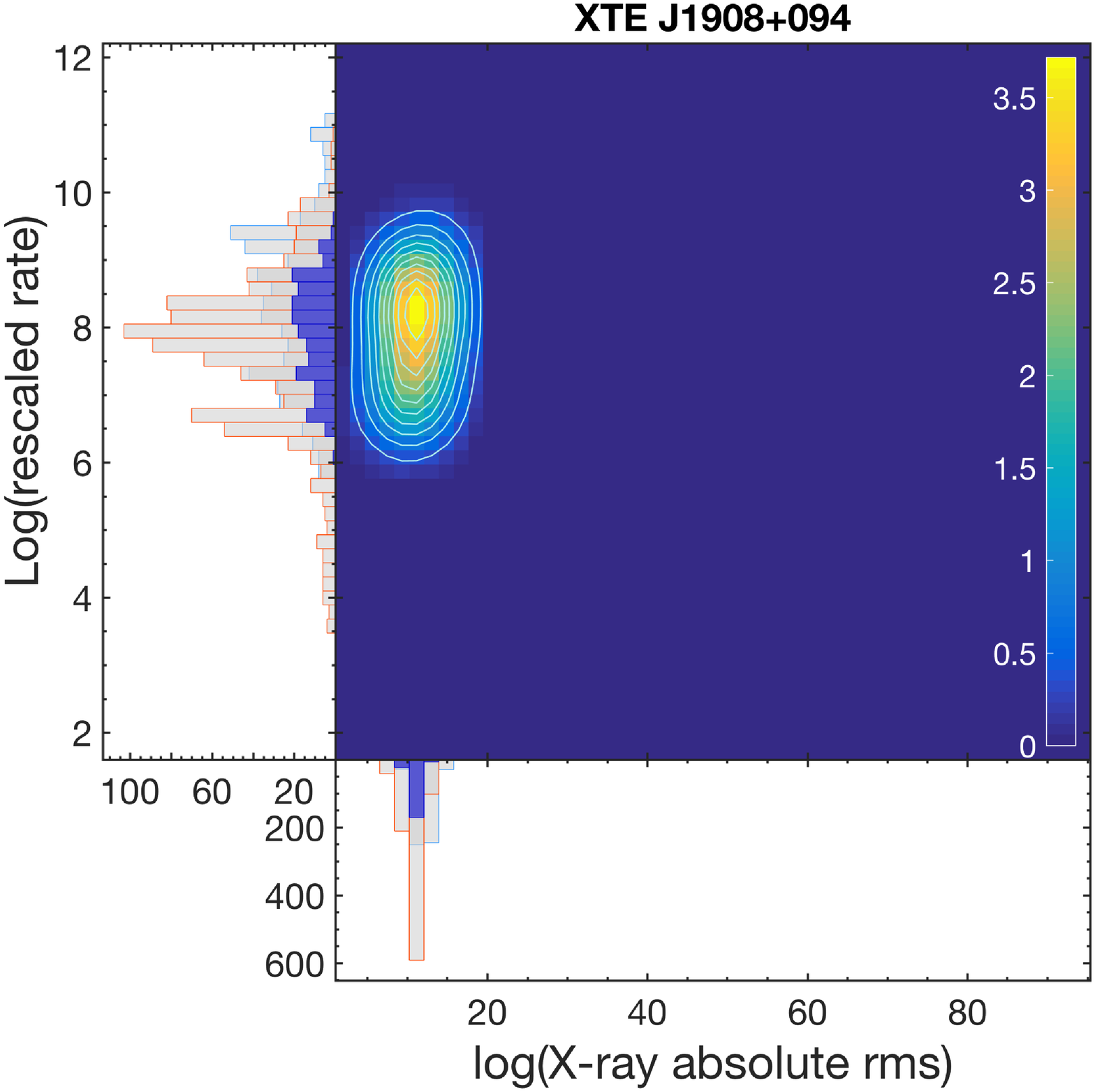} \\

\includegraphics[width=0.3\textwidth]{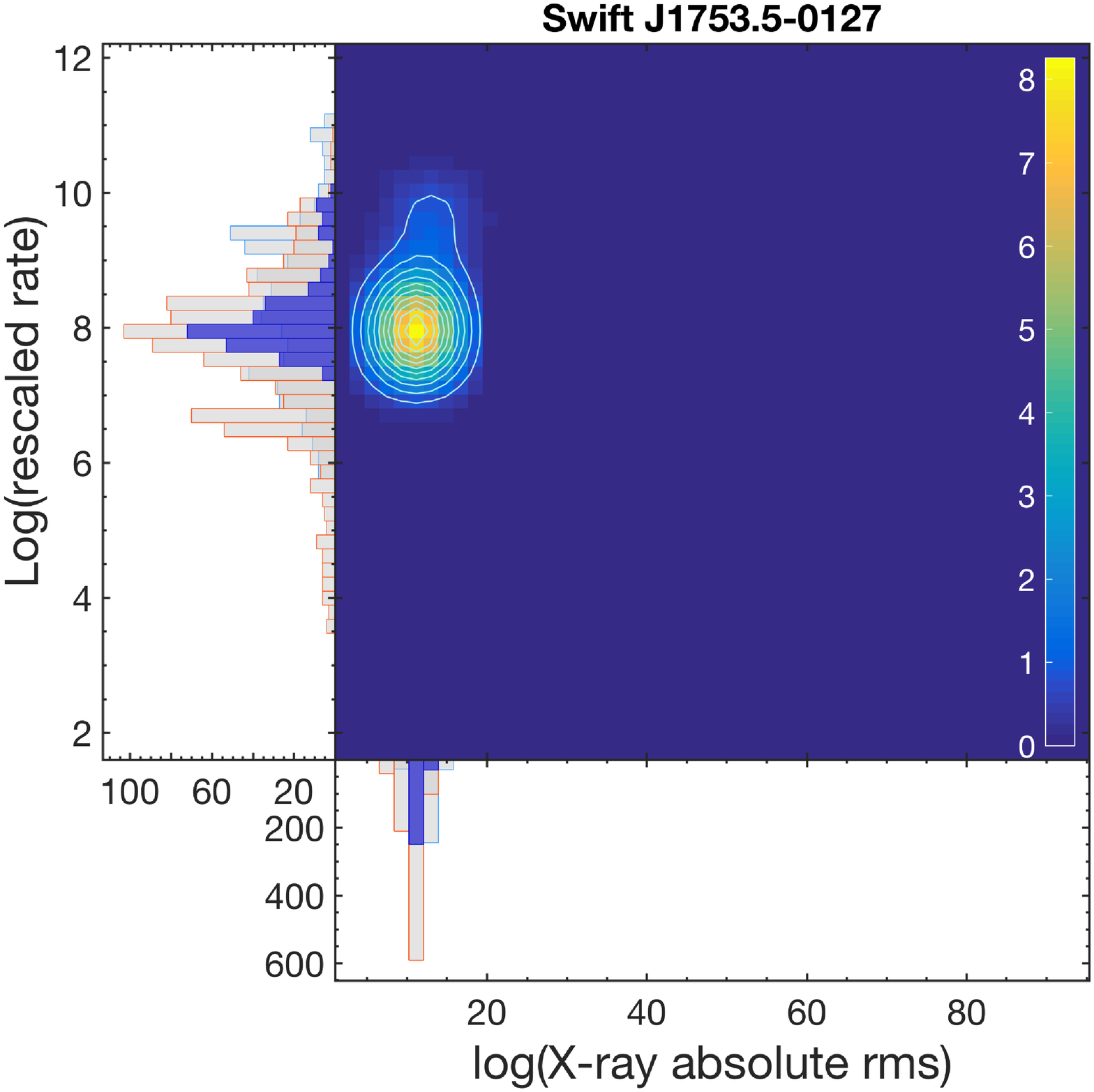} &
\includegraphics[width=0.3\textwidth]{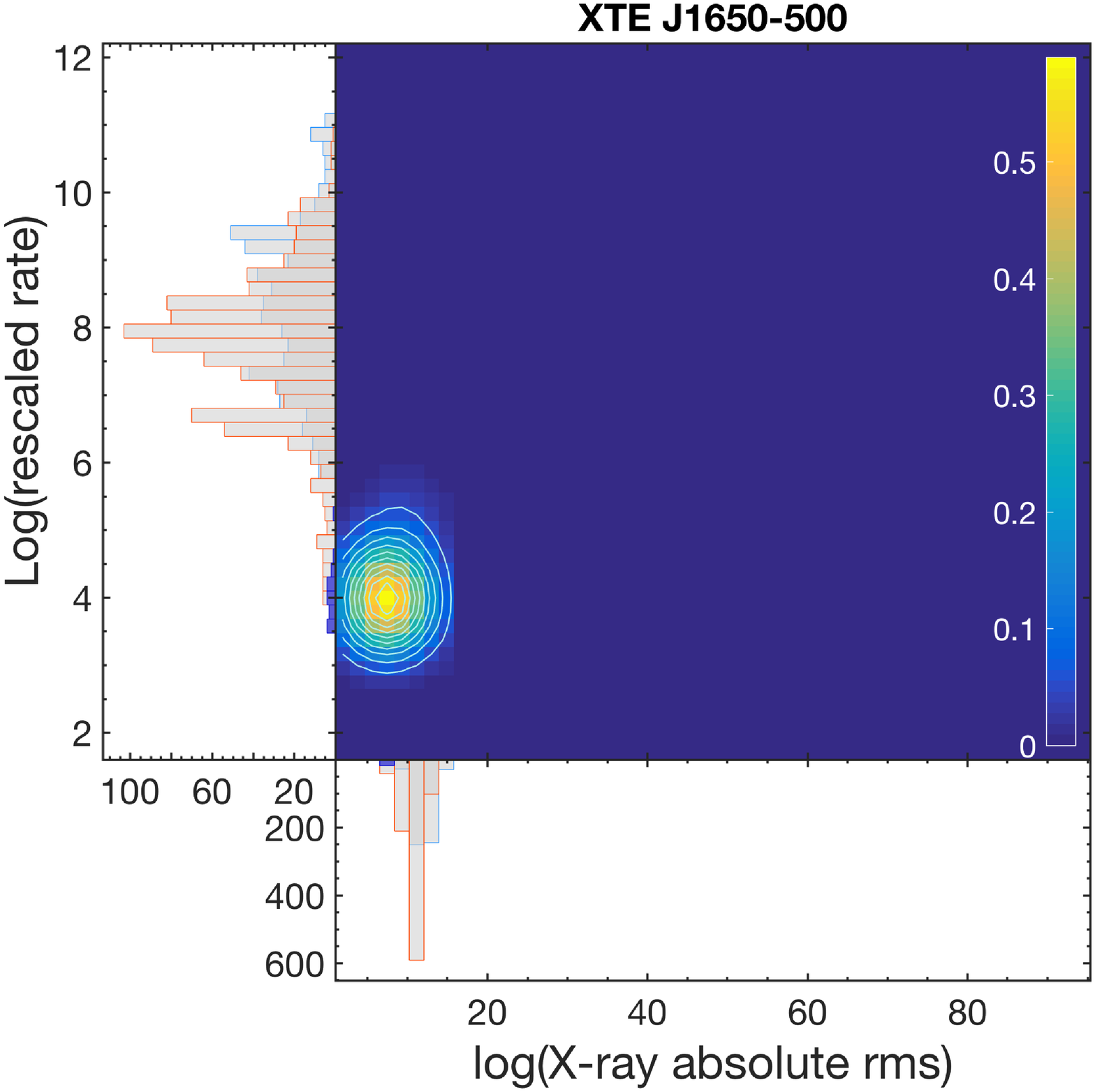} &
\includegraphics[width=0.3\textwidth]{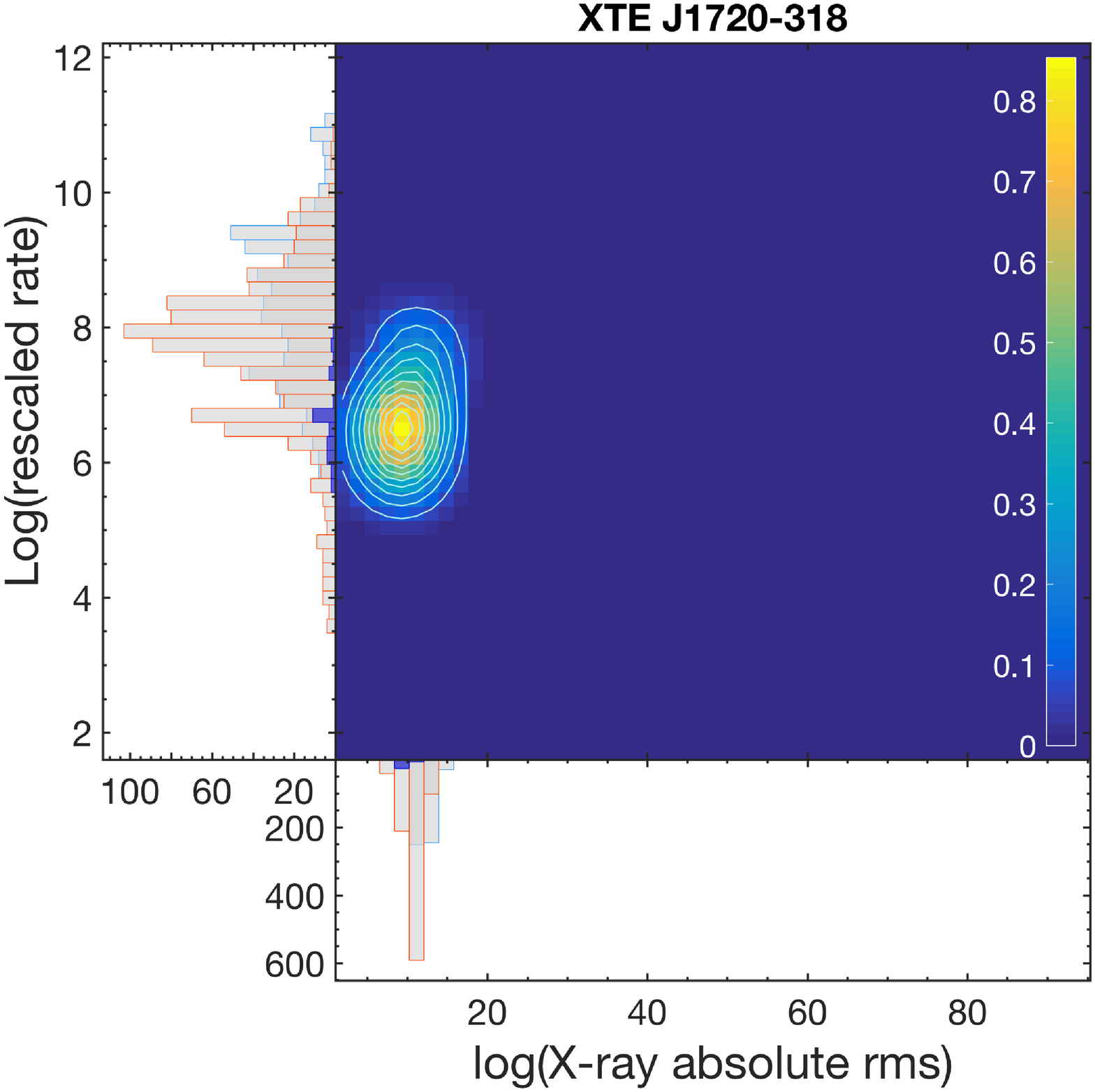} \\

\includegraphics[width=0.3\textwidth]{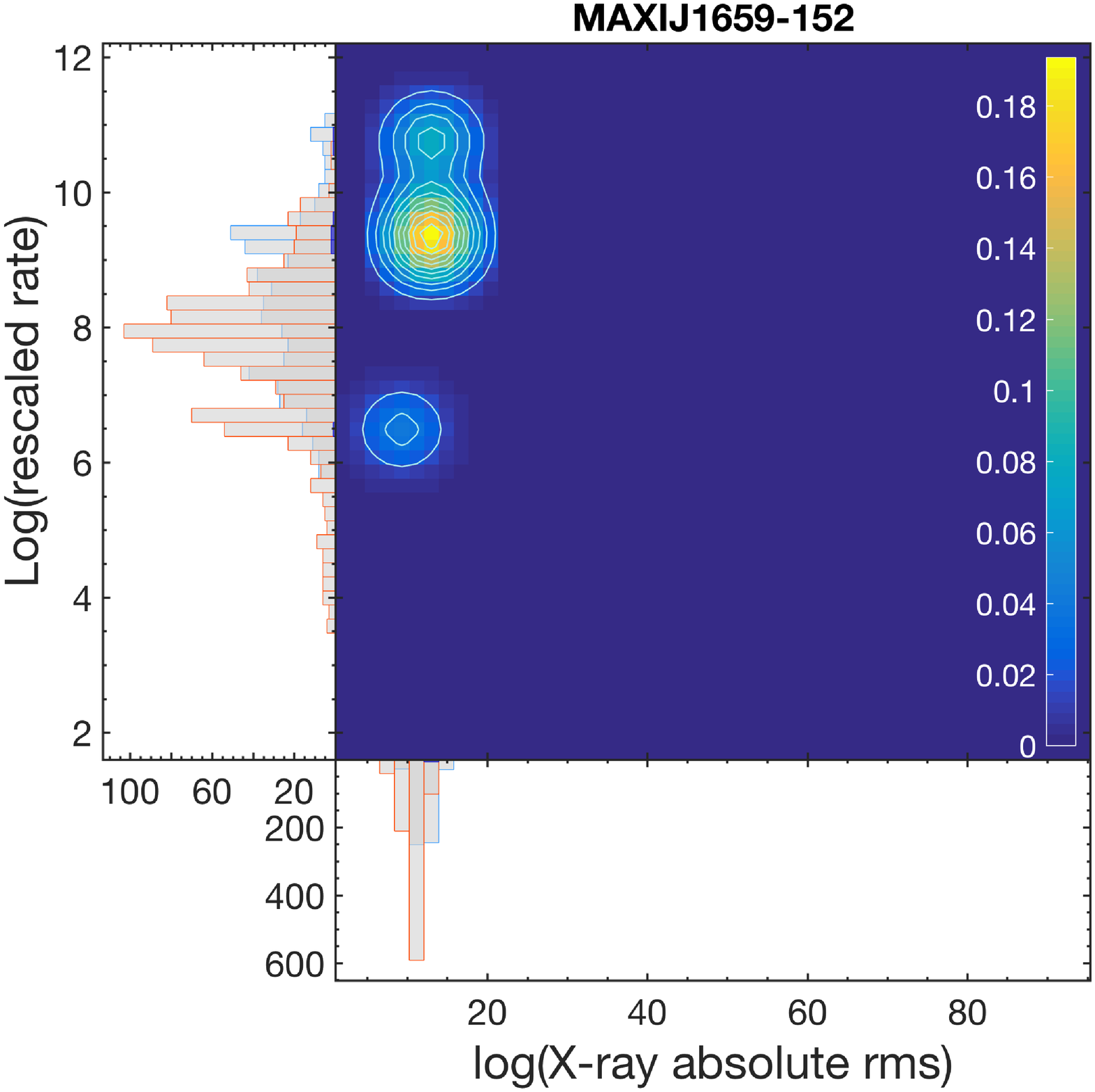} & 
\includegraphics[width=0.3\textwidth]{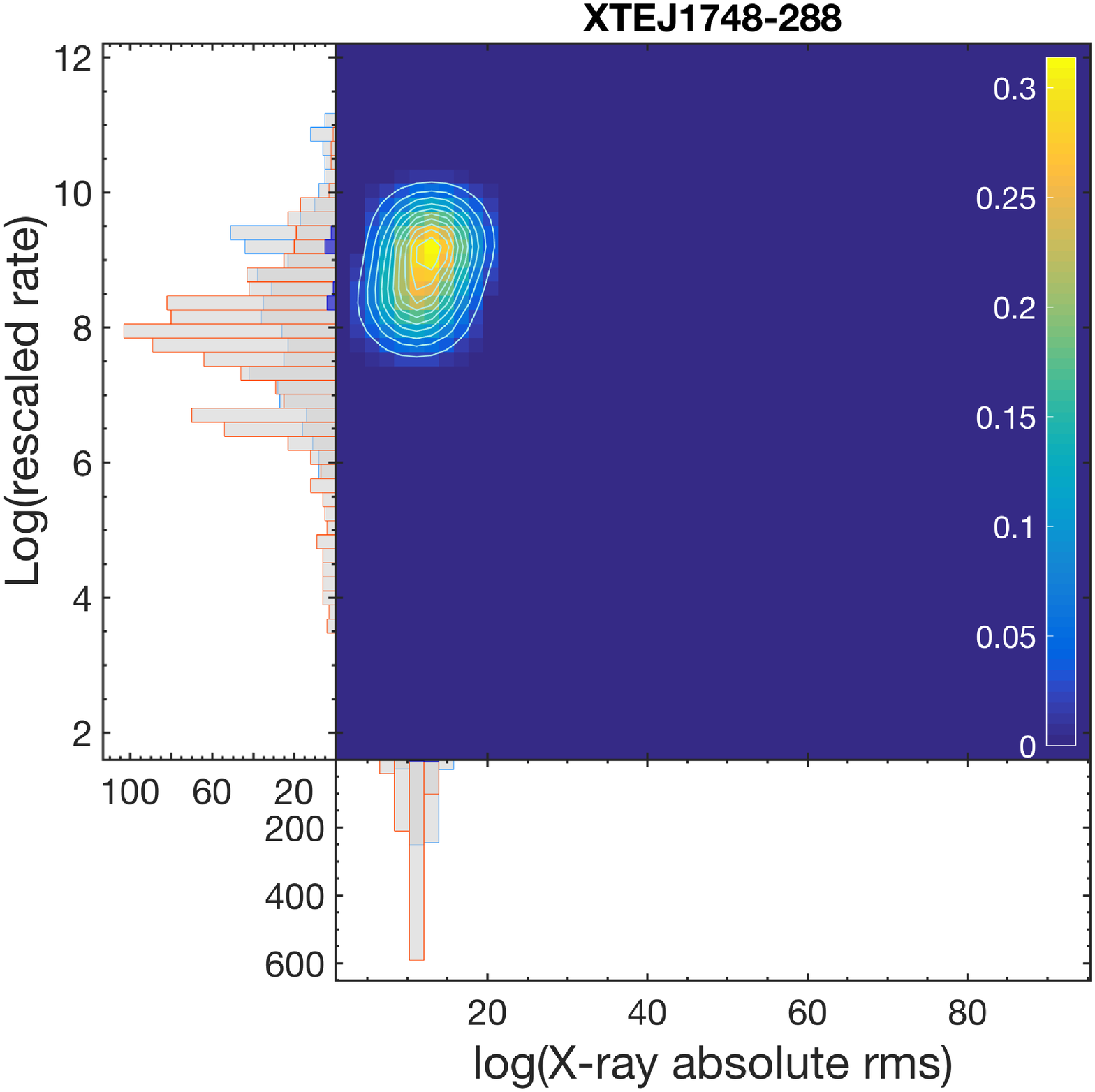} & 
\includegraphics[width=0.3\textwidth]{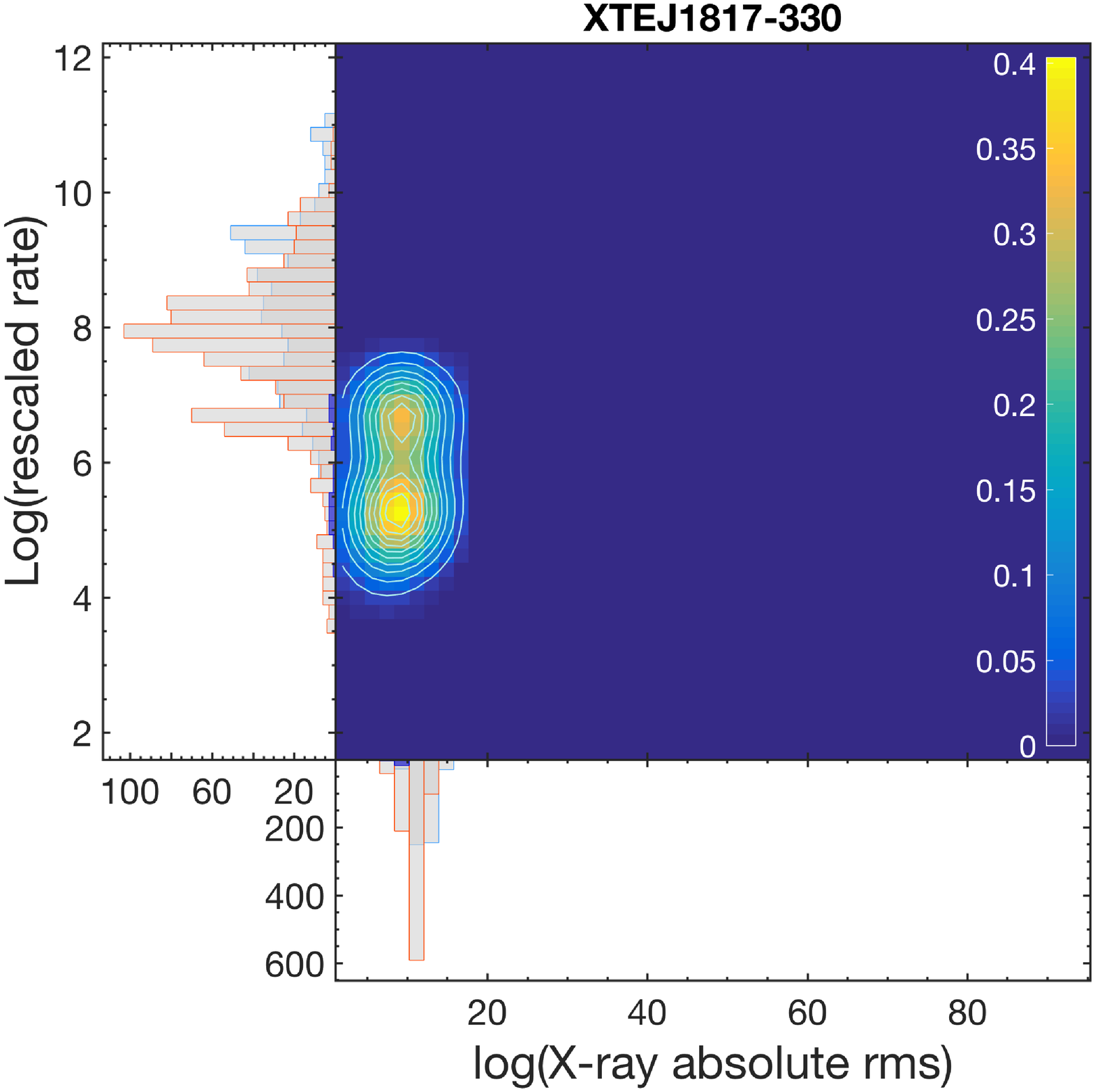} \\ 

\end{tabular}
\end{figure}

\newpage

\begin{figure*}
\centering
\begin{tabular}{c c c}
\includegraphics[width=0.3\textwidth]{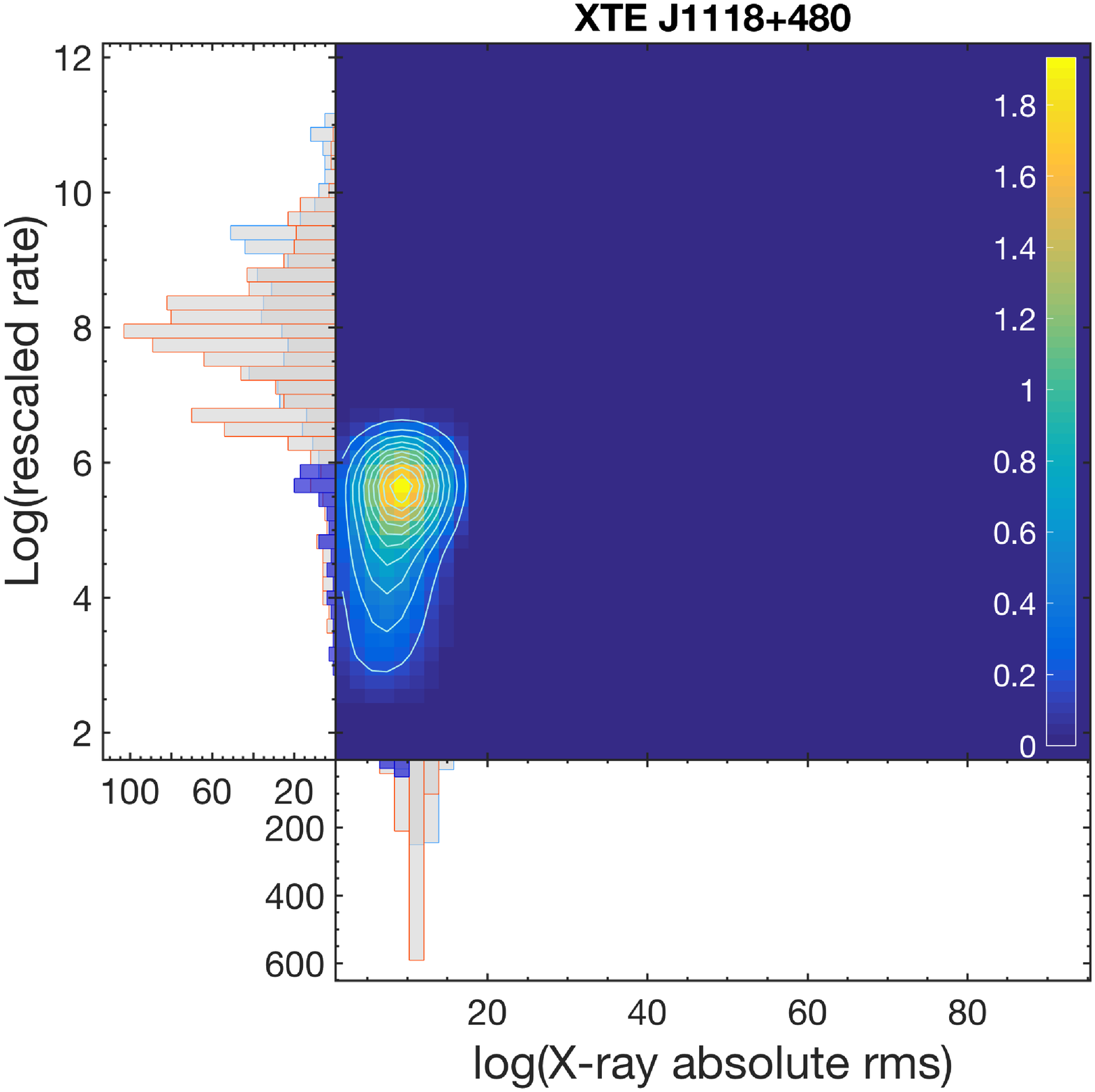} & 
\includegraphics[width=0.3\textwidth]{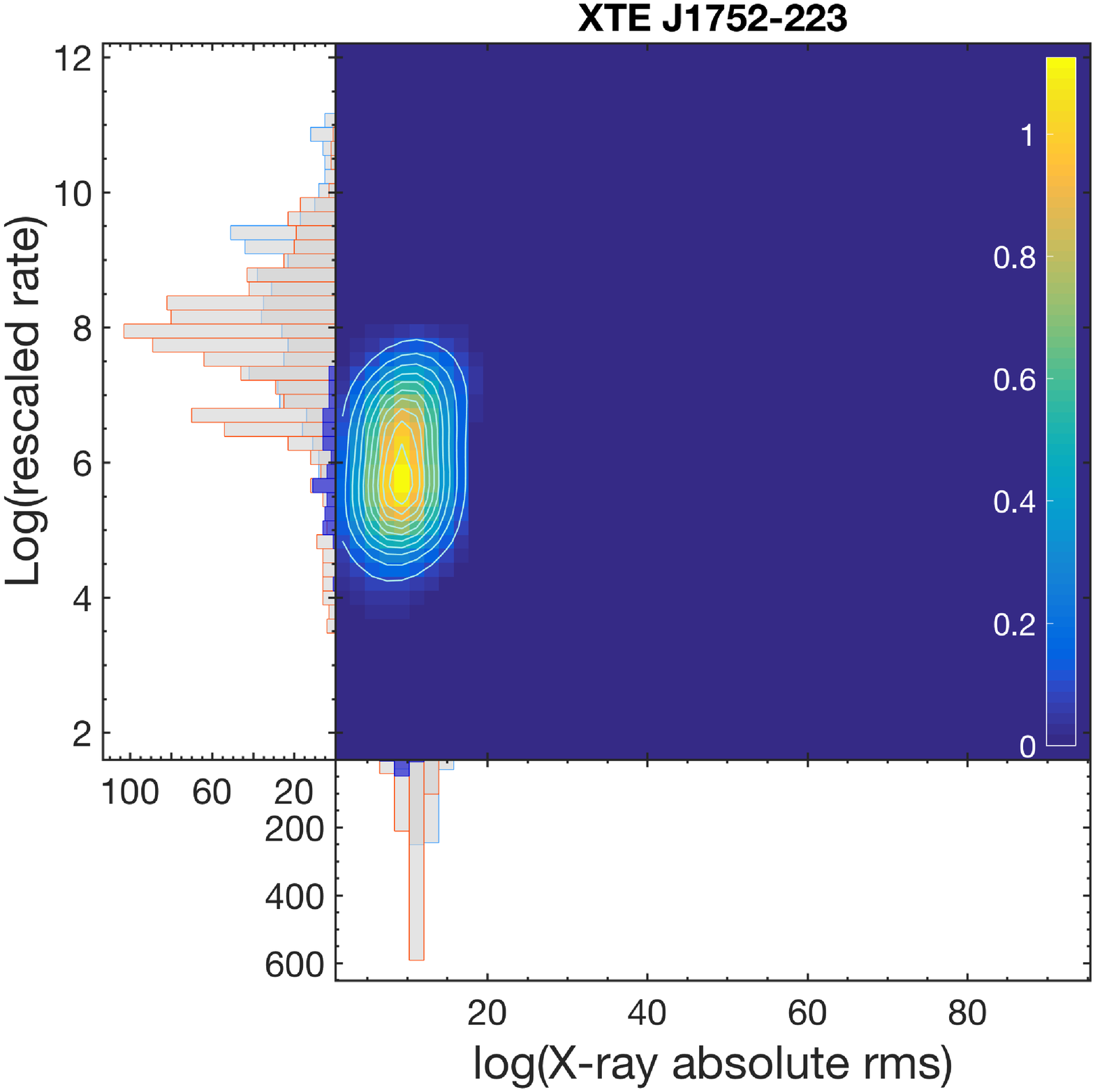} & 
\includegraphics[width=0.3\textwidth]{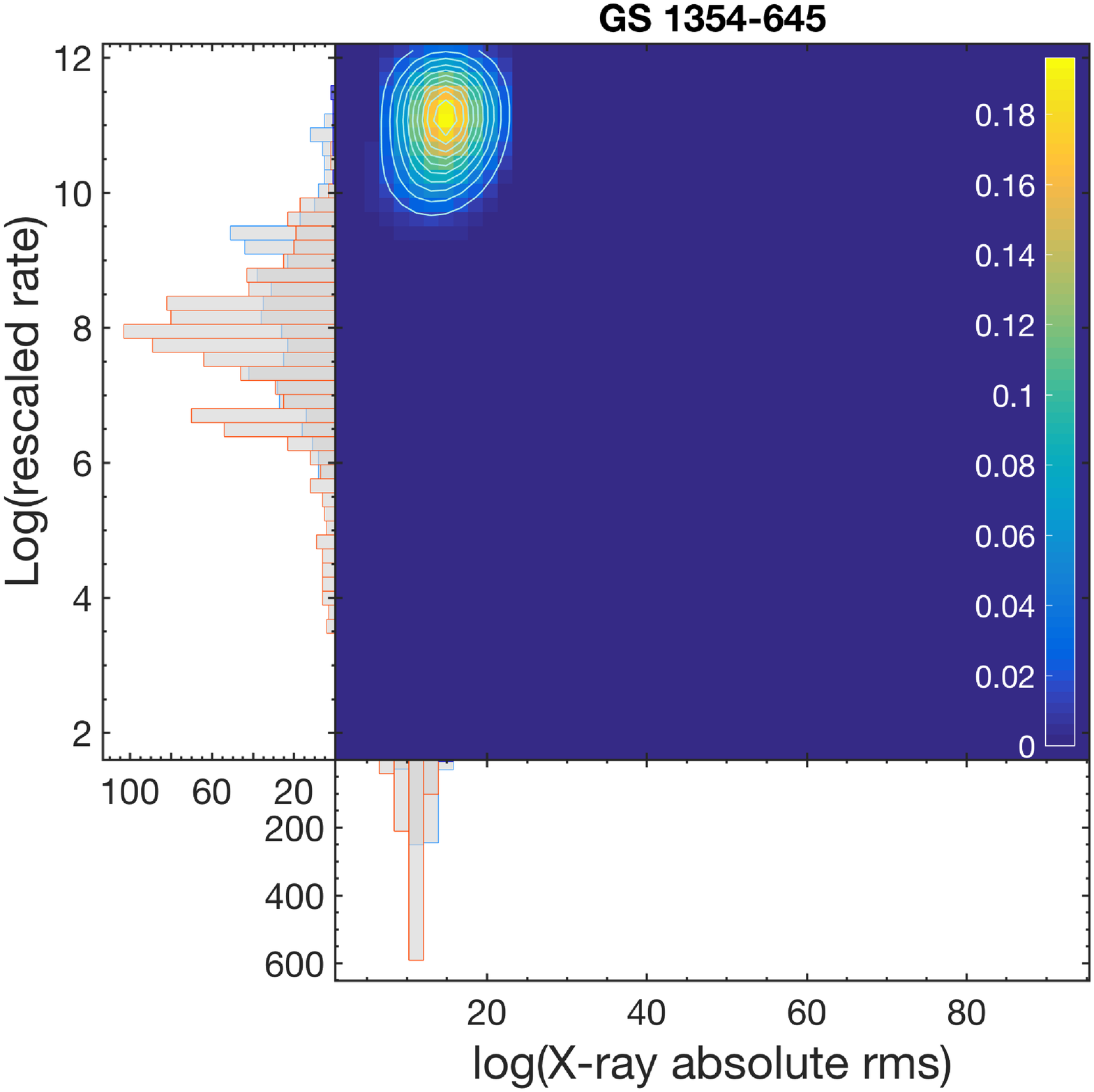} \\

\includegraphics[width=0.3\textwidth]{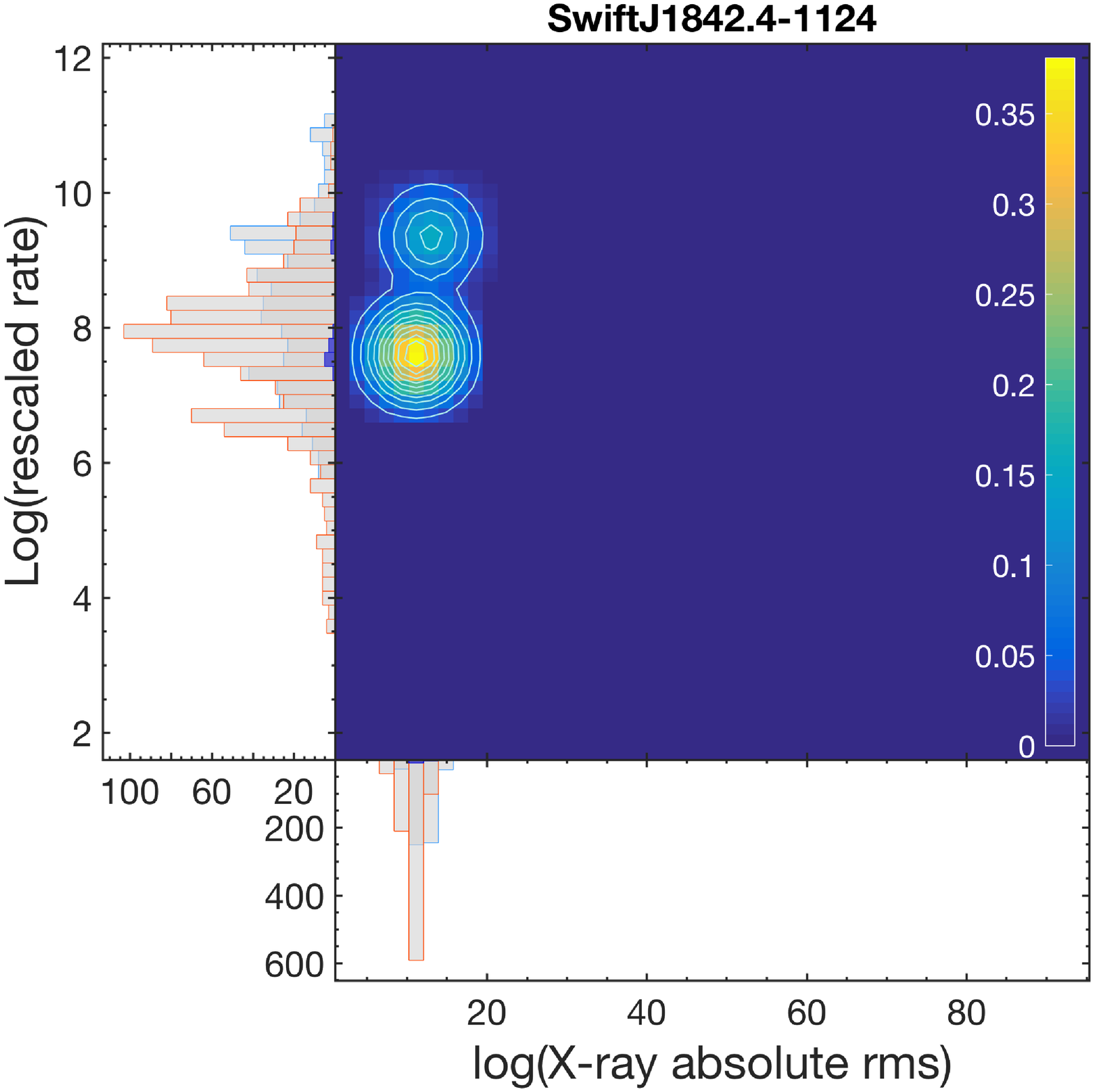} & 
\includegraphics[width=0.3\textwidth]{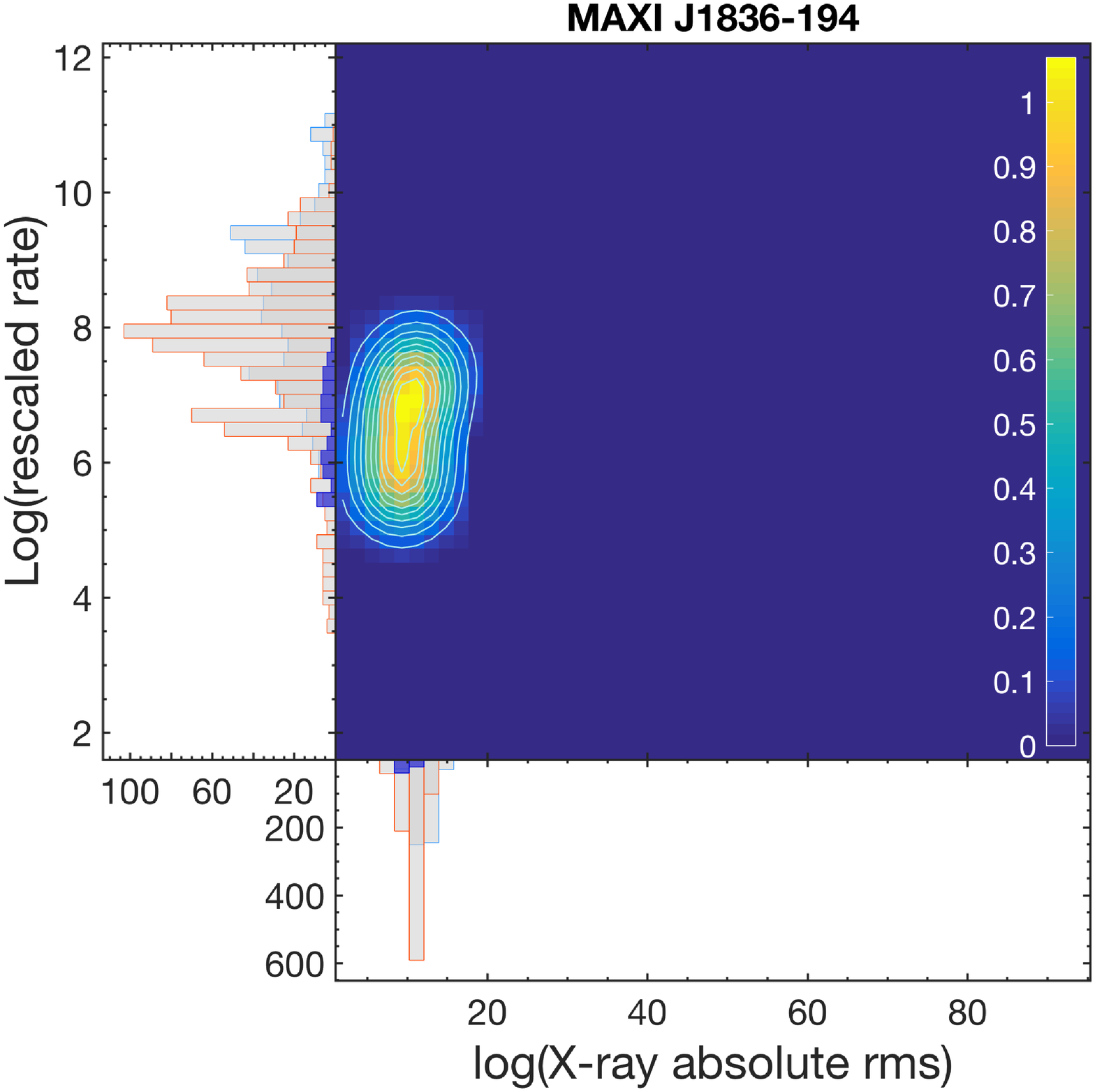} &
\includegraphics[width=0.3\textwidth]{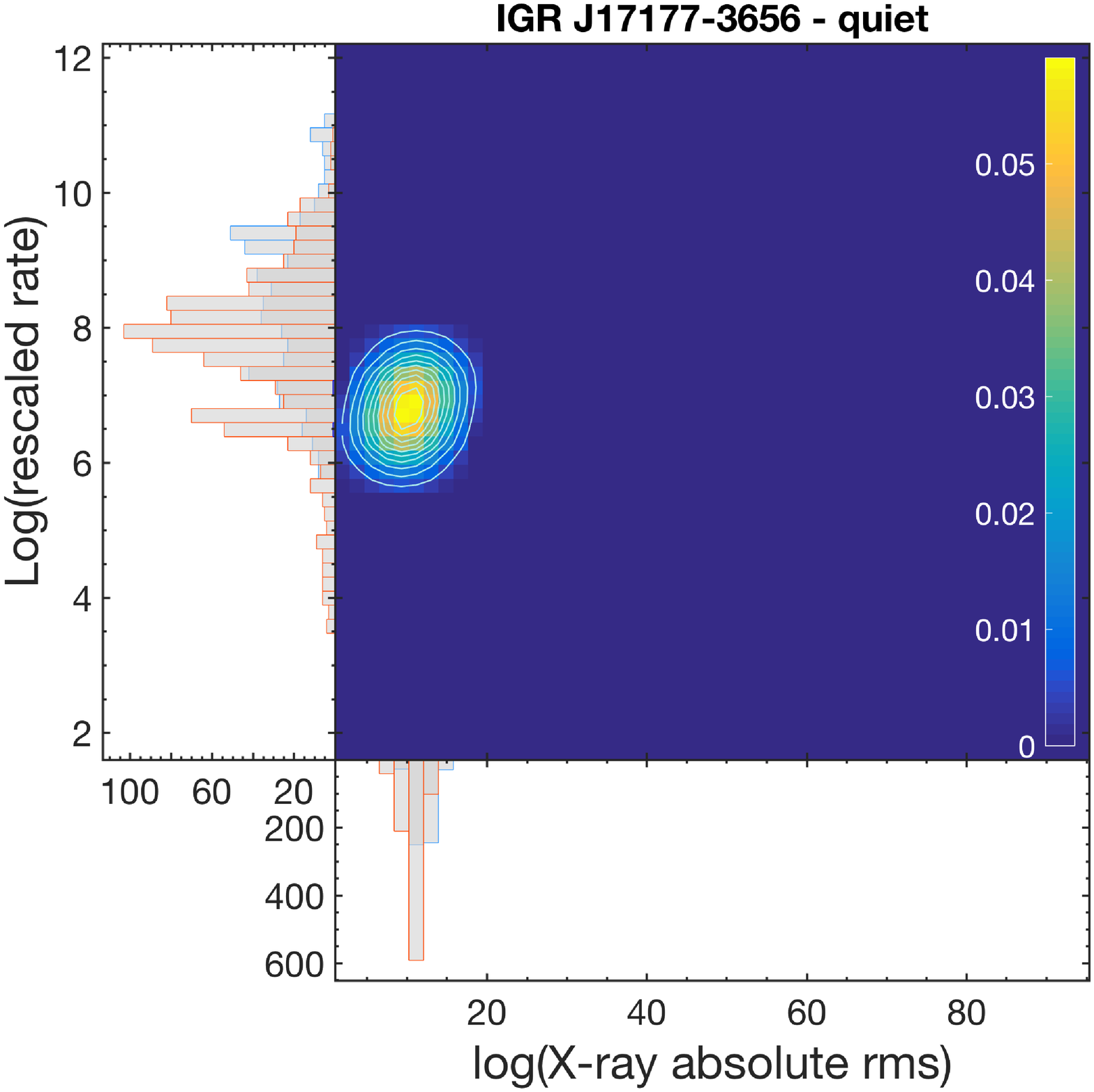} \\

\end{tabular}
\end{figure*}

\begin{figure*}
\centering
\caption{Rms-Intensity diagrams, showing the hard-line(s). The dashed lines marks constant levels of fractional rms, from top left to bottom right at 1\%, 5\%, 10\% and up to 100\% at steps of 10\%.}
\label{fig:individual_hardlines}
\begin{tabular}{c c c}
\includegraphics[width=0.3\textwidth]{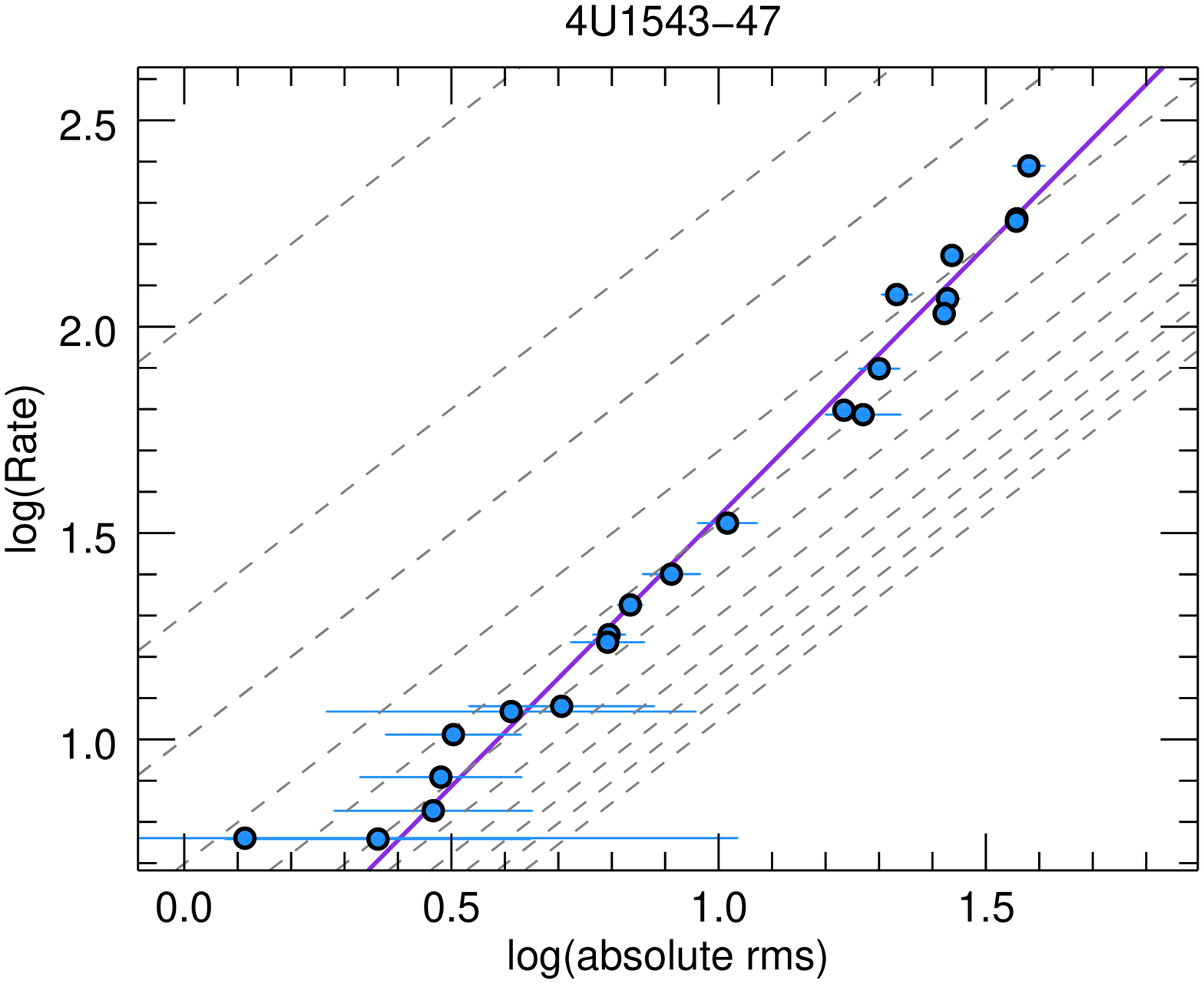} &
\includegraphics[width=0.3\textwidth]{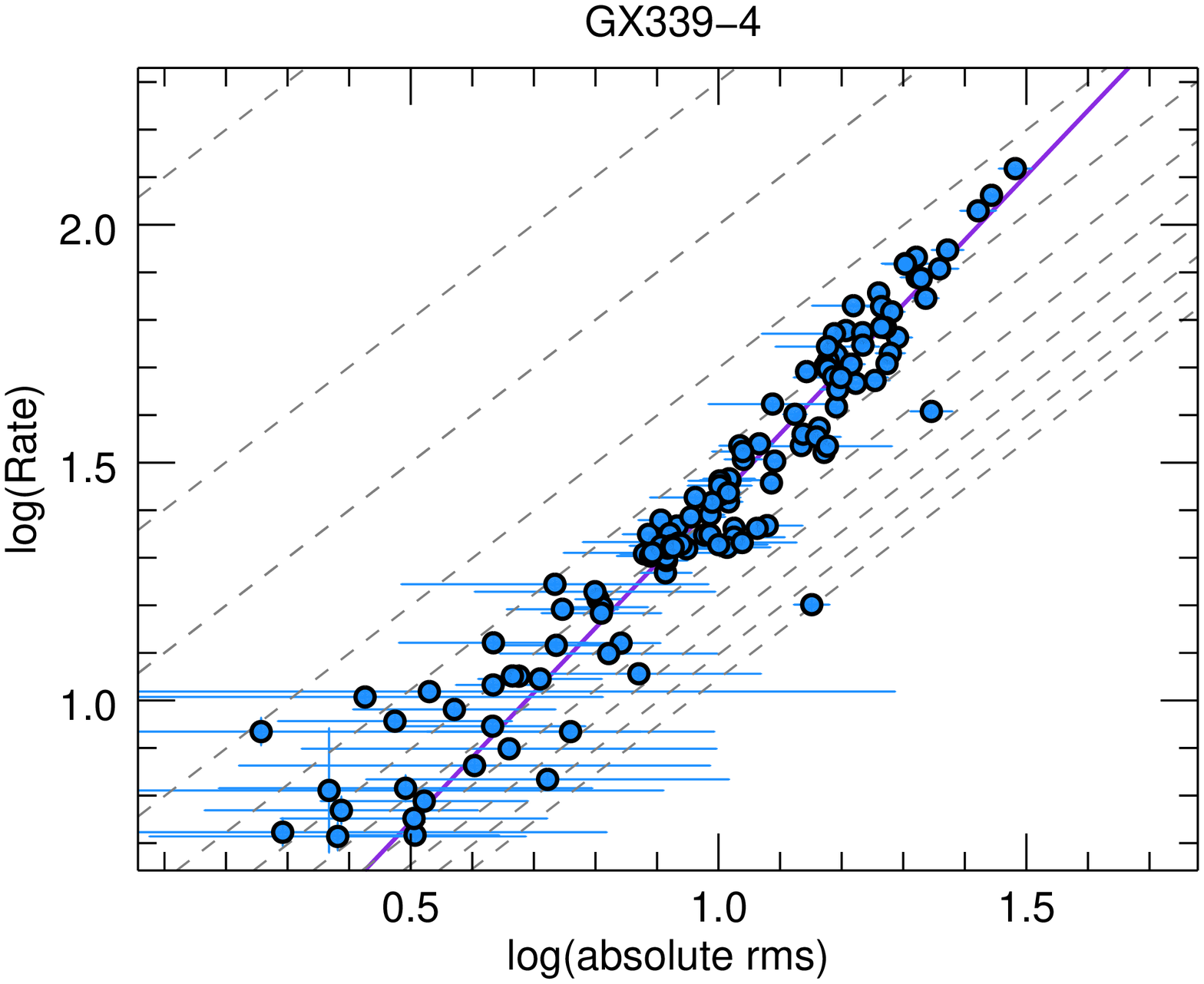} &
\includegraphics[width=0.3\textwidth]{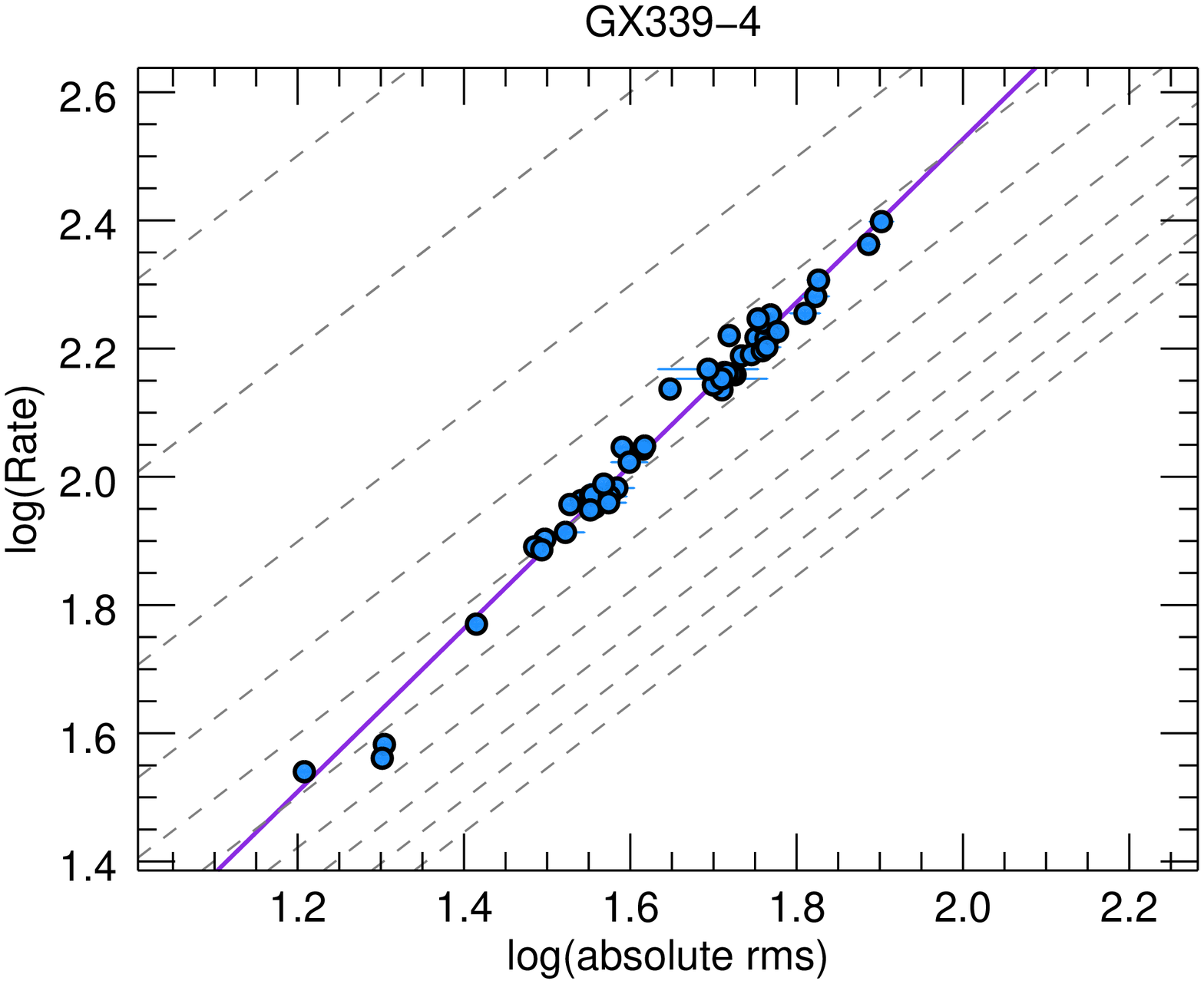} \\

\includegraphics[width=0.3\textwidth]{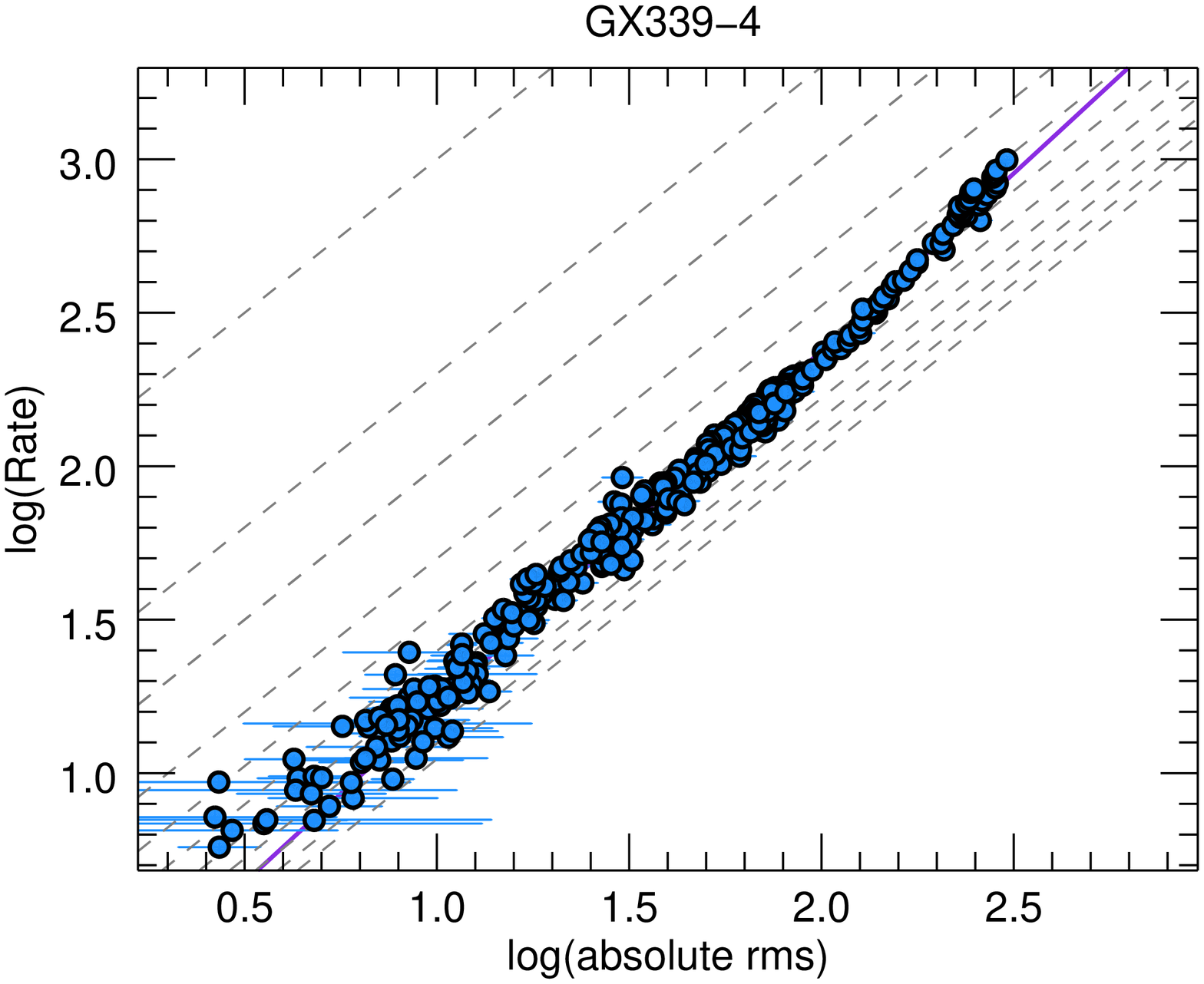} &
\includegraphics[width=0.3\textwidth]{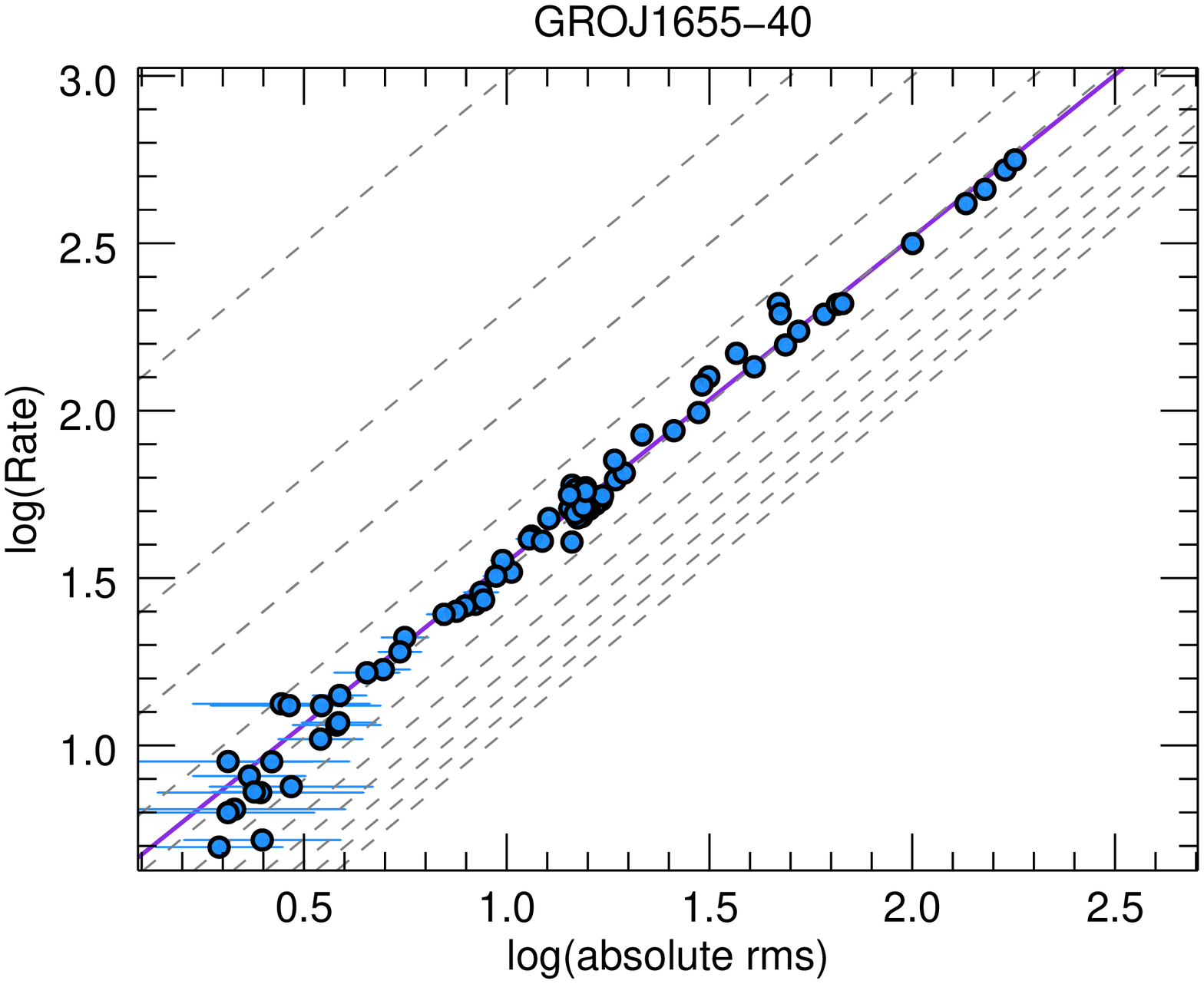} &
\includegraphics[width=0.3\textwidth]{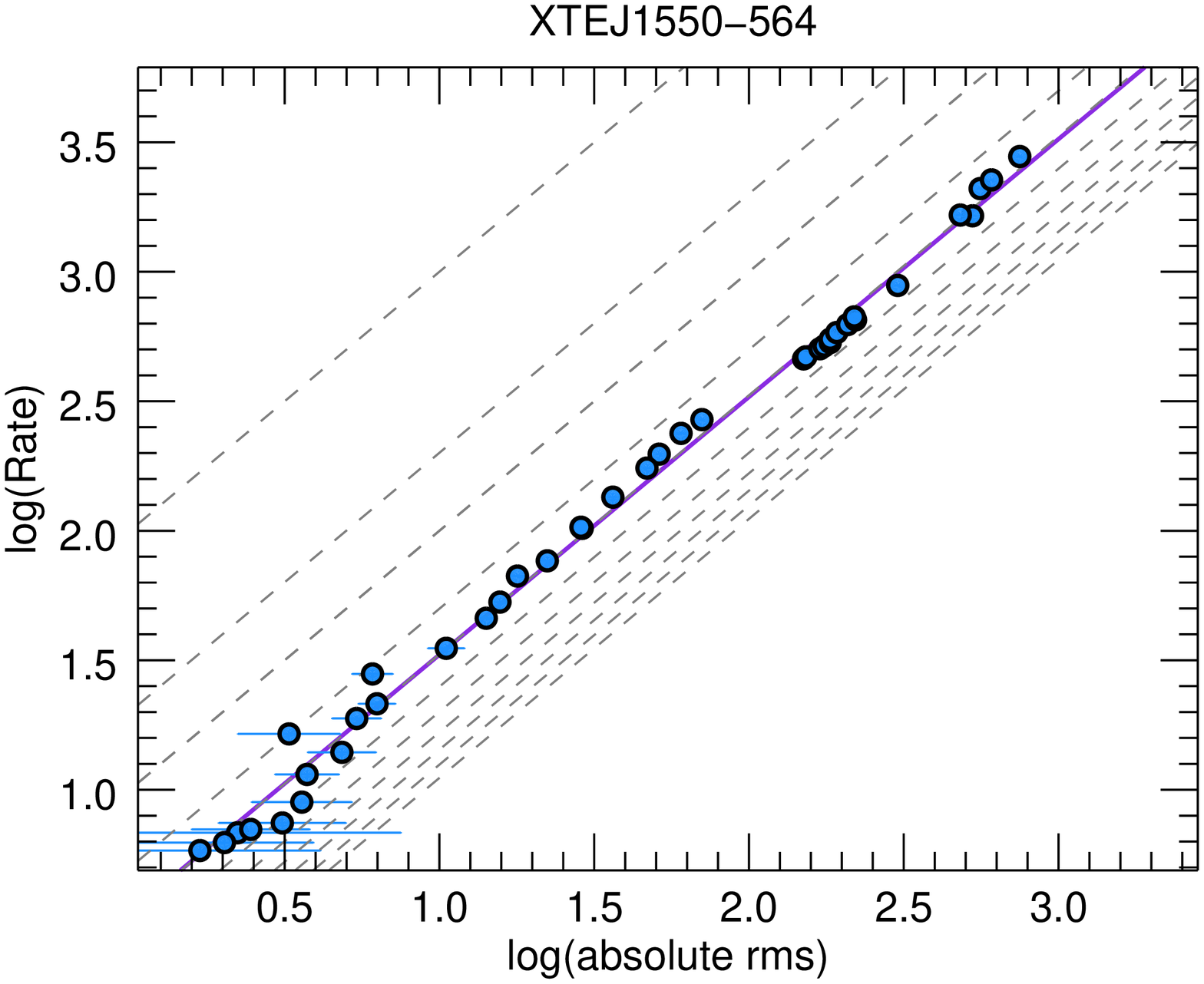} \\
 
\includegraphics[width=0.3\textwidth]{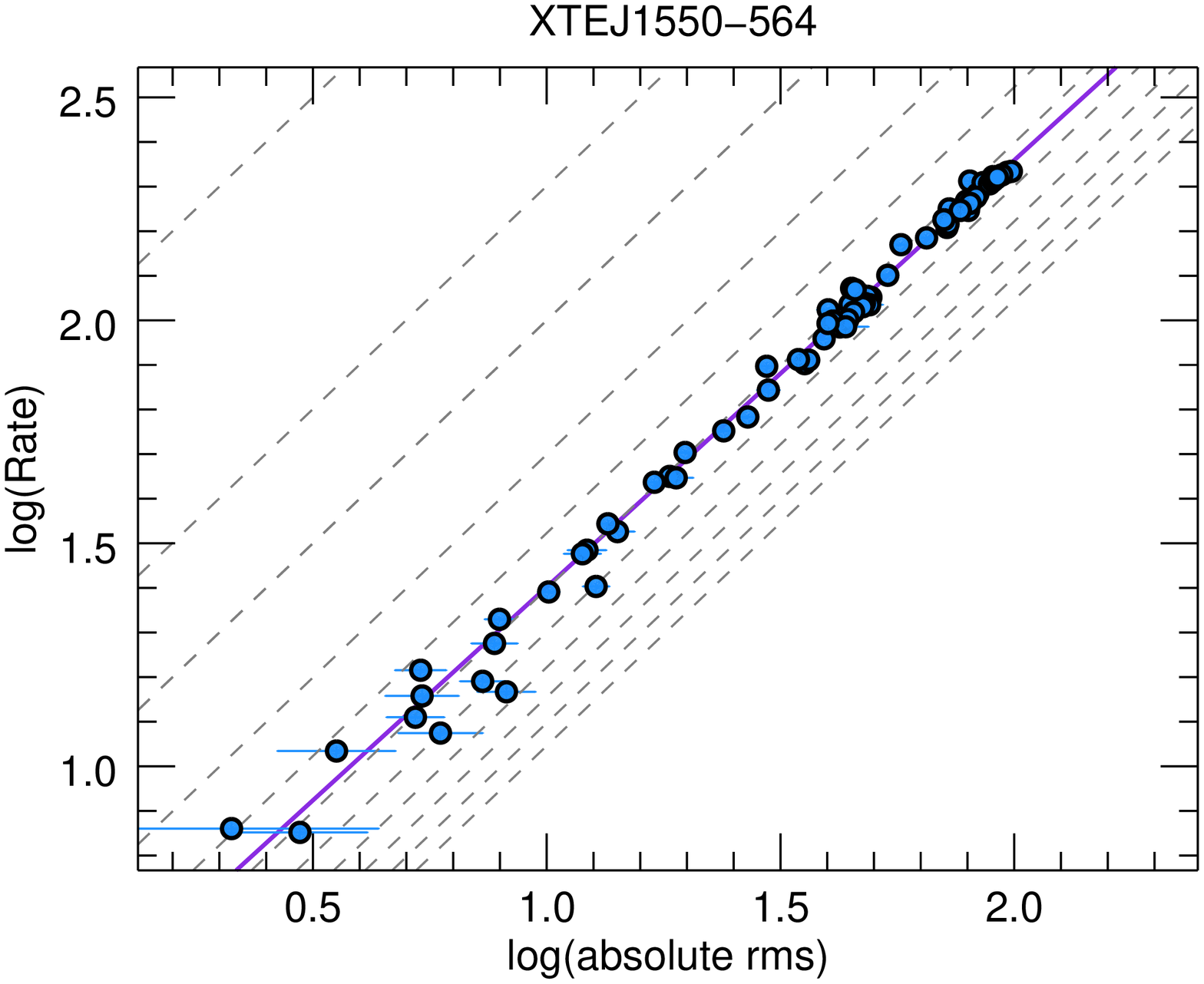} & 
\includegraphics[width=0.3\textwidth]{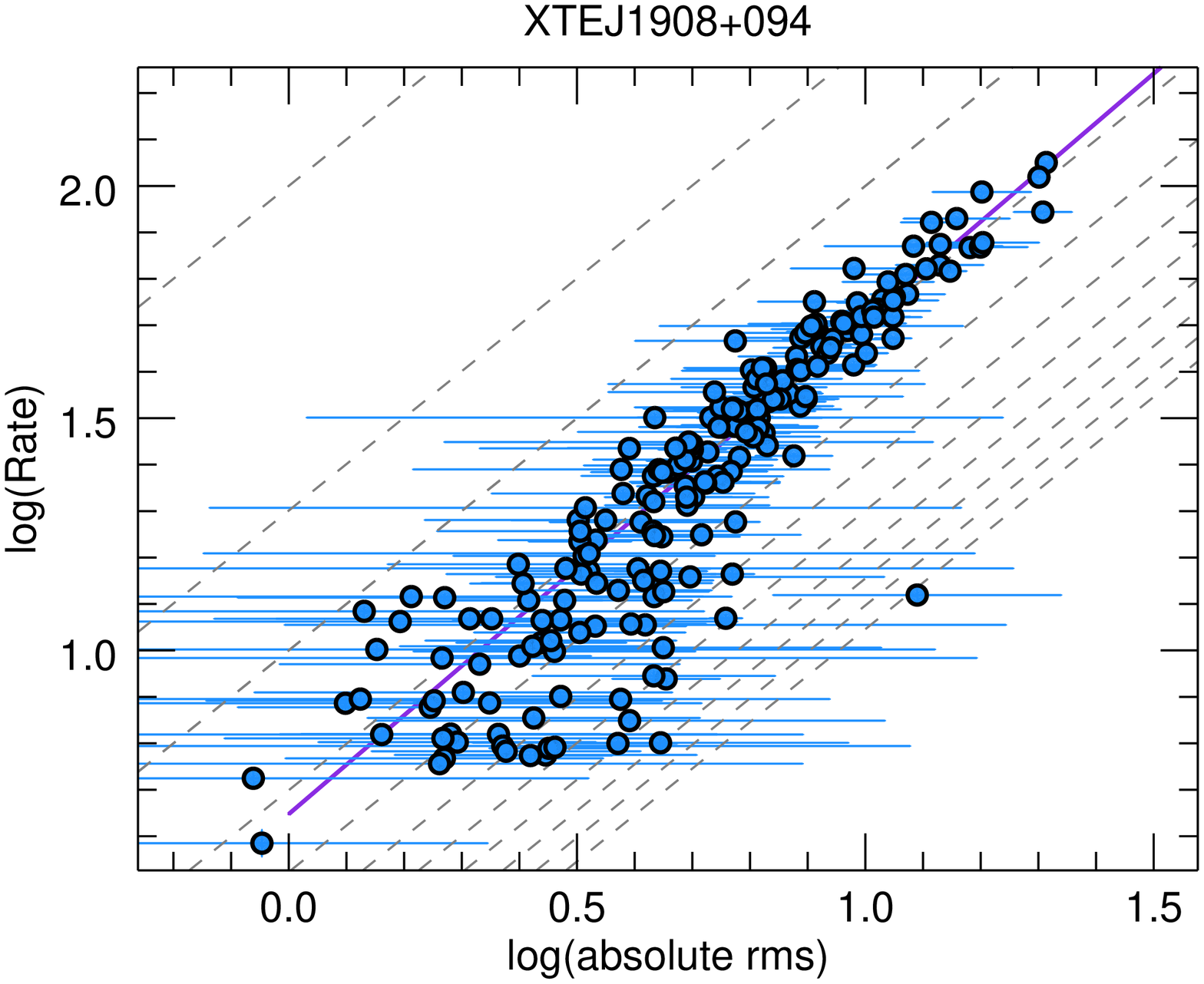} &  
\includegraphics[width=0.3\textwidth]{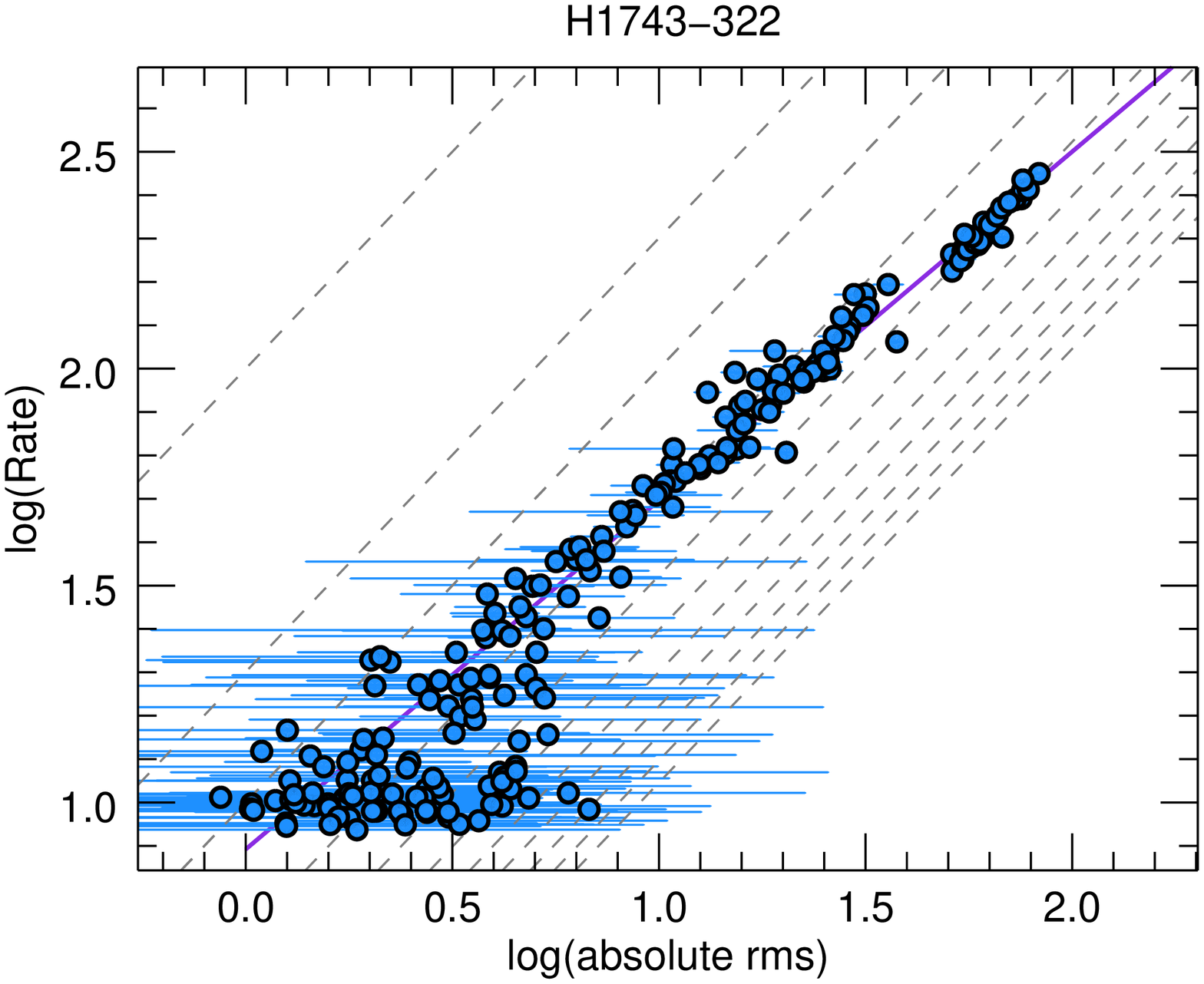} \\

\includegraphics[width=0.3\textwidth]{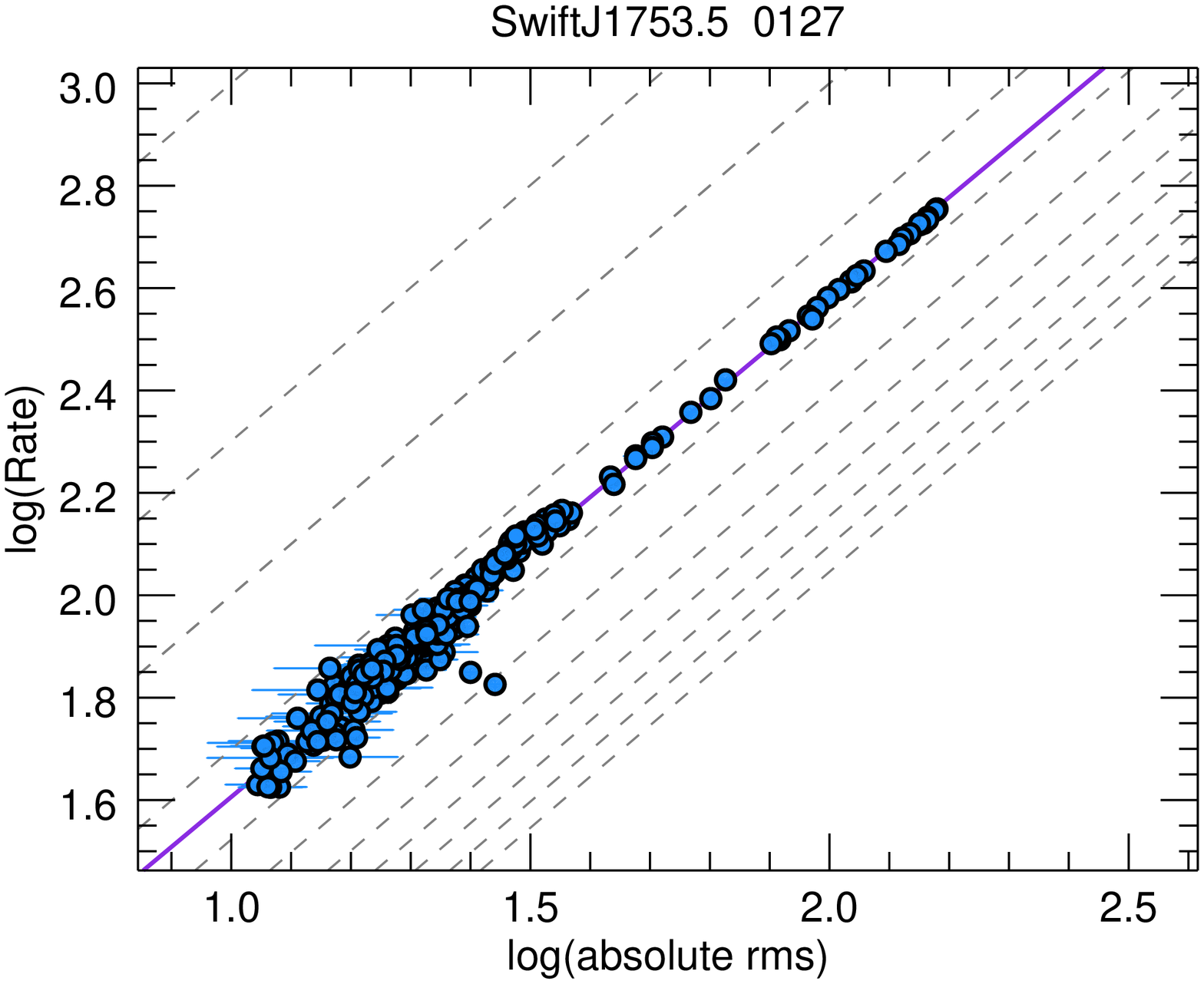} & 
\includegraphics[width=0.3\textwidth]{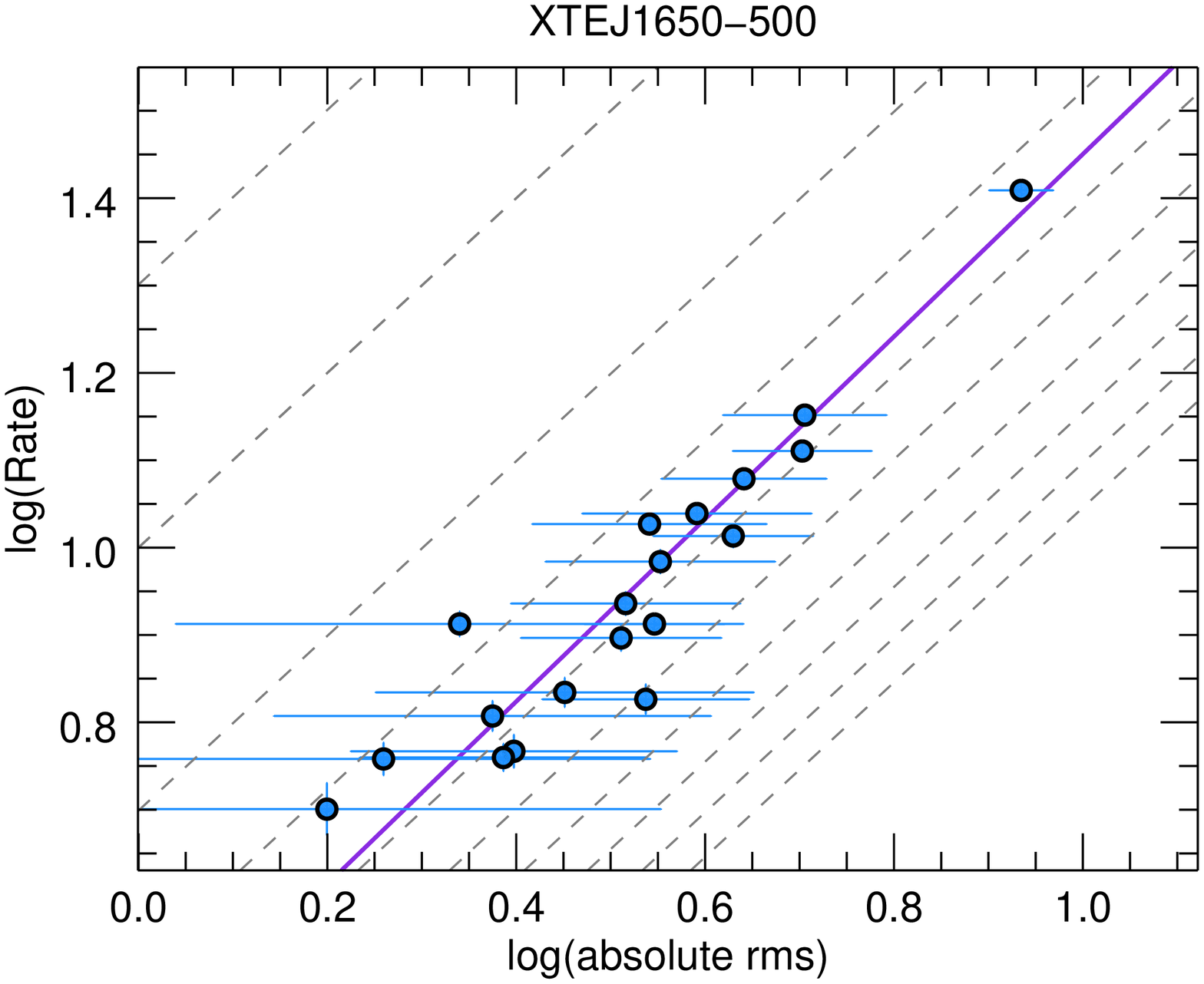} &
\includegraphics[width=0.3\textwidth]{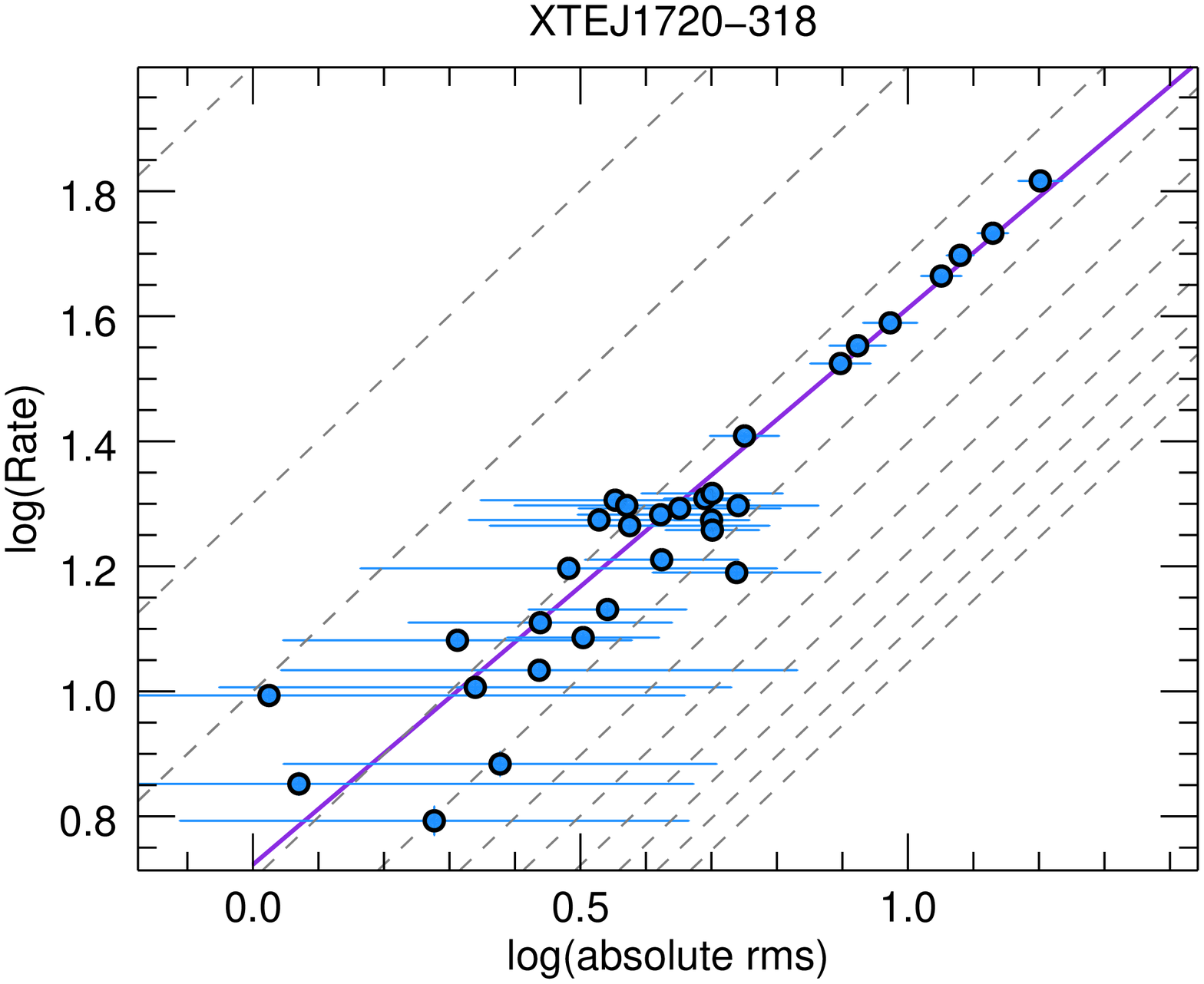} \\
 
\includegraphics[width=0.3\textwidth]{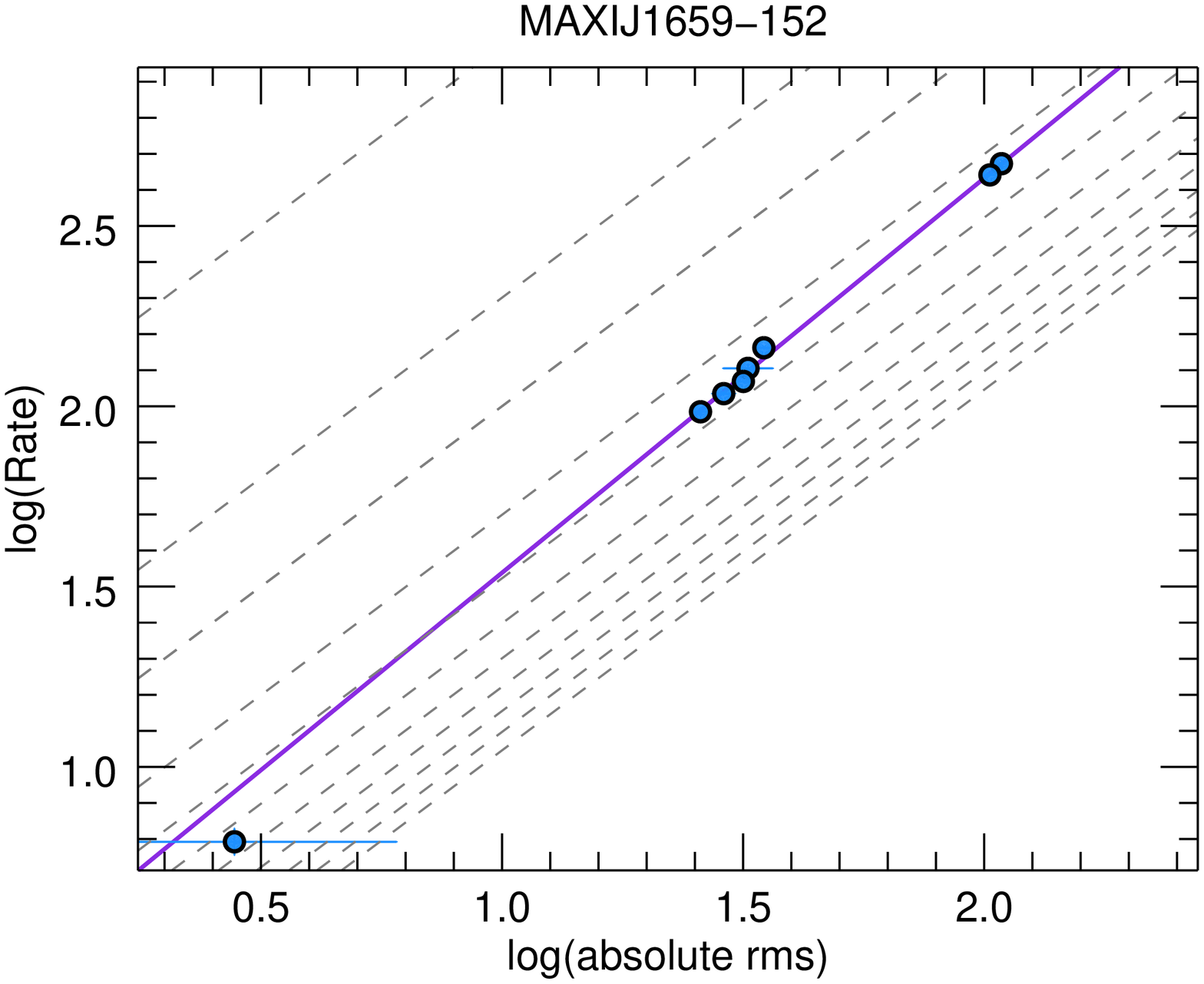} & 
\includegraphics[width=0.3\textwidth]{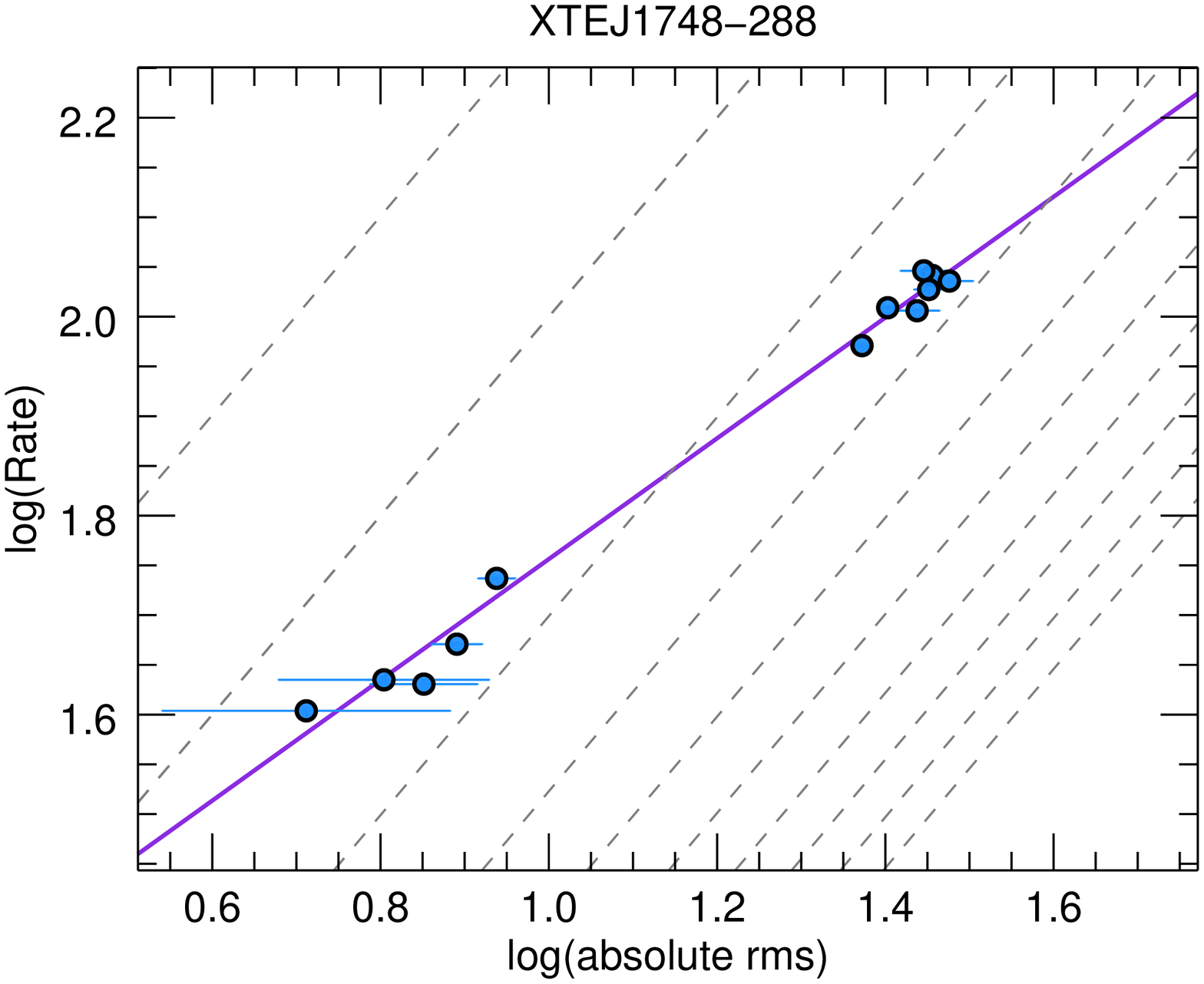} & 
\includegraphics[width=0.3\textwidth]{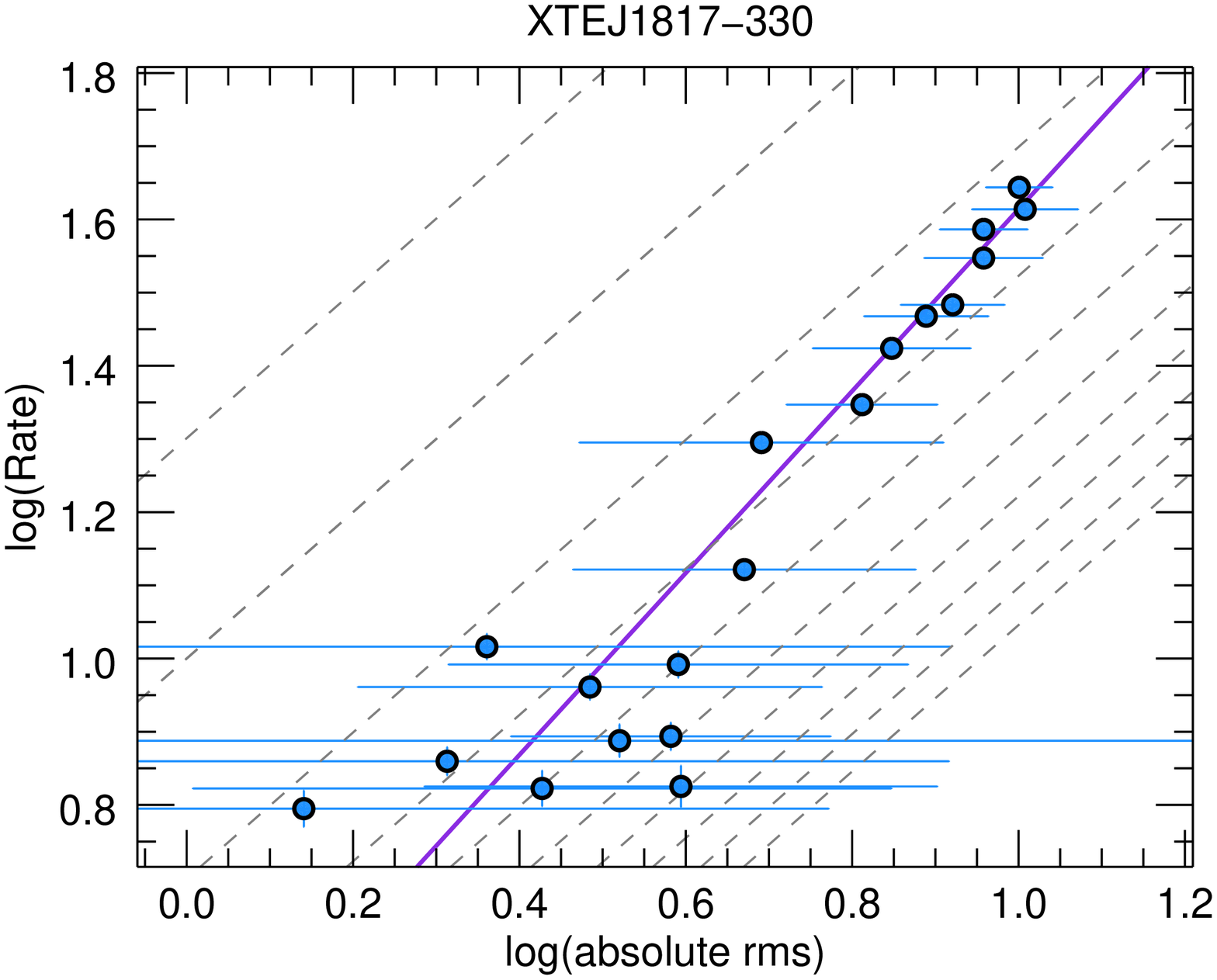} \\ 
\end{tabular}
\end{figure*}

\begin{figure*}
\centering
\begin{tabular}{c c c}
\includegraphics[width=0.3\textwidth]{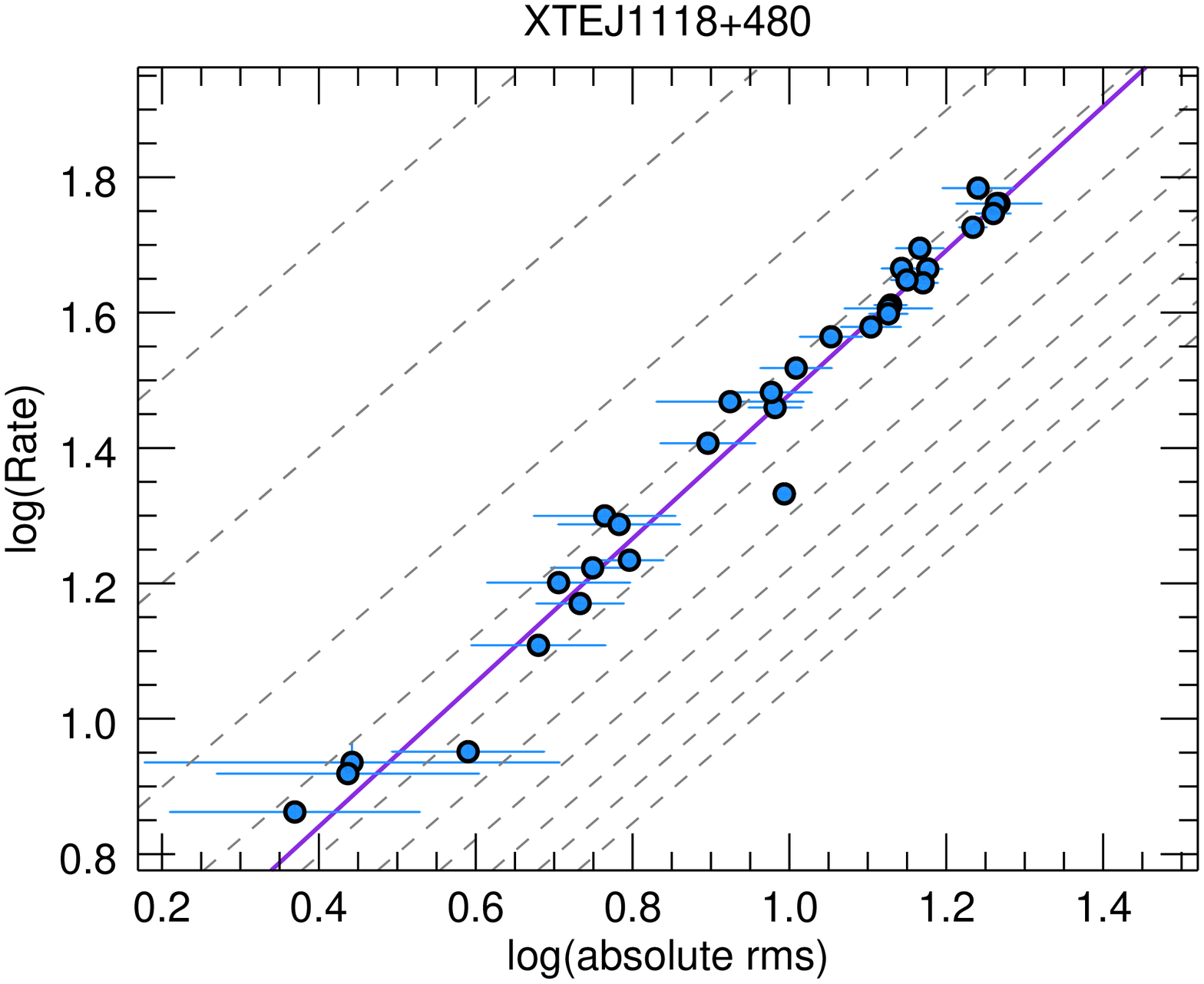} & 
\includegraphics[width=0.3\textwidth]{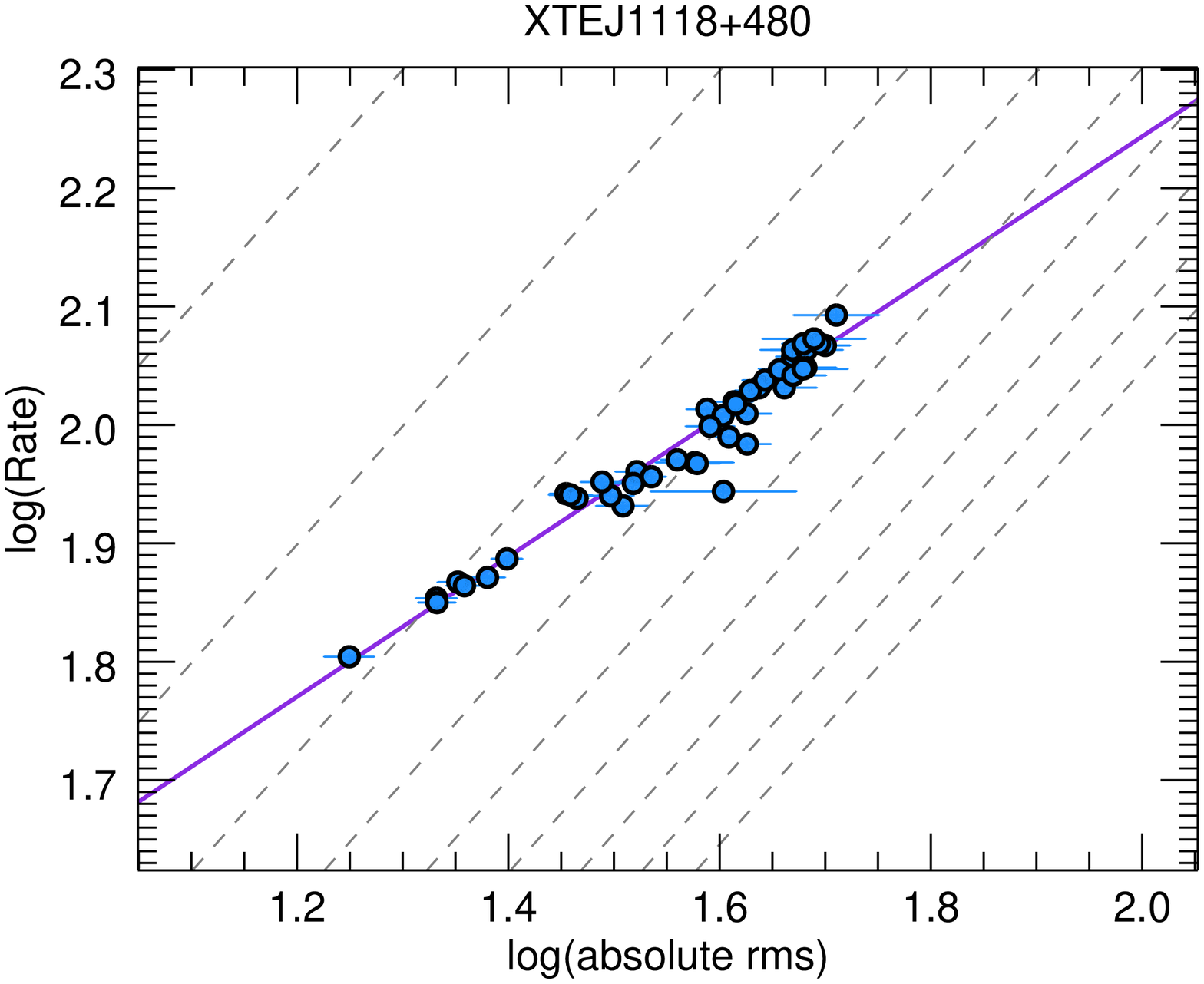} &
\includegraphics[width=0.3\textwidth]{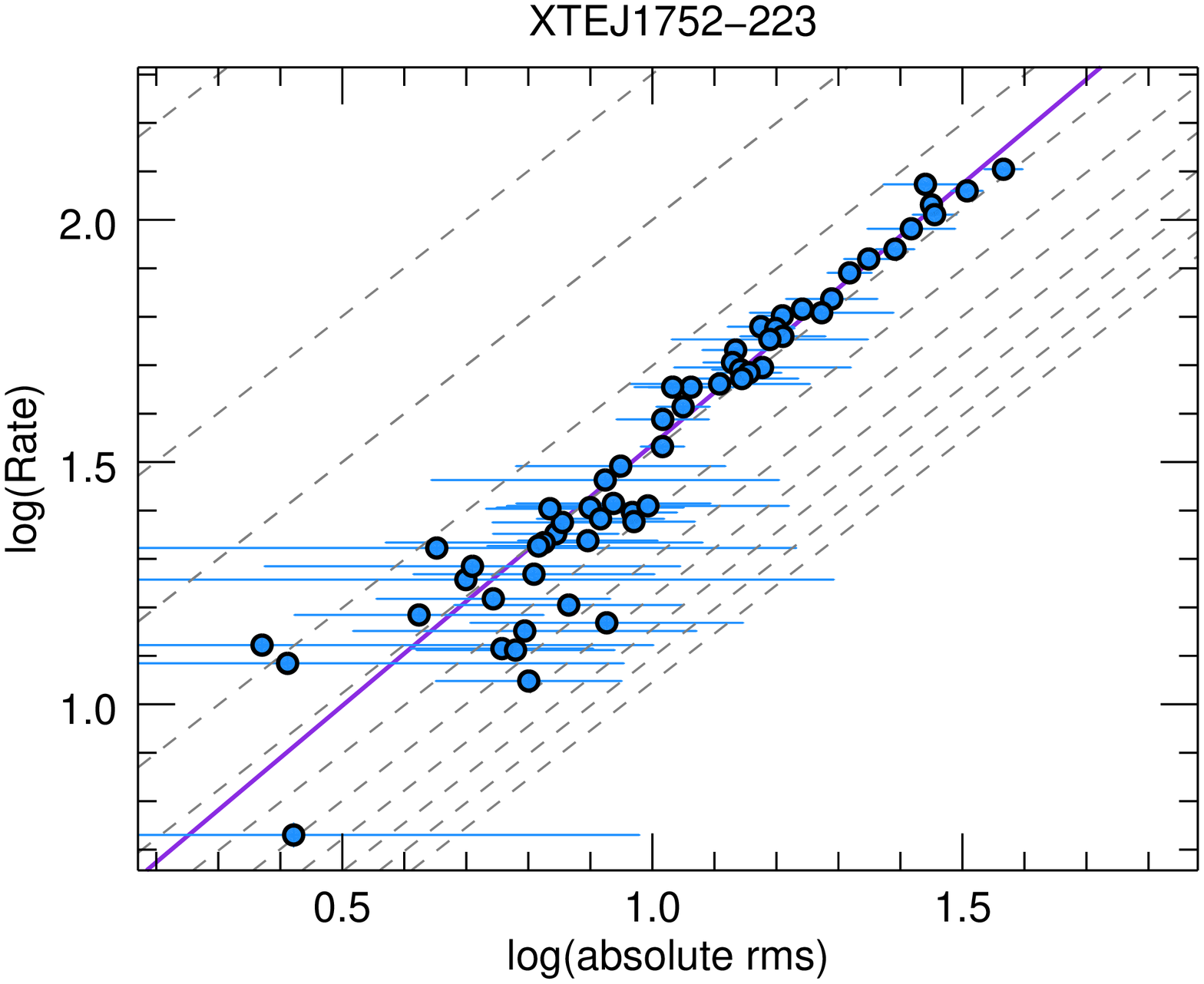} \\

\includegraphics[width=0.3\textwidth]{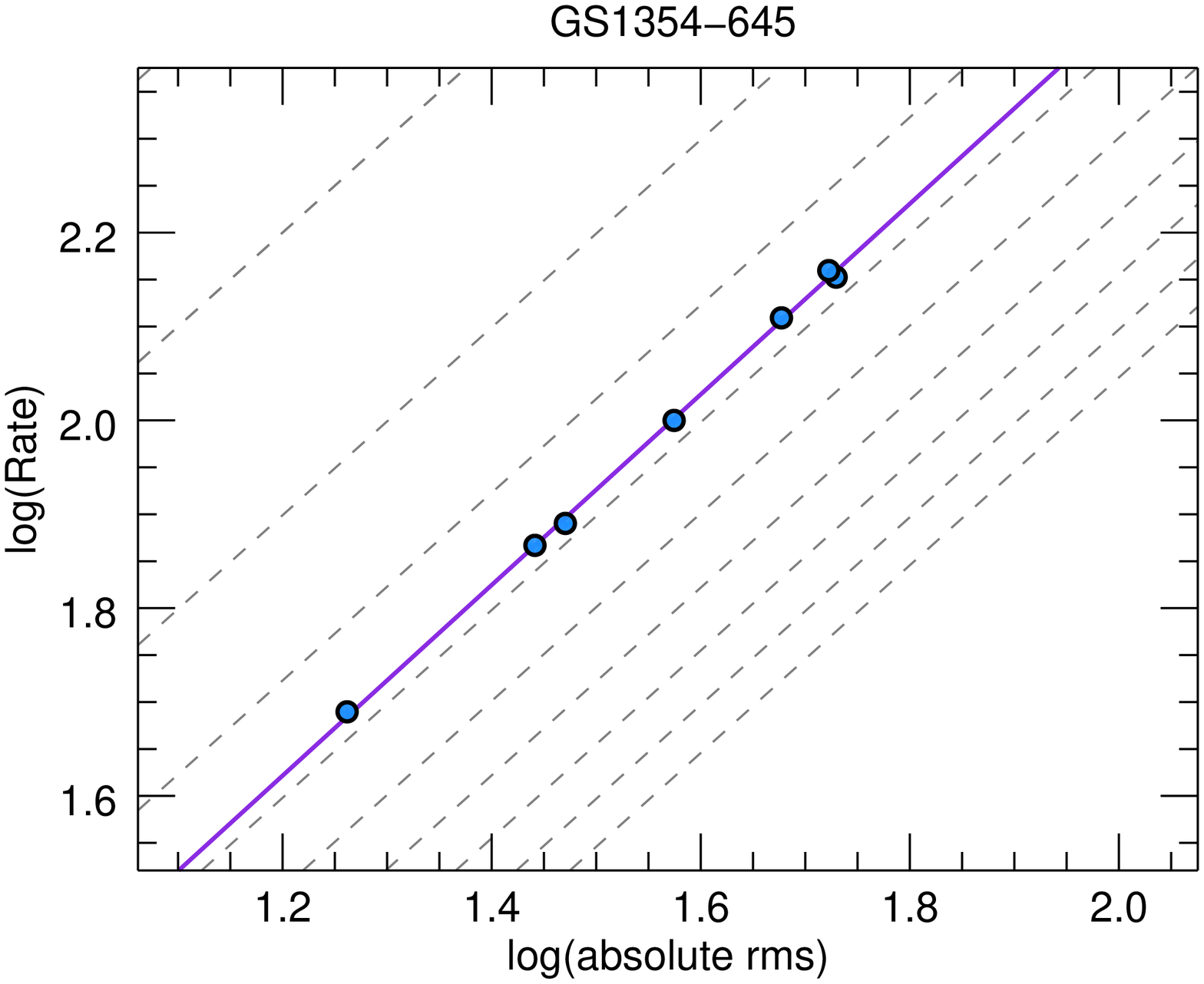} &
\includegraphics[width=0.3\textwidth]{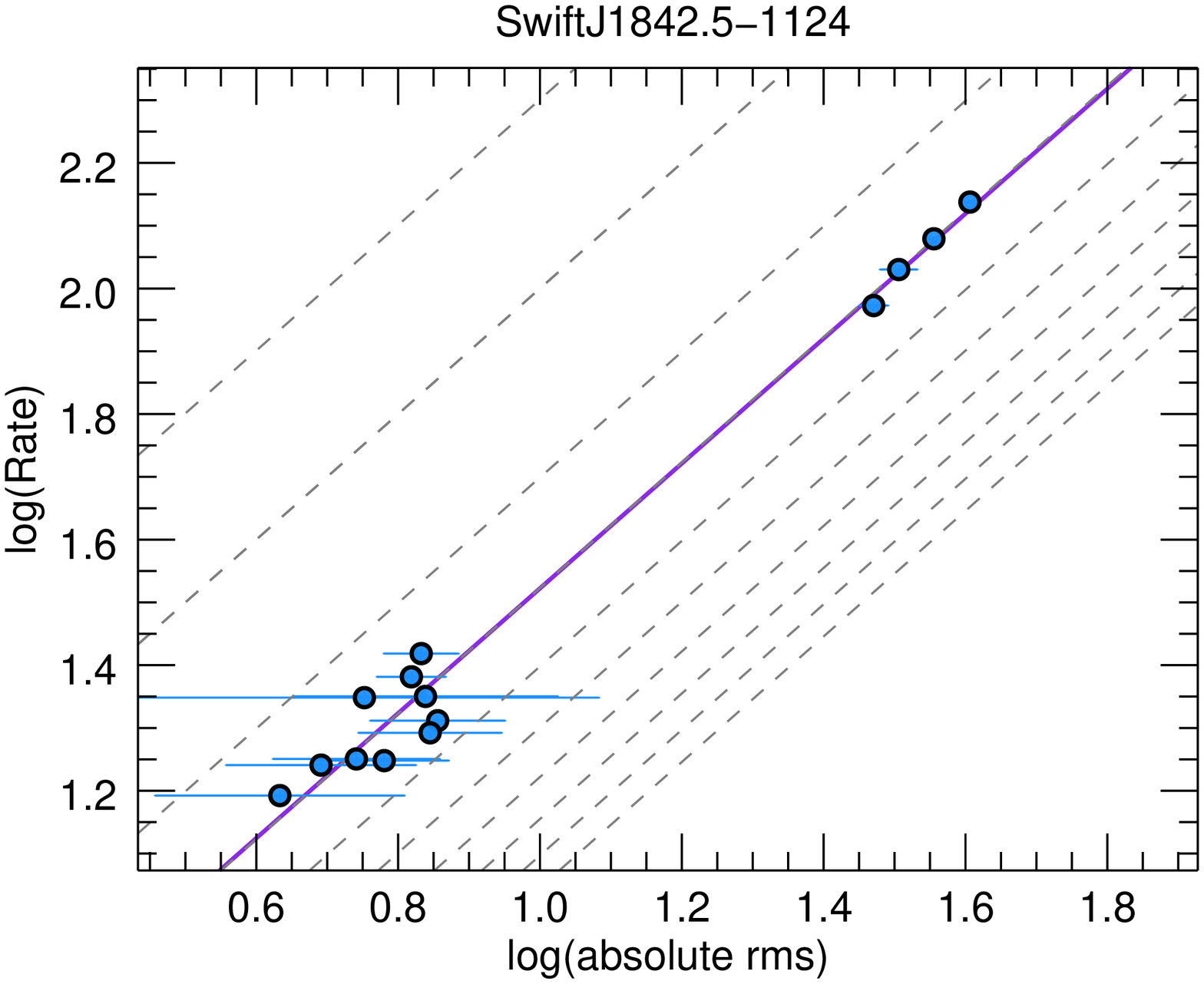}  &
\includegraphics[width=0.3\textwidth]{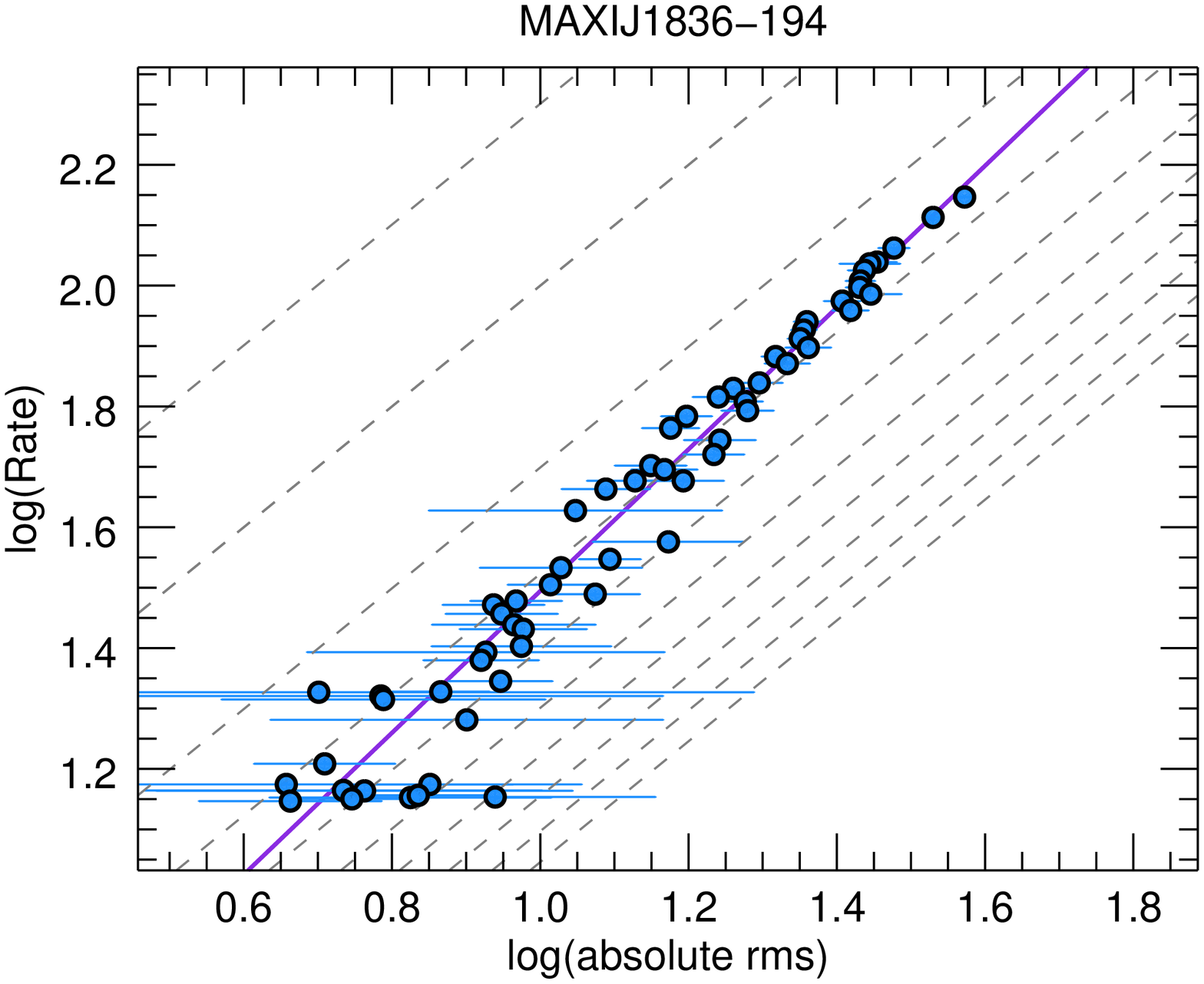} \\
\end{tabular}
\end{figure*}

\label{lastpage}
\end{document}